\documentclass[a4paper,12pt]{article}
\usepackage{subfigure}
\usepackage{wrapfig}
\usepackage[utf8]{inputenc}
\usepackage{authblk}
\usepackage{amsmath}
\usepackage{amssymb}
\usepackage{amsfonts}
\usepackage{mathtools}
\usepackage{chngcntr}
\usepackage{graphicx}
\usepackage{parallel}
\usepackage{lipsum}
\usepackage{multicol}
\usepackage[paperwidth=8.5in, paperheight=11in,top=2.5cm,bottom=2.5cm,left=2.5cm,right=2.5cm]{geometry}
\usepackage{multirow}
\usepackage[table]{xcolor}
\usepackage{natbib}
\usepackage{listings}
\usepackage{lscape}
\usepackage{caption}
\usepackage{ulem}
\usepackage[title]{appendix}
\usepackage[doublespacing]{setspace}

\title{Measuring price impact and information content of trades in a time-varying setting}

\author[1]{Francesco Campigli}
\author[2]{Giacomo Bormetti}
\author[3,*]{Fabrizio Lillo}

\affil[1]{\footnotesize Scuola Normale Superiore, Pisa, Italy. Email: francesco.campigli@sns.it}
\affil[2]{Department of Economics and Management and CAM-Risk, University of Pavia, Via San Felice al Monastero 5, 27100, Pavia, Italy. Email: giacomo.bormetti@unipv.it}
\affil[3]{Scuola Normale Superiore, Pisa, Italy. Email: fabrizio.lillo@sns.it}

\date{}

\begin{document}

\maketitle
\begingroup
\renewcommand{\thefootnote}{\fnsymbol{footnote}}
\footnotetext[1]{The authors thank the participants of the Sofie 2023 Conference (Seoul), the International Association for Applied Econometrics Conference - IAAE 2023 (Oslo), and 1st CAM-Risk Conference 2024 (Pavia) for useful comments and suggestions. GB and FL acknowledge financial support from the Italian Ministry MUR under the PRIN project \textit{Dynamic models for a fast-changing world: An observation driven approach to time-varying parameters} (grant agreement no. 20205J2WZ4). The authors thank Dr. German Rodikov, Dipartimento di Matematica, Universit\`a di Bologna german.rodikov@unibo.it for the important support to this work.}
\endgroup

\begin{abstract}
We propose a non-linear observation-driven version of the Hasbrouck vector auto-regression framework for dynamically estimating trades' market impact and information content. We find that market impact displays an intraday pattern superimposed with large fluctuations. Some of them are exogenous, and, as an example, we investigate market impact dynamics around FOMC announcements. Unlike the vector autoregression results, we reveal that the information content of the trades depends on the local liquidity level and the recent history of prices and trades. Finally, we use the model to estimate the time-varying permanent impact parameter, which allows performing a dynamic transaction cost analysis.
\end{abstract}



\section{Introduction}
\label{section:intro}

One of the most important challenges in market microstructure is the modeling and empirical estimation of how information is impounded into prices (\cite{o2003presidential}). Financial markets are characterized by an asymmetry between informed traders and uninformed liquidity providers. A series of seminal stylized models, starting from those of \cite{glosten1985bid} and \cite{kyle1985continuous}, clarified that traders' information is slowly incorporated into prices through trading, which progressively reveals information to the market and leads to changes in prices and in the order flow. Thus, the price discovery process depends on how prices react to trades, an important quantity termed {\it price impact}, and on the market's reaction to the past price and trade dynamics. At least two types of price impact emerge from this stylized description: the {\it instantaneous impact}, which describes the immediate price change caused by trade, and the {\it long term impact}, which describes the long-term price change taking into account the subsequent order flows and price changes triggered by the trade. \cite{hasbrouck1991measuring} identified the long-term impact as a measure of the information content of trade. 

Understanding and modeling price impact is also of paramount importance in several financial applications. Market impact is the main source of transaction cost for large investors and optimal execution schemes (\cite{almgren2001optimal}), aiming at minimizing impact costs depending crucially both on the impact model and on the reliability of the estimated model parameters. From an empirical perspective, understanding the price discovery process is tantamount to modeling the joint dynamics of price changes and trades (or, more generally, orders). In the early Nineties, the pioneering work of \cite{hasbrouck1991measuring} paved the way for a modeling framework of the relationships between trades and quotes by proposing a Structural Vector Auto-Regressive (SVAR) model of price changes and trade signs\footnote{It is widely accepted that trade size has a small incremental explanatory power beyond trade sign when estimating the SVAR model (see \cite{hasbrouck2007empirical}).} for studying how prices incorporate information. In the model, the instantaneous impact of the market is described by a parameter that gives the proportionality factor between the sign of the trade at a given time and the contemporaneous price change; the long-term impact, and thus the information content of a trade, is estimated as the cumulative long-term impulse response function of the SVAR. It is important to notice, also to better understand our contribution below, that Hasbrouck's model is linear and with constant parameters, thus impacts and information content are constant and do not depend on time or on market conditions. Hasbrouck's findings show that trades convey substantial information (\cite{hasbrouck1988trades}) that causes a persistent impact on prices (\cite{hasbrouck1991summary}). Analysis shows that long-term impact arrives only with long delays and varies between securities (\cite{hasbrouck1991measuring}). \cite{hautsch2012market} show that limit orders also have a long-term effect on prices, while \cite{dufour2012permanent} document the long-term impact of trade on the bond market. Extending the VAR models, \cite{engle2004impacts} proposed a non-linear VAR to describe the possible role of volume in the information content of the trade, and \cite{barclay2003competition} developed a multiple equation VAR system, which includes a trade indicator variable for each of the market participants.

Although groundbreaking, subsequent research has highlighted some important limitations of Hasbrouck's SVAR approach. \cite{dufour2000time} generalized the original VAR model by introducing a dummy temporal variable, measuring the effect of trade time on permanent impact. They argued that the price impact and liquidity depend on the frequency of the trading activity, which varies throughout the day. In this regard, the work challenges the assumption of constant liquidity. Using a different setting, \cite{cont2014price} and \cite{mertens2022liquidity}  find that price changes are mainly driven by liquidity, which exhibits significant intraday variations, challenging the constant impact assumption. More recently,  \cite{philip2020estimating} proposed a Reinforcement Learning framework to reinterpret an extended version of the original Hasbrouck model. \cite{philip2020estimating} documents a significant non-linear relation between permanent price impact and trade size and demonstrates that the traditional VAR framework is often misspecified due to the non-linear relation. Using a sample of two equally informed market participants who trade with different trade sizes or timings, Philip concludes that the VAR model can lead to incorrect inferences providing different permanent price impacts.

In this paper, we propose an extension of Hasbrouck's framework which fixes several of the above limitations, postulating a non-linear relation between trade signs and price changes. Specifically, our framework (i) focuses on the \textit{probability} of a buy trade, (ii) exploits the aggregation over heterogeneous time scales introduced by \cite{corsi2009simple} in the context of realized volatility modeling to consider hundreds of lags, and (iii) models a time-varying instantaneous market impact reacting to the recent price and trade history. The time-varying setting is based on the Dynamic Conditional Score-Driven Approach (DCS) of \cite{creal2013generalized} and \cite{harvey2013dynamic}. Among the observation-driven models, DCS provides a general framework for time-varying parameter models. The key ingredient in DCS models is that the scaled score updates the time-varying parameters. This update rule leverages the observation density and includes many existing models~\footnote{For instance, see Normal GARCH by \cite{bollerslev1986generalized}, ACD by \cite{engle1998autoregressive}, and MEM by \cite{engle2006multiple}.}. Due to ease of estimation, high flexibility, and wide application range, DCS models represent a valuable alternative to other observation- and parameter-driven models.  Moreover, as will become clear for our models, the DCS approach can be seen as a filter of misspecified dynamics of the time-varying parameters. 

The present work brings different contributions to the literature, developing a modified score-driven version of the SVAR model by \cite{hasbrouck1991measuring}. First, we do not model the evolution of the trade sign process directly, but we focus on the evolution of the occurrence probability of a buy trade. We employ an inverse logistic function to describe the dynamics of the trading indicator, consistent with its binary nature. Our specification provides a better-specified description of the trade-sign time series, as confirmed by several tests run on the model's residuals. Second, we aggregated lagged values of the order flow variable of the modified model on three different scales following the technique introduced in \cite{corsi2009simple} to mimic the persistence of the realized volatility. The aggregation allows us to mitigate the curse of dimensionality while capturing the well-known persistence of the order flow (see \cite{lillo2004long,toth2015equity}), while at the same time improving the quality of the model specification. As the last and most significant modification in our model, the instantaneous impact parameter of trades on quotes and the volatility of the return idiosyncratic component are time-varying and follow score-driven integrated dynamics. We derive the score-driven aggregated modified Hasbrouck model (SDAMH) equations by computing the log-likelihood function and the scaled score needed for the DCS updating of the instantaneous impact and volatility.

We empirically test our approach by estimating the SDAMH model on a data sample from Apple Inc (AAPL), Amazon (AMZN), Benefitfocus Inc. (BNFT), CleanSpark Inc. (CLSK), Extreme Networks (EXTR), Fulgent Genetics (FLGT), Alphabeth Inc. (GOOG), Intel (INTC), Altria Group (MO$^*$\footnote{We denote it with MO$^*$ instead of MO because we use the latter notation when referring to Market Orders.}), and Microsoft (MSFT) stocks listed on NASDAQ. We focus on these stocks, as they show different microstructural properties. AAPL, AMZN, GOOG, INTC, and MSFT are very liquid stocks, while BNFT, CLSK, EXTR, FLGT, and MO$^*$ are much less liquid. Our choice allows to analyze the effect of liquidity (measured in terms of the amount of market orders) on the price impact and on the information content of stock trades. Since the results for stocks with similar liquidity are homogeneous, we only report estimates for MSFT, AMZN, CLSK, and FLGT, while the results for the remaining stocks are available in the online supplementary material in Online Appendix F. Estimates show that the integrated DCS dynamics reproduces the relevant features of volatility and instantaneous impact. Concerning the latter, we find that it displays an intraday decreasing pattern, being large at the beginning of the day and declining afterward, and displaying some large localized fluctuations, possibly related to endogenously or exogenously driven liquidity crises. As a specific case of the latter, we study in detail the days of the Federal Open Market Committee (FOMC) announcements. To estimate the time-varying information content of trades, we estimate the long-run value of the Cumulative Impulse Response Function (CIRF) computed using Monte Carlo simulation. The analysis reveals that the information content of the stock trades varies mainly according to two factors: The instantaneous impact parameter and the \textit{state} of the market in a specific trade. The \textit{state} corresponds to the conditional mean of the SDAMH model, ignoring the impact of the last trade. We further unravel these findings using linear regressions, which confirm that the permanent impact positively depends on the instantaneous time-varying impact and negatively on the conditional mean of the model. These findings are in clear contrast to the \cite{hasbrouck1991measuring} SVAR model, where the information content is constant and independent of the state of the market. Analysis of the CIRF dynamics around FOMC announcements confirms the ability of the impact parameter in the DCS specification to react quickly to changing market conditions. 

A potentially important objection to non-linear score-driven models is that a simpler rolling-window estimation of Hasbrouck's original SVAR model might achieve a similar time-varying characterization of price impact. We emphasize that a rolling-window Hasbrouck approach is intrinsically inadequate for two main reasons. First, within each estimation window, the SVAR remains linear, implying that the long-run CIRF cannot depend on the instantaneous market state. Second, and more importantly, when calculating impulse responses over future horizons in Hasbrouck's framework, the instantaneous impact is held strictly fixed at its window-estimated level. This ignores the dynamic feedback loops from incoming trades and price updates that occur along the IRF horizon. By contrast, the SDAMH framework continuously updates the time-varying parameters along the impulse trajectory, explicitly capturing the market's dynamic adjustment to order flow shock updates.

Last but not least, we advocate the relevance of our approach in transaction cost analysis. This task requires a careful estimation of the permanent impact of trade since the implementation shortfall depends crucially on this parameter. Standard practice for its estimation (see, for example, \cite{cartea2016incorporating})  regresses the price increments aggregated over a macroscopic time scale (typically 5 minutes) on aggregated transaction volumes across a long period (typically one day). This leads to a static estimation, which does not use all available data and does not dynamically adapt to changing market conditions. At variance, we exploit the complete information content of tick-by-tick price and trade time series and derive the (approximate) analytic permanent impact function of the SDAMH model. Our formula quantifies the effect of a single transaction on mid-quote returns using the model's estimated coefficients.
However, the SDAMH model avoids data aggregation and takes advantage of the information content of the time series of the price and the trade more efficiently, thus providing a lower standard error. We tested the two approaches on all the stocks. 
The permanent impact coefficient estimated using SDAMH is more accurate than the one obtained with \cite{cartea2016incorporating}. This fact opens the possibility of running a dynamical transaction cost analysis and performing optimal execution conditionally on the intraday time. 

Before moving to the main sections of the paper, it is worth mentioning another key finding of \cite{philip2020estimating}. The recursive model provides empirical evidence that agents adapt trade sizes to available liquidity: ``\textit{Agents trade large volumes when the market is liquid, resulting in small price impacts per unit traded, and small volumes when the market is illiquid, resulting in large price impacts per unit traded.}'' Our approach extends the pioneering work by Hasbrouck investigating the effect of a non-linear model specification on the permanent price impact but does not relax the assumption of the trade sign as a proxy for the trade activity. A natural future extension, necessary to move in the direction of the above remark by Philip, will consider the interplay between mid-quote returns, the signed trade sizes, and the quote sizes outstanding at the opposite side of the limit order book. As a preliminary analysis, we have run several tests regressing the long-run CIRF on the istantaneous impact, the market state, and the trade signed volume finding no evidence of significance for the latter regressors at 5\% confidence level (see Online Appendix D for details). Past volumes may play a role in revealing the information content of a trade. However, the results in the online appendix show that modeling this effect is delicate and likely requires simultaneously modeling the level of liquidity available on the opposite side of the order book. Online Appendix F also reports the results from an estimation exercise where we augment the SDAMHM model including the depth imbalance and the bid-ask spread as exogenous regressors in the log-return and time-varying liquidity parameter equations, respectively. The coefficients remain statistically not significant at the 5\% level across the largest majority of trading days both for liquid and illiquid stocks.

The remaining of the paper is organized as follows. In section \ref{section:model}, we derive the equations of the SDAMH model and describe the estimation procedure of the model's parameters. Section \ref{section:app} presents an empirical application of the SDAMH model to actual data. We report the ML estimates of the model's static parameters and plot the filtered time-varying impact and of the volatility over the entire sample. We investigate the implications of a time-varying impact and market \textit{state} on the conditional information content of stock trades. In Section \ref{section:fomc}, we focus on the behavior of the permanent impulse in the case of the FOMC announcement of June 16, 2021 for the most liquid stocks: MSFT and AMZN. In section \ref{section:trancost}, we show the relevance of our approach for transaction cost analysis and optimal execution, deriving the approximate permanent impact expression from the model parameters and comparing it with the alternative method discussed in \cite{cartea2016incorporating}. Section \ref{section:conclusion} concludes. 

\section{The SDAMH model}
\label{section:model}

To study the joint dynamics of trades and prices and to measure the information content of a trade, \cite{hasbrouck1991measuring} proposed the following model
\begin{equation}\label{hasbrouckmodel}
\begin{split}
r_t = \sum_{i=1}^{p}a_i r_{t-i} + \sum_{i=0}^{p}b_i x_{t-i} + v_{1,t}, \\
x_t = \sum_{i=1}^{p}c_i r_{t-i} + \sum_{i=1}^{p}d_i x_{t-i} + v_{2,t},
\end{split}
\end{equation}
where $t\in \mathbb{N}$ is the trade time, $r_t=(q_t-q_{t-1})/q_{t-1}$ is the mid-point price return\footnote{Other choices are possible. The original \cite{hasbrouck1991measuring} paper uses the mid-quote price change, while \cite{philip2020estimating} uses the change in the natural logarithm of the mid-point. On these very short timescales, the different definitions lead to negligible differences.}, $q_t$ being the mid-point just after the trade at time $t$, and $x_t \in \{-1,+1\}$ is a trading dummy indicator for sell/buy orders of the trade at time $t$. The noise terms $v_{1,t}$ and $v_{2,t}$ have variances $\sigma_1^2$ and $\sigma_2^2$, respectively, and $\rho$ is their correlation, which is set to zero to make the model identifiable. Eq.~(\ref{hasbrouckmodel}) describes a SVAR model where $b_0$, the structural parameter, describes the contemporaneous effect between $r_t$ and $x_t$. From a financial point of view, $b_0$ describes the instantaneous impact of a trade. SVAR model parameters can be estimated using ordinary least square (OLS) and their analytical tractability allows to derive the impulse response function (IRF) expressed as a function of the estimated parameters in closed form solution (see \cite{lutkepohl2005new}). Moreover, following the interpretation of Hasbrouck, the asymptotic cumulative IRF measures the long-term effect of trade on prices and thus represents the information content of trade. Hasbrouck's model assumes that the information content of a trade is constant in the investigated period. This is a consequence of the fact that the parameters in Eq. \eqref{hasbrouckmodel} are constant and of the linear structure of the model. Indeed, it is well known that for a VAR model, the IRF does not depend on the history of the process, but only on the parameters of the model (\cite{lutkepohl2005new}). However, the information content of trade is unlikely to be constant throughout the day, and one can expect that the arrival of news, for example, an FOMC meeting, or the hectic intraday dynamics can dramatically modify it. Furthermore, the fluctuating nature of liquidity makes the response of price to trade not constant, which cannot be captured by Eq.~(\ref{hasbrouckmodel}). 
In this paper, we propose a model for measuring the {\it dynamics} of liquidity and of the information content of a trade by proposing a time-varying parameter version of Hasbrouck's model. To this end, we rely on the DCS approach of \cite{creal2013generalized} and \cite{harvey2013dynamic}. To introduce it, consider a sequence of observations $\{z_t\}_{t=1}^T$, where each $z_t \in\mathbb{R}^M$,  and a conditional probability density $P(z_t\vert f_t)$, that depends on a vector of time-varying parameters $f_t \in \mathbb{R}^W$. Defining the score as $\nabla_t := \partial \log{P(z_t | f_t)}/\partial f'_t$, a score-driven model assumes that the dynamics of the parameters follows
\begin{equation}\label{eq:gasupdaterule}
f_{t+1} = \omega +\ \beta f_t  + \alpha s_t\,\quad\text{where}\quad
s_t := (\mathcal{I}_{t|t-1})^{-1} \nabla_t\, 
\end{equation}
and $\mathcal{I}_{t|t-1}$ is the Fisher information matrix.
In Eq.~(\ref{eq:gasupdaterule}) $\omega$ is a vector of dimensions $W$, $\alpha$ and $\beta$ are matrices $W\times W$ of static parameters. Interestingly, one can consider a score-driven model as either a data-generating process (DGP) or a filter of unknown dynamics. In both cases, the most important feature of Eq. \eqref{eq:gasupdaterule} is the role of the score as the driver of the dynamics of $f_t$. The dependence of $f_{t+1}$ on the vector of observations $z_t$ follows from the conditional density specification, which determines the score. Thanks to the random occurrence of $z_t$, the updating rule provides stochastic dynamics when one looks at the model as a DGP. When regarded as a filter, the update rule in Eq.~\eqref{eq:gasupdaterule} delivers a sequence of filtered parameters $\{\hat{f}_t\}_{t=1}^T$. Maximizing the log-likelihood of the whole sequence of observations provides an estimate of the static parameters. Score-driven models have seen an explosion of interest in recent years~\footnote{See for example http://www.gasmodel.com.} due to their flexibility and ease of estimation. Many state-of-the-art popular econometric models belong to the family of score-driven models. Examples are the GARCH model, the Exponential GARCH model, the Autoregressive Conditional Duration model, and the Multiplicative Error Model. In this paper, we develop a DCS version of Hasbrouck's model to describe and filter the time-varying dynamics of market impact, return volatility, and information content of a trade. To this end, we assume that the parameters $b_0$ and $\sigma_1^2$ of Eq.~\eqref{hasbrouckmodel} are time-varying according to DCS dynamics. Our choice is minimal because one could build a model where all the parameters of Eq.~\eqref{hasbrouckmodel} follow a DCS model. However, the complexity of the model would grow significantly and the estimation of the model would be computationally more difficult. We believe that $b_0$ captures an important part of the relation between trades and prices, and for this reason, in this paper, we explore its dynamics. To build a DCS version of the Hasbrouck model it is necessary to specify the conditional probability density $P(z_t\vert f_t)$ of the model. In the original formulation of Eq. \eqref{hasbrouckmodel} the distribution of the noise terms $v_{1,t}$ and $v_{2,t}$ is not specified~\footnote{Remember that the estimation is performed via OLS.}. Moreover, the binary nature of $x_t$ requires a complicated temporal dependence structure for $v_{2,t}$.  
Therefore, we modify the original model to specify the probability density of the observables. In particular, we replace the second equation in Eq. \eqref{hasbrouckmodel} by using an inverse logistic map for $\mathbb{P}(x_t = +1| \mathcal{F}_{t-1} )$, where $\mathcal{F}_{t}$ is the available information set at time $t$
\begin{align}\label{modhasbrouckmodel}
r_t &= \mu_{1} + \sum_{i=1}^{p}a_i r_{t-i} + \sum_{i=0}^{p}b_i x_{t-i} + v_{1,t},& \nonumber \\ 
\mathbb{P}(x_t = +1 \big| \mathcal{F}_{t-1}) &:= \pi_t = Logit^{-1} \Big( \mu_{2} + \sum_{i=1}^{p}c_i r_{t-i} + \sum_{i=1}^{p}d_i x_{t-i}\Big),
\end{align}
 where $Logit^{-1}(z) = (1+e^{-z})^{-1}$ is the inverse logistic map of $z$ and we assume that $v_{1,t}\sim {\mathcal N}(0,\sigma^2)$. Notice that we have also added two constant parameters, $\mu_1$ and $\mu_2$, to include possible drift effects on price returns and trade signs. The Modified Hasbrouck's (MH) model of Eq.~\eqref{modhasbrouckmodel} explicitly provides the form of the conditional density, as required to build a DCS time-varying parameter version, and can be estimated by using the Maximum Likelihood Estimator (MLE). The transformation we use provides two main advantages. First, the inverse logistic transformation is consistent with the binary nature of the trading dummy indicator. Second, it introduces nonlinearity in the model. \cite{philip2020estimating} demonstrates that when the permanent impact function has a linear relationship with the trade sign, the model is highly misspecified. The logistic transformation makes the relationship between the permanent impact function (we will see this later in detail) and the trade sign variable non-linear. As pointed out later, these modifications make the IRF non-linear and no longer available in a closed-form solution. Before presenting the DCS version of Hasbrouck's model, we present a second modification of Eq.~\eqref{hasbrouckmodel}. The number of parameters to estimate in Hasbrouck's model depends on the number of lags $p$ in the VAR equations. The original \cite{hasbrouck1991measuring} uses $p=5$, but several more recent papers found this number too small to reproduce the dynamics of trades and prices. For example, a vast body of literature found that the order flow process $x_t$ is consistent with a long memory process 
\cite{bouchaud2003fluctuations,lillo2004long,bouchaud2009markets,toth2015equity,bouchaud2018trades}. To avoid the use of a large $p$, and thus the estimation of a large number of parameters, we employ the technique developed in \cite{corsi2009simple} for the time aggregation of regressors in realized volatility models and used in the market microstructure by \cite{hasbrouck2021price}. Let define $L_1, L_2 \in \mathbb{N}$ two integers indicating the number of past lags we consider for the aggregation. Consequently, we introduce the aggregated variables $x_t^{(L_1)} := \frac{1}{L_1-1}\sum_{i=2}^{L_1}x_{t-i}$ and $x_t^{(L_2)} := \frac{1}{L_2-L_1}\sum_{j=L_1+1}^{L_2}x_{t-j}$ and use them as regressors of the VAR together with $r_{t-1}$ and $x_{t-1}$~\footnote{It is not necessary to define aggregated regressors for the $r_t$ process as the midpoint price returns do not show long memory property.}. In the following, we choose $L_1=10$ and $L_2=100$. Then, the model reads
\begin{align}\label{aggmodhsbrouckmodel}
r_t & = \mu_{1} + a_1 r_{t-1} +  b_{0} x_t + b_1 x_{t-1} + \bar{b}_{10} x_t^{(10)} + \bar{b}_{100} x_t^{(100)} + v_{1,t}, & \nonumber \\ 
\pi_t & = Logit^{-1}\Big(\mu_{2}  + c_1 r_{t-1} +  d_1 x_{t-1} + \bar{d}_{10} x_t^{(10)} + \bar{d}_{100} x_t^{(100)}\Big),
\end{align}
 where $\mu_{1}$, $\mu_{2}$, $a_1$, $b_0$, $b_1$, $\bar{b}_{10}$, $\bar{b}_{100}$, $c_1$, $d_1$, $\bar{d}_{10}$, $\bar{d}_{100}$ and $\sigma^2$ are constant. We term this model the Aggregated Modified Hasbrouck (AMH) model. In Online Appendix \ref{section:rescheck} we run misspecification tests on the original Hasbrouck model of Eq.~\eqref{hasbrouckmodel}, the modified model of Eq.~(\ref{modhasbrouckmodel}), and the aggregated version of Eq.~\eqref{aggmodhsbrouckmodel}.  We find that the MH model of Eq.~(\ref{modhasbrouckmodel}) provides more Gaussian residuals than the original Hasbrouck model and the AMH model provides even better results. Furthermore, using the Bayesian Information Criteria (BIC), we find that the AMH model of Eq.~(\ref{aggmodhsbrouckmodel}) provides the best fit to the data (please refer to Online Appendix \ref{section:modelselstat}).  Consistently, we modify the latter to build a market impact model with time-varying parameters. Although the model of Eq.~(\ref{aggmodhsbrouckmodel}) depends on $12$ parameters, in this paper, we consider a minimal modification by assuming that only $b_0$ and $\sigma^2$ vary through time. We define 
\begin{align}\label{infoset}
		\mu^{(1)}_{t} & := \mu_{1} + a_1 r_{t-1} + b_{0,t} x_t + b_1 x_{t-1} + \bar{b}_{10} x_t^{(10)} + \bar{b}_{100} x_t^{(100)}, & \\ 
		\mu^{(2)}_{t} & := \mu_{2}  + c_1 r_{t-1} +  d_1 x_{t-1} + \bar{d}_{10} x_t^{(10)} + \bar{d}_{100} x_t^{(100)}. \nonumber
\end{align}
Then, according to the DCS specification, we assume that $b_{0,t}$ and $\sigma_t^2$ follow the dynamics
\begin{equation*}
\begin{bmatrix}
b_{0,t+1}  \\
\log(\sigma^2_{t+1}) \\
\end{bmatrix} 
= 
\begin{bmatrix}
\omega \\
\varrho \\
\end{bmatrix} 
+
\begin{bmatrix}
 \beta & 0 \\
0  & \lambda \\
\end{bmatrix}
\begin{bmatrix}
b_{0,t}  \\
\log(\sigma^2_{t}) \\
\end{bmatrix} 
+
\begin{bmatrix}
\alpha & 0 \\
0 & \gamma \\
\end{bmatrix}
\begin{bmatrix}
s_{t,1} \\
s_{t,2} \\
\end{bmatrix}\,, 
\end{equation*}
where the scaled score $s_t$ is defined in Eq.~(\ref{eq:gasupdaterule}). We compute it from the  conditional densities, that read as follows~\footnote{For sake of clarity, we explicitly include $b_{0,t}$ and $\sigma^2_{t}$ in the conditioning information set, even though $b_{0,t}$ and $\sigma^2_{t}$ belong to $\mathcal{F}_{t-1}$.}
\begin{align*}
p_{r}(r_t |x_t,  { b_{0,t},  \sigma^2_{t}, \mathcal{F}_{t-1}}, \theta) & = \frac{e^{ -\frac{1}{2 \sigma^2_t}(r_t - \mu^{(1)}_{t})^2}}{\sqrt{2 \pi \sigma_t^2}}\,,& \\ 
p_{x}(x_t | {\mathcal{F}_{t-1}}, \theta) & = Logit^{-1}(\mu_{t}^{(2)})^{\frac{1+x_t}{2}}\big( 1-Logit^{-1}(\mu_{t}^{(2)})\big)^{\frac{1-x_t}{2}}\,. 
\end{align*}
After some straightforward computations, the log-likelihood function is given by 
\begin{equation}\label{likeGAS}
\begin{split}
\mathcal{L} := \sum_{t=1}^{T}\mathcal{L}_t (r_t,x_t | \mathcal{F}_{t-1}, \theta)= -\frac{T}{2}\log 2\pi - \frac{1}{2} \sum_{t=1}^{T} \log \sigma_t^2 - \frac{1}{2} \sum_{t=1}^{T}\frac{\big(r_t-\mu_{t}^{(1)}\big)^2}{\sigma_t^2} + \\ \sum_{t=1}^{T}\frac{1+x_t}{2}\log \big((1+e^{-\mu_{t}^{(2)}})^{-1}\big) + \sum_{t=1}^{T}\frac{1-x_t}{2}\log \big(1-(1+e^{-\mu_{t}^{(2)}})^{-1}\big)\,,
\end{split}
\end{equation}
thus \begin{equation*}
\nabla_t = 
\begin{bmatrix}
 \frac{\partial \mathcal{L}_t}{\partial b_{0,t}} \\
 \frac{\partial \mathcal{L}_t}{\partial \sigma^2_{t}} \\
\end{bmatrix} 
=
\begin{bmatrix}
 \frac{x_t}{\sigma_t^2}(r_t - \mu^{(1)}_{t}) \\
\frac{(r_t - \mu^{(1)}_{t})^2 - \sigma_t^2 }{2 \sigma_t^4}  \\
\end{bmatrix}
\text{and} \ \mathcal{I}_{t|t-1}^{-1} = \mathbb{E}_{t-1}[\nabla_t \nabla_t^\intercal]^{-1} =
\begin{bmatrix}
\sigma^2_{t} & 0 \\
0 & \sigma^2_{t}/2 \\
\end{bmatrix}. 
\end{equation*}
Hence, the dynamics of $b_{0,t}$ is 
\begin{equation}\label{qtdyn}
b_{0,t+1} = \omega + \beta b_{0,t} + \alpha  x_t \big(r_t - \mu_{t}^{(1)}\big),  
\end{equation} 
while we restrict that of $\log(\sigma^2_{t+1})$ to 
\begin{equation}
    \log(\sigma^2_{t+1}) = \log(\sigma^2_{t}) + \gamma \Bigg[\frac{(r_t - \mu^{(1)}_{t})^2}{\sigma^2_{t}}-1\Bigg]\,,
\end{equation}
where the integrated autoregressive specification is motivated by the well-known return volatility time variation at the intra-day scale (\cite{buccheri2021score}).
The DCS update in Eq.~\eqref{qtdyn} is the baseline autoregressive recursion for the time-varying impact parameter obtained from a general derivation. We run several tests using in-sample and out-of-sample criteria to select the most suitable DCS specification. In particular, we evaluate both stationary and integrated specifications for the DCS recursion. For selection, we rely on the BIC and the Out-of-Sample Likelihood (OSL). The analysis reveals that for our dataset, the most suitable DCS specification is the integrated one, corresponding to $\omega = 0$ and $\beta = 1$. Further details are available in Online Appendix \ref{section:modelselGAS}. 
In conclusion, the SDAMH model reads as 
\begin{align}\label{model1eq1}
r_t & = \mu^{(1)}_{t} + v_{1,t}, & \\
\label{model1eq2}
\pi_t  &= Logit^{-1} \big( \mu^{(2)}_{t} \big), \\
\label{model1eq3}
b_{0,t+1} & =  b_{0,t} + \alpha x_t (r_t - \mu^{(1)}_{t}), \\
\label{model1eq4}
\log(\sigma^2_{t+1}) & = \log(\sigma^2_{t}) + \gamma \Bigg[\frac{(r_t - \mu^{(1)}_{t})^2}{\sigma^2_{t}}-1\Bigg]\,
\end{align}
 where $\mu^{(1)}_{t}$ and $\mu^{(2)}_{t}$ are defined in Eq.~(\ref{infoset}) and $v_{1,t} \sim N(0,\sigma_t^2)$. 
 
It is plausible to argue that the trading environment is richer than the state captured by $\mu_t^{(1)} - b_{0,t} x_t$, which only reflects past returns and trade signs. To test for the effect of omitted explanatory varables, we have considered the SDAMH model incorporating external regressors to account for market conditions in a more detailed way. We included in $\mu_t^{(1)}$  a term proportional to the depth imbalance and in the right hand side of Eq.~(\ref{model1eq3}) the bid-ask spread at time $t$. The empirical results suggest that these variables do not significantly alter the findings. The impact on the evolution of $b_{0,t}$ and $\sigma^2_t$ is marginal and the core insights of our model remain robust. Consequently, while incorporating a broader set of market state variables might theoretically enrich the model, the practical benefits in explaining price impact and volatility time variation appear limited in our setting. The interpretation of the score-driven update of Eq.~\eqref{model1eq3} is enlightening. The update of the instantaneous impact coefficient $b_{0,t}$ is proportional to the difference between the observed return ($r_t$) and the expected return~\footnote{Note that $\mu^{(1)}_{t}$ is the conditional mean of $r_t$ given the past information $\mathcal{F}_{t-1}$ and $x_t$.} ($\mu^{(1)}_{t}$).  Then, suppose a buy market order arrives at time $t$ ($x_t = +1$). If the observed return $r_t$ is larger than the conditional expectation $\mu^{(1)}_{t}$, one concludes that the market is less liquid than originally thought and the estimation of the instantaneous impact should be increased, i.e. $b_{0,t+1} > b_{0,t}$. In the opposite case, when $r_t$ is smaller than the conditional expectation, we have $b_{0,t+1} < b_{0,t}$.  
In other words, the score-driven update provides real-time learning of the (il)liquidity (as measured by the market impact). The adjustment of $b_{0,t+1}$ is proportional to the ``surprise" brought about by a new trade with respect to the conditional expectation based on past prices and the order flow. The speed of adjustment is given by the parameter $\alpha$, estimated from the data via MLE. 

As mentioned above for generic DCS models, one can regard our model either as a DGP or as a filter of misspecified dynamics. When used as a DGP, it describes the stochastic dynamics of market impact. One does not know the instantaneous impact parameter $b_{0,t}$ in advance, until one randomly samples past returns and market orders. When used as a filter, after observing a sequence of prices and orders, one applies the model to dynamically learn the time-varying parameter trajectory. In the latter setting, the integrated scheme of Eq.~\eqref{model1eq3} is more suitable for modeling the price impact and capturing the non-stationary behavior, possibly due to intra-day periodicity or to news announcements. As we show in the empirical section, the integrated scheme quickly reacts in the case of abrupt changes in market conditions.

As with most DCS models, our model can be estimated using MLE. Given the vector of unknown parameter $\theta \in \Theta \subset \mathbb{R}^{D} $, we maximize the log-likelihood function in Eq.~\eqref{likeGAS} to find the ML estimates $\hat{\theta}$, as
\begin{equation}\label{maxprob}
 \hat{\theta} =	\operatorname*{arg\,max}_{\theta \in \Theta} \sum_{t=1}^{T} \mathcal{L}_t (r_t,x_t | \mathcal{F}_{t-1}, \theta)\,.
\end{equation}
Despite being quite straightforward, some care must be devoted to the initialization of the parameters in the recursion to obtain a faster and accurate convergence to the optimal solution. In Online Appendix \ref{app:estimation} we describe in more detail the estimation scheme and the initialization procedure. We also present the results of some numerical experiments on simulated data to show the effectiveness of the initialization procedure we adopt.

\subsection{Impulse Response Analysis}
\label{section:irf}

In the original \cite{hasbrouck1991measuring} paper, the SVAR modeling of the joint dynamics of returns and trades is used to compute the information content of a trade by using IRF analysis. In particular, the cumulative impulse response function (CIRF) associated with a trade of a given sign provides its information content or the long-term impact. More precisely, the IRF is defined as
\begin{equation}\label{irf}
\text{IRF}(h; \delta_x) =  \mathbb{E}[r_{t+h}|r_t, x_t+\delta_x,r_{t-1},x_{t-1},\dots]-\mathbb{E}[r_{t+h}|r_t, x_t,r_{t-1},x_{t-1},\dots],
\end{equation}
where the shock on quotes is given by $\delta_x = +1$ and the number of periods (horizons) $h = 1, \dots, H$ is measured in ``trade time". The CIRF is then defined as
\begin{equation}\label{cirf}
\text{CIRF}(H;\delta_x) = \sum_{h = 1}^{H} \text{IRF}(h;\delta_x).
\end{equation}
SVAR models are linear models where $b_0$ is a constant parameter. For this reason, the shape of the IRF at each trade horizon depends only on the estimated parameters (see for instance \cite{lutkepohl2005new}). For instance, a Hasbrouck model of Eq.~\eqref{hasbrouckmodel} with $p=3$ can be written in matrix form as
\begin{equation*}
\begin{bmatrix}
1 & -b_0 \\
0 & 1\\
\end{bmatrix}
\begin{bmatrix}
r_t \\
x_t \\
\end{bmatrix} 
= 
\begin{bmatrix}
\mu_{1} \\
\mu_{2} \\
\end{bmatrix} 
+
\begin{bmatrix}
a_1 & b_1 \\
c_1 & d_1 \\
\end{bmatrix}
\begin{bmatrix}
r_{t-1} \\
x_{t-1} \\
\end{bmatrix} 
+
\begin{bmatrix}
{a}_{2} & {b}_{2} \\
{c}_{2} & {d}_{2} \\
\end{bmatrix}
\begin{bmatrix}
r_{t-2} \\
x_{t-2} \\
\end{bmatrix} 
+
\begin{bmatrix}
{a}_{3} & {b}_{3} \\
{c}_{3} & {d}_{3} \\
\end{bmatrix}
\begin{bmatrix}
r_{t-3} \\
x_{t-3} \\
\end{bmatrix} 
+
\begin{bmatrix}
v_{1,t} \\
v_{2.t} \\
\end{bmatrix},
\end{equation*}
or, more compactly, as $B y_t = \mu + A_{1} y_{t-1} + A_{2} y_{t-2} + A_{3} y_{t-3} + v_t$, where $y_t = (r_t,x_t)'$, $\mu = (\mu_{1}, \mu_{2})'$ and $v_t = (v_{1,t}, v_{2,t})'$ are two-dimensional real vectors. Premultiplying the previous equation by $B^{-1}$, one obtains a reduced-form VAR(3) model from which the IRF easily follows as $\text{IRF}(h;\delta_x) = J \textbf{A}^h J' \delta_x$, where $J:= [I_2,\textbf{0},\textbf{0}]$ is a $2 \times 6$ selection matrix and \textbf{A} is the $6 \times 6$ matrix of the companion-form VAR(1) model. It is worth noticing that for Hasbrouck's model, the IRF and thus the CIRF depend only on the estimated parameters and not on the recent history of prices and trades. On the contrary, the MH, AMH, and SDAMH models do not admit close-form solutions for the IRF, both because the models are non-linear and because in the SDAMH two parameters are time-varying. To compute the IRF, we follow the procedure of \cite{gallant1993nonlinear} and \cite{koop1996impulse} which use Monte Carlo simulations to evaluate the Eq.~\eqref{cirf}. Notice that, as a consequence of the nonlinearity, for constant parameter models (MH and AMH) the CIRF depends both on the estimated static parameters and the ``state" of the market. The history of prices and trades affects the CIRF in each trade $t$ and, as a consequence, it is possible to compute at each point in time the contribution of the market conditions to the CIRF. The SDAMH model of Eqs. \eqref{model1eq1}, \eqref{model1eq2} and \eqref{model1eq3} take a step ahead because it is a non-linear time-varying parameter model. Therefore, the CIRF depends on the estimated static parameters, the ``state" of the market, and the time-varying parameters $b_{0,t}$ and $\sigma_t$. As a direct consequence, in the non-linear models (AMH and SDAMH) the CIRF at a given time is different for a buy and for a sell trade, while in the linear models (e.g., Hasbrouck) the CIRF is symmetric. In the empirical analyses below we will make the arbitrary choice of measuring the CIRF of a buy trade, but the results can also be easily obtained for a sell trade. The procedure we adopt is as follows. We first estimate the static parameters $\theta$ of the SDAMH model on a sample day, maximizing Eq.~\eqref{maxprob}. We then obtain the filtered estimates of the time-varying parameters $b_{0,t}$ and $\sigma_t$. For each trade $t$, we simulate $S$ times the model, using the available information set~\footnote{Since the aggregation of the HAR model of \cite{corsi2009simple} requires a pre-sample of $L_2 = 100$ observations, the daily computation of the IRF starts from trade 100.}. Specifically, for each horizon $h=1,..,H$, we compute conditional expectations as in Eq.~\eqref{irf}. By doing this, we generate a series of $T - L_2$ IRFs denoted by $\text{IRF}_{t,h}$. Applying Eq.~\eqref{cirf}, we determine the cumulative impulse response function indicated as $\text{CIRF}_{t,h}$. Finally, the series of the long-run (or permanent) value of the IRF is given by the cumulative response at the last horizon lag $\text{CIRF}_{t,H}$ and its average across the $S$ simulation is denoted by $\text{LRCIRF}_t$. 

\section{Empirical analysis}
\label{section:app}

This section presents the application of the SDAMH model to real data. The aim is to show the performance of the score-driven model to capture the time-varying features of the return volatility, the instantaneous impact, and thus the temporal variations of the information content of stock trades. 
In Online Appendix \ref{app:mc}, instead, we present the results obtained with numerical simulations, where we will show the finite sample performance on Monte Carlo simulations and in filtering misspecified dynamics.

\subsection{Dataset}
\label{section:dataset}
The dataset in our study consists of stocks listed on NASDAQ, one of the largest and most liquid electronic exchanges globally. NASDAQ’s electronic limit order book (LOB) plays a key role in price formation, with the best bid and ask prices determining the top of the book. Using high-frequency data from LOBSTER (lobsterdata.com), we can precisely track movements in the mid-point of quotes and analyze trade impacts with very accurate timestamping at the nanosecond level. Such a precision is particularly valuable in a high-frequency context, where trades often occur in rapid succession, and even slight inaccuracies in timestamping can lead to misleading estimates of price impact. Specifically, we consider Apple Inc., Amazon, Benefitfocus Inc., CleanSpark Inc, Extreme Networks, Fulgent Genetics, Alphabeth Inc., Intel, Altria Group, and Microsoft during the entire month of June, 2021 (22 trading days).   This sample month provides a good balance between capturing a quite rich market activity and a reasonable computational burden for the estimation process. Table \ref{stats} shows some summary statistics of the quote mid-point return $r_t$ and the trade indicator $x_t$ for the investigated stocks. Stocks have different microstructural characteristics. As we can see from the number of market orders (\# MO), we have five more liquid stocks (AAPL, AMZN, GOOG, INTC, and MSFT) and five less liquid stocks (BNFT, CLSK, EXTR, FLGT, and MO$^*$). The daily number of market orders spans almost two orders magnitude, highlighting the heterogeneity of our sample. We select these stocks to analyze how liquidity (in terms of daily number of market orders) affects the estimated price impact and the information content of stock trades. Our dataset shows a few consecutive transactions with the same timestamp (at nanosecond resolution). These are likely transactions triggered by the same market order executed against different limit orders. We decided to aggregate them~\footnote{As a robustness check we have considered also the non-aggregated case, finding very similar results.}. Specifically, we selected only the last execution price and the corresponding trade sign. 
For our analysis, we report the results for the MSFT, AMZN, CLSK, and FLGT tickers~\footnote{Results for the other tickers can be found in the online supplementary material in the Appendices B and F.}.
\begin{table}[t!]
\centering
\resizebox{\textwidth}{!}{
\begin{tabular}{llllllllll}
\textbf{Ticker} & 
  \textbf{Price} &
  \textbf{Volume} &
  \textbf{\# MO} &
  \textbf{Buy \%} &
  \textbf{Spread} &
  \textbf{Mean} &
  \textbf{Std} &
  \textbf{Sk} &
  \textbf{Ex Kurt} \\ \hline
\textbf{AAPL}  & 129.767 & 201.448  & 43,392   & 47.5\% & 1.382  & 8.875$\cdot 10^{-8}$ & 3.578$\cdot 10^{-5}$ & 0.140  & 6.536   \\
\textbf{AMZN}  & 3,364.428 & 22.327   & 15,018   & 48.6\% & 71.781 & 1.506$\cdot 10^{-7}$ & 6.307$\cdot 10^{-5}$ & 0.112  & 7.026   \\
\textbf{BNFT}  & 17.671   & 114.046  & 1,791.5  & 46.6\% & 3.076  & -8.233$\cdot 10^{-7}$ & 1.070$\cdot 10^{-3}$ & 0.572  & 11.753  \\
\textbf{CLSK}  & 79.583   & 57.094   & 2,348.318 & 51.6\% & 3.291  & 3.925$\cdot 10^{-6}$  & 7.193$\cdot 10^{-4}$ & 0.250  & 15.120  \\
\textbf{EXTR}  & 11.194   & 177.058  & 857.818  & 50.3\% & 3.288  & -9.790$\cdot 10^{-7}$ & 6.111$\cdot 10^{-4}$ & -0.053 & 9.948   \\
\textbf{FLGT}  & 70.225   & 49.603   & 1,175.818 & 40.2\% & 3.875  & -2.474$\cdot 10^{-6}$ & 6.035$\cdot 10^{-4}$ & 0.020  & 20.133  \\
\textbf{GOOG}  & 2,498.587 & 27.787   & 7,286    & 50.6\% & 72.299 & 9.993$\cdot 10^{-8}$  & 7.646$\cdot 10^{-5}$ & 0.027  & 19.094  \\
\textbf{INTC}  & 56.770   & 296.247  & 11,155.045 & 50.3\% & 2.733  & -6.140$\cdot 10^{-8}$ & 1.003$\cdot 10^{-4}$ & -0.441 & 32.087  \\
\textbf{MO}    & 48.345   & 203.682  & 3,205.045 & 50.6\% & 2.943  & -4.985$\cdot 10^{-7}$ & 1.366$\cdot 10^{-4}$ & 0.062  & 5.799   \\
\textbf{MSFT}  & 258.770  & 78.232   & 49,199   & 51.5\% & 1.926  & 5.220$\cdot 10^{-8}$  & 3.078$\cdot 10^{-5}$ & -0.034 & 22.521  \\ \hline
\end{tabular}}%
\caption{\footnotesize Average  price (\$), average  volume ($\times 10^3$), number of market orders (\# MO), percentage of buy MO (Buy \%), bid-ask spread (Spread) in ticks, mean, standard deviation (Std), skewness (Sk), and excess kurtosis (Ex Kurt) of $r_t$. Results are monthly averages on the days of June, 2021. }
	\label{stats}
\end{table}

\subsection{Estimation results}
\label{section:estres}

As a preliminary analysis, we consider different DCS specifications of the AMH model with the aim of assessing whether a stationary or integrated score-driven model better describes the data. We compare the two different DCS models using the BIC and OSL.    
Tables~\ref{tab:biccompgas} shows that the integrated specification (SD-INT) is almost always better performing than the stationary one in terms of the in-sample criterion. The stationary specification (SD-AR) outperforms the integrated one only on a single day out of 22 for CLSK. Please refer to Online Appendix B.4 for the details concerning the full sample, where the stationary model outperforms the integrated one a single day for INTC and EXTR, and two days for BNFT. The liquidity level for these stocks is the lowest in the sample.  Tables~\ref{tab:oslcompgas} reports the percentage of days, out of 21, when the SD-INT specification performs better than the SD-AR in terms of the  OSL. The prevalence of the integrated model is striking. The stationary specification is occasionally selected for all assets, even the most liquid ones. A more detailed daily-level analysis reveals that the integrated specification is preferred in 90.95\% cases out of 210. Including the cumulated non-signed volume as a proxy for liquidity in the analysis, no systematic relationship emerges between the level of liquidity and the days on which the AR model is selected.
\begin{table}[htt]
\centering
\begin{tabular}{cccccccc}
           & \textbf{MSFT} &  & \textbf{AMZN} &  & \textbf{CLSK} &  & \textbf{FLGT} \\ \hline
SD-AR      & 0.00\%        &  & 0.00\%        &  & 4.55\%        &  & 0.00\%        \\
SD-INT     & 100\%         &  & 100\%         &  & 95.45\%       &  & 100\%         \\
\end{tabular}
	\caption{\footnotesize 
    Percentage of days out of 22 a DCS specification performs better than the other one according to the BIC.} 
	\label{tab:biccompgas}
\end{table}

\begin{table}[htt]
\centering
\begin{tabular}{cccccccc}
           & \textbf{MSFT} &  & \textbf{AMZN} &  & \textbf{CLSK} &  & \textbf{FLGT} \\ \hline
SD-AR      & 9.6\%         &  & 14.3\%        &  & 9.6\%         &  & 4.8\%         \\
SD-INT     & 90.4\%        &  & 85.7\%        &  & 90.4\%        &  & 95.2\%        \\
\end{tabular}
	\caption{\footnotesize 
    Percentage of days out of 21 a DCS specification performs better than the other one according to the OSL.} 
	\label{tab:oslcompgas}
\end{table}
Tables \ref{avpar} and \ref{avpar1} show the monthly averages of the estimated static parameters (average absolute $t$-stats in parentheses) for the four stocks. The findings are quite similar across stocks. The estimated intercept of the return equation ($\mu_{1}$) is statistically significant, negative for MSFT and CLSK, and positive for AMZN and FLGT, indicating the presence of a global drift. The level of the logit equation, $\mu_{2}$, is positive for MSFT, CLSK, and FLGT, and negative for AMZN. In all cases it is close to zero, indicating that unconditional buy trades are as frequent as sell trades, consistently with Table~\ref{stats}. Concerning the coefficients measuring the lagged impact of the midpoint return on itself, we notice that $a_1$ is positive for MSFT, while it is negative for AMZN, CLSK, and FLGT,  indicating in the latter cases a short-term mean reversion of mid-returns. The coefficients $b_1$, $\bar{b}_{10}$, and $\bar{b}_{100}$ in the equation of $r_t$ measure the lagged effect of past trades on future returns. The absolute value of the first coefficient is generally larger than the others at higher lags, and all are strongly significant, confirming that the pressure of buy/sell orders on returns is strong and persistent at a high-frequency scale. The estimated absolute values of the $b$ coefficients are higher for less liquid stock than those of more liquid stock, indicating that the lagged impact of trades on returns is greater for stocks with lower number of trades. The impact of the past $r_{t-1}$ on $x_t$ is measured by the coefficient $c_1$, which is negative and highly significant. This coefficient captures the adjustment of the order flow to the recent price trend, and the negative sign indicates that more buyers (sellers) enter the market when the price decreases (increases). Another important feature emerging from the estimation is the strong persistence of trade signs. The positive values of $d_1$, $\bar{d}_{10}$, and $\bar{d}_{100}$ reflect the positive impact of the lagged trade signs on the current one. Consistently with the well-known persistence of the order flow (see, for example, \cite{toth2012does}, \cite{toth2015equity}, and \cite{lillo2023order}), we confirm the strong significance of the coefficients of the aggregated signs, meaning that buy (sell) orders tend to follow buy (sell) orders. Then, $\alpha$, the unique static parameter of the recursion in Eq.~(\ref{model1eq3}), measuring the level of fluctuation of the score, is statistically significant for all the stocks. A similar result holds for $\gamma$. These findings support the choice of a time-varying parameter approach to model the volatility and instantaneous impact behaviors. Finally, the initial values of the filtered instantaneous impact and volatility are positive and statistically significant. We conclude by noticing that the value of the daily estimates of the static parameters is stable over time. Table \ref{avpar} indeed shows that the standard deviation computed on the 22 trading days in our sample ("Std") is very small compared to the temporal average of the parameter estimates ("Est").
\begin{table}[htt]
	\centering
    \resizebox{\textwidth}{!}{
	\begin{tabular}{cccccccccccccccccc}
	&&&\textbf{MSFT}&&&&&&\textbf{AMZN}&& \\ \hline
	& Est & Std & & Est & Std & & Est & Std & & Est & Std \\ \hline
	$\mu_{1}$&-2.180 $\cdot 10^{-7}$ & 5.079 $\cdot 10^{-7}$ & $\mu_{2}$& 0.025 & 0.050 & $\mu_{1}$& 6.750 $\cdot 10^{-7}$ & 1.028 $\cdot 10^{-7}$ & $\mu_{2}$&-0.019 & 0.055 \\
	&(24.429) &  & &(9.500) &  & &(28.566) &  & &(25.345) &  \\ 
	$a_1$ &0.015 & 0.025   &$c_1$ & -9,119  & 1,943     & $a_1$   &-0.112 & 0.019 &$c_1$& -5,210 & 904 \\
	&(20.600) &  & &(11.693) &  & &(16.895) &  & &(30.763) &  \\
	$b_1$ &1.031 $\cdot 10^{-6}$ & 8.512 $\cdot 10^{-7}$  &$d_1$& 1.149 & 0.064   & $b_1$& 1.176 $\cdot 10^{-5}$ & 1.995 $\cdot 10^{-6}$ &$d_1$& 1.711 & 0.070 \\
	&(20.211) &  & &(36.953) &  & &(11.895) &  & &(32.443) &  \\
	$\bar{b}_{10}$&-2.057 $\cdot 10^{-6}$ & 7.347 $\cdot 10^{-7}$ & $\bar{d}_{10}$& 0.320 & 0.056 & $\bar{b}_{10}$& -3.849$\cdot 10 ^{-6}$ & 1.693 $\cdot 10 ^{-6}$ & $\bar{d}_{10}$& 0.327 & 0.064 \\
	&(15.800) &  & &(29.714) & & &(9.273) & & &(12.109) & \\
	$\bar{b}_{100}$& -2.965 $\cdot 10^{-6}$ & 1.697 $\cdot 10 ^{-6}$ & $\bar{d}_{100}$& 0.188 & 0.104 & $\bar{b}_{100}$&  -2.155 $\cdot 10^{-6}$ & 1.693 $\cdot 10^{-6}$  & $\bar{d}_{100}$& 0.160 & 0.103 \\
	&(8.747) &  & &(25.808) &  & &(15.273) &  & &(9.553) &  \\ 
	$\gamma$& 7.099$\cdot 10^{-3}$ & 2.565$\cdot 10^{-3}$  & $\sigma_{0}$& 5.430$\cdot 10^{-5}$ & 1.725 $\cdot 10^{-5}$ & $\gamma$&1.463 $\cdot 10^{-2}$ & 0.506$\cdot 10^{-2}$ & $\sigma_0$&  9.463$\cdot 10^{-5}$& 4.049$\cdot 10^{-5}$ \\
	&(10.768) &  & &(11.148) &  & &(12.891) &  & &(19.337) &   \\ 
	$\alpha$& 2.360$\cdot 10^{-3}$ & 6.152 $\cdot 10^{-4}$  & $b_{0,0}$  &  7.825 $\cdot 10^{-6}$ & 2.646 $\cdot 10^{-6}$& $\alpha$&4.929 $\cdot 10^{-3}$ &1.548$\cdot 10^{-3}$ & $b_{0,0}$ & 2.118 $\cdot 10^{-5}$ & 1.105 $\cdot 10^{-5}$ \\
	&(18.836) &  & &(9.957)  & &  &(28.184)  &  & &(22.917) &  \\ 
	\hline
	\end{tabular}
 }
	\caption{\footnotesize Point estimates of the SDAMHM model. Temporal daily average of the coefficients ("Est"), standard deviation computed on the 22 trading samples, and average absolute $t$-stats (in brackets) of the estimates. Results reported for MSFT and AMZN.} 
	\label{avpar}
\end{table}

\newpage

\begin{table}[htt]
    \centering
    \resizebox{165mm}{35mm}{
    \begin{tabular}{cccccccccccccccccc}
    &&&\textbf{CLSK}&&&&&&\textbf{FLGT}&& \\ \hline
               & Est & Std         &   & Est & Std &              &Est & Std &          & Est & Std \\ \hline
    $\mu_{1}$&-1.134 $\cdot 10^{-5}$ & 3.171 $\cdot 10^{-5}$ & $\mu_{2}$& 0.025 & 0.050 & $\mu_{1}$& 1.699 $\cdot 10^{-6}$ & 1.525 $\cdot 10^{-5}$ & $\mu_{2}$&0.009 & 0.137 & \\
               &(27.034) &  &                                      &(4.336)  &  &                   &(5.529) &  &                                          &(8.674)  &  \\ 
        $a_1$         &-0.127 & 0.044   &$c_1$          &  -147.961  & 266.276     & $a_1$   &-0.056 & 0.027 &$c_1$& -12.907 & 124.174 \\
                       &(12.338) &  &                     & (67.324) &  &                                  &(16.385) &         & & (58.079) &  \\
        $b_1$    &-1.606 $\cdot 10^{-4}$ & 4.478 $\cdot 10^{-5}$  &$d_1$& 1.608 & 0.182   & $b_1$&1.460 $\cdot 10^{-4}$ & 4.596 $\cdot 10^{-5}$ &$d_1$& 1.505 & 0.134 \\
            &(18.194) &                                       && (23.847) &  &             &(57.958) &                                      && (16.115) &  \\
    $\bar{b}_{10}$&1.603 $\cdot 10^{-5}$ & 2.124 $\cdot 10^{-5}$ & $\bar{d}_{10}$& 0.320 & 0.056 &    $\bar{b}_{10}$& -6.061 $\cdot 10^{-6}$ & 1.756 $\cdot 10^{-5}$ &    $\bar{d}_{10}$& 0.220 & 0.126 \\
                          &(31.435) &                                                   && (7.478) & &                   &(4.865) &    &&                                (17.382) &  \\
    $\bar{b}_{100}$& -1.334 $\cdot 10^{-5}$ & 9.501 $\cdot 10^{-5}$ & $\bar{d}_{100}$& 0.188 & 0.104 & $\bar{b}_{100}$& -2.866 $\cdot 10^{-5}$ & 4.210 $\cdot 10^{-5}$  & $\bar{d}_{100}$& 0.243 & 0.297 \\
            &(28.954) &                                                   && (8.701) & &                        &(6.319) &             && (10.933)  &  \\ 
    $\gamma$& 0.017 & 0.014  & $\sigma_{0}$& 4.435 $\cdot 10^{-4}$ & 9.869 $\cdot 10^{-5}$ &    $\gamma$&0.0137 & 0.0057 & $\sigma_0$&  4.435 $\cdot 10^{-4}$&9.869 $\cdot 10^{-5}$ \\
        &(19.563) & & & (10.506)&   &         &(25.515)  &&&(14.519)&   \\ 
    $\alpha$& 0.017 & 0.013  & $b_{0,0}$  &  7.825 $\cdot 10^{-6}$ & 2.646 $\cdot 10^{-6}$&     $\alpha$&0.0101 &0.0064 & $b_{0,0}$ & 4.435 $\cdot 10^{-4}$ & 9.869 $\cdot 10^{-5}$ \\
        &(16.605) &  & &(17.243)  & &  & (16.315)  &  & &(14.519)      \\     \hline
    \end{tabular}
 }
    \caption{\footnotesize  Point estimates of the SDAMHM model. Temporal daily average of the coefficients ("Est"), standard deviation computed on the 22 trading samples, and average absolute $t$-stats (in brackets) of the estimates. Results reported for CLSK and FLGT.} 
    \label{avpar1}
\end{table}
We conclude this section by providing some operational guidance, when applying the SDAMH framework across securities with diverse liquidity and microstructural features.
While the integrated DCS specification dominates in terms of BIC for highly liquid stocks, the OSL analysis occasionally selects the autoregressive DCS specification. Unfortunately, there is not a clear connection with the liquidity level of the stock on a given day. For illiquid stocks where trade frequency is low, the 100-lag HAR aggregation component $x_t^{(100)}$ can be omitted without loss of model consistency, reducing the pre-sample burn-in requirement from 100 trades to 10 trades ($x_t^{(10)}$). The static parameter initialization should leverage a two-step linear approximation: first estimating the linear AH model via OLS, and then using scaled estimates to initialize the inverse-logistic log-likelihood maximization. This guarantees robust numerical convergence across both large-tick and small-tick assets.

\subsection{Time-varying price impact and volatility}
\label{section:tvimp}

After the discussion on the static parameters, we illustrate the behavior of the filtered time-varying parameters $b_{0,t}$ and $\sigma_t$. 

\begin{figure}[htt]
	\centering
         \textbf{All sample time-varying estimates of the volatility}\par\medskip
	\subfigure[June, 2021 - MSFT]
	{\includegraphics[width=7cm]{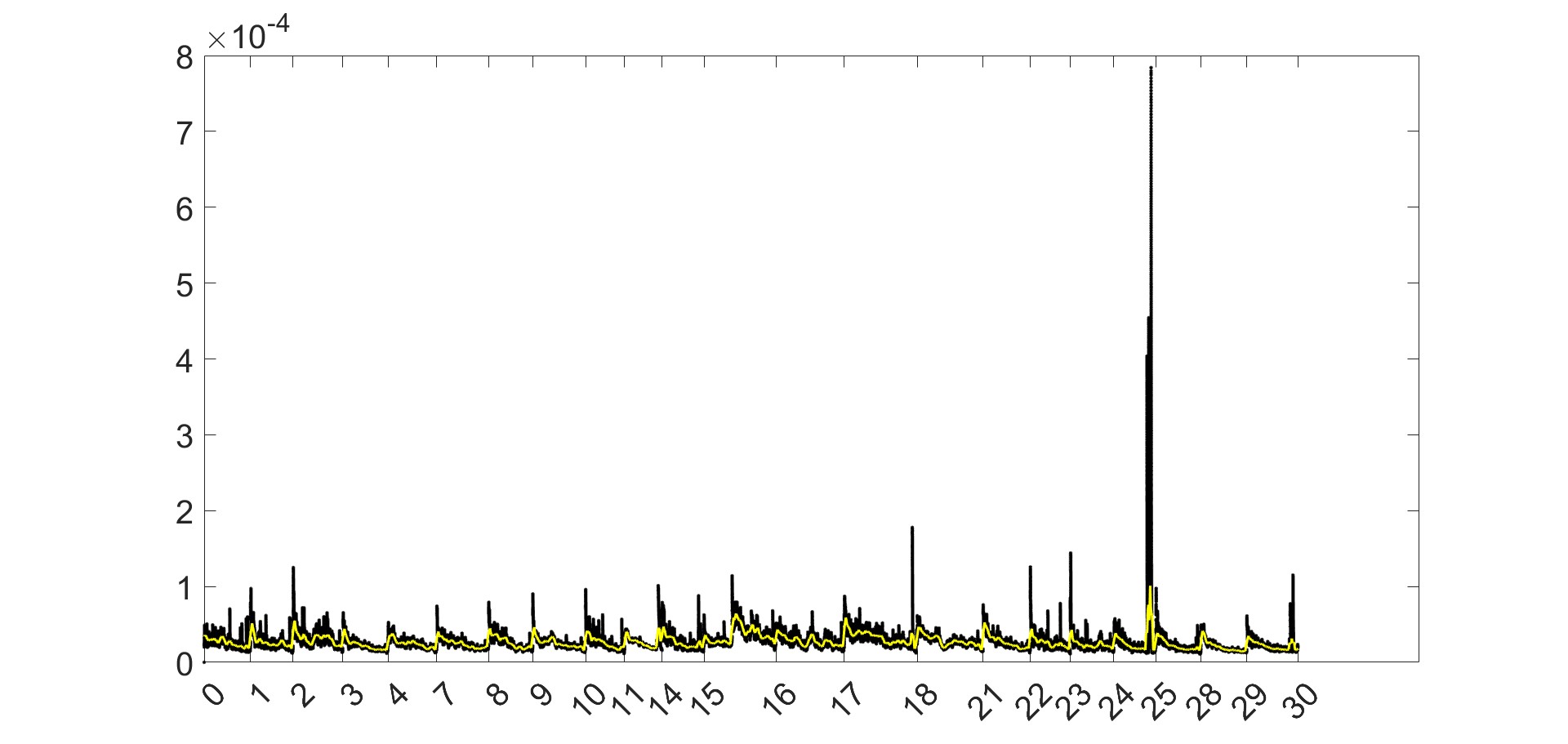}} 
        \hspace{1mm}
	\subfigure[June, 2021 - AMZN]
	{\includegraphics[width=7cm]{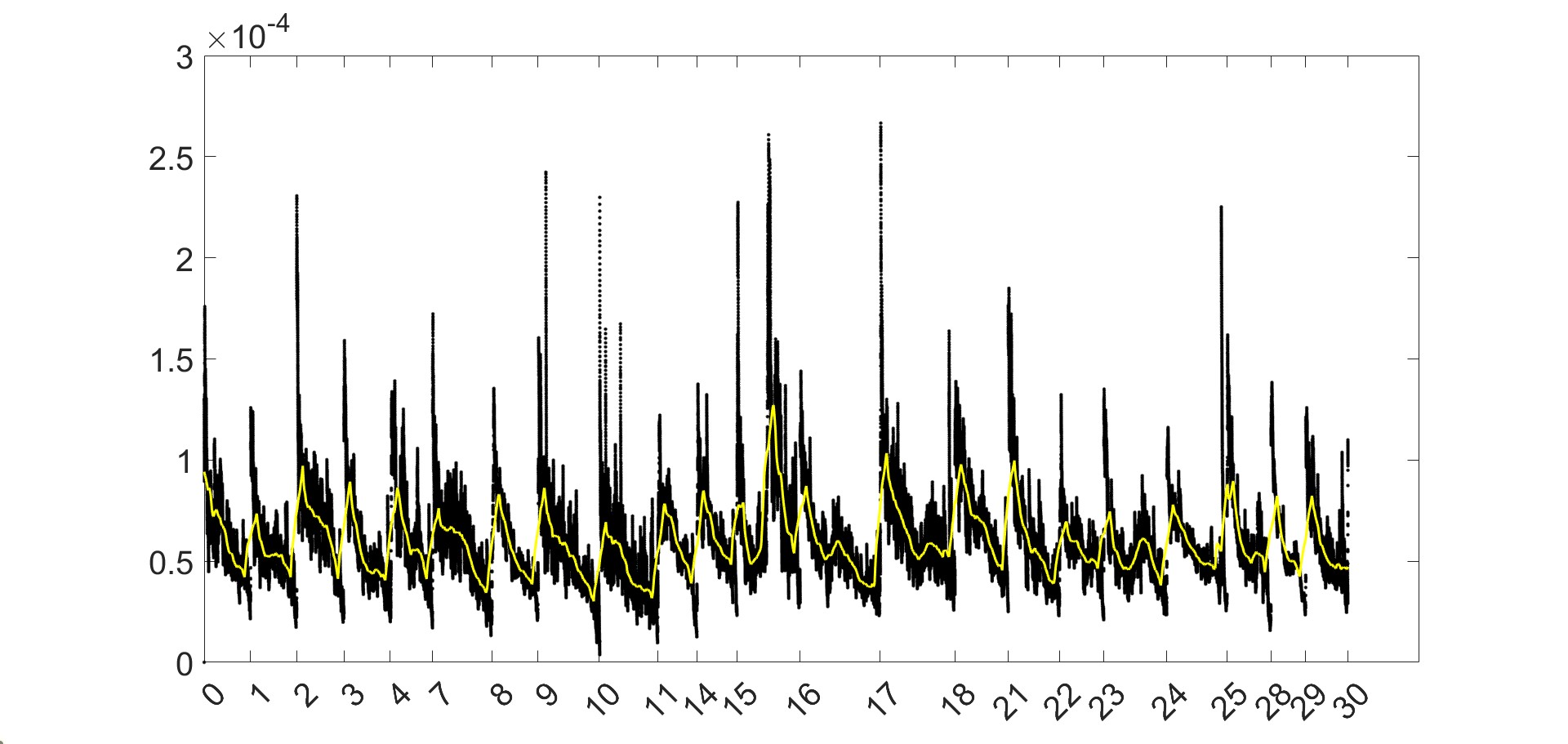}}
         \hspace{1mm}
	\subfigure[June, 2021 - CLSK]
	{\includegraphics[width=7cm]{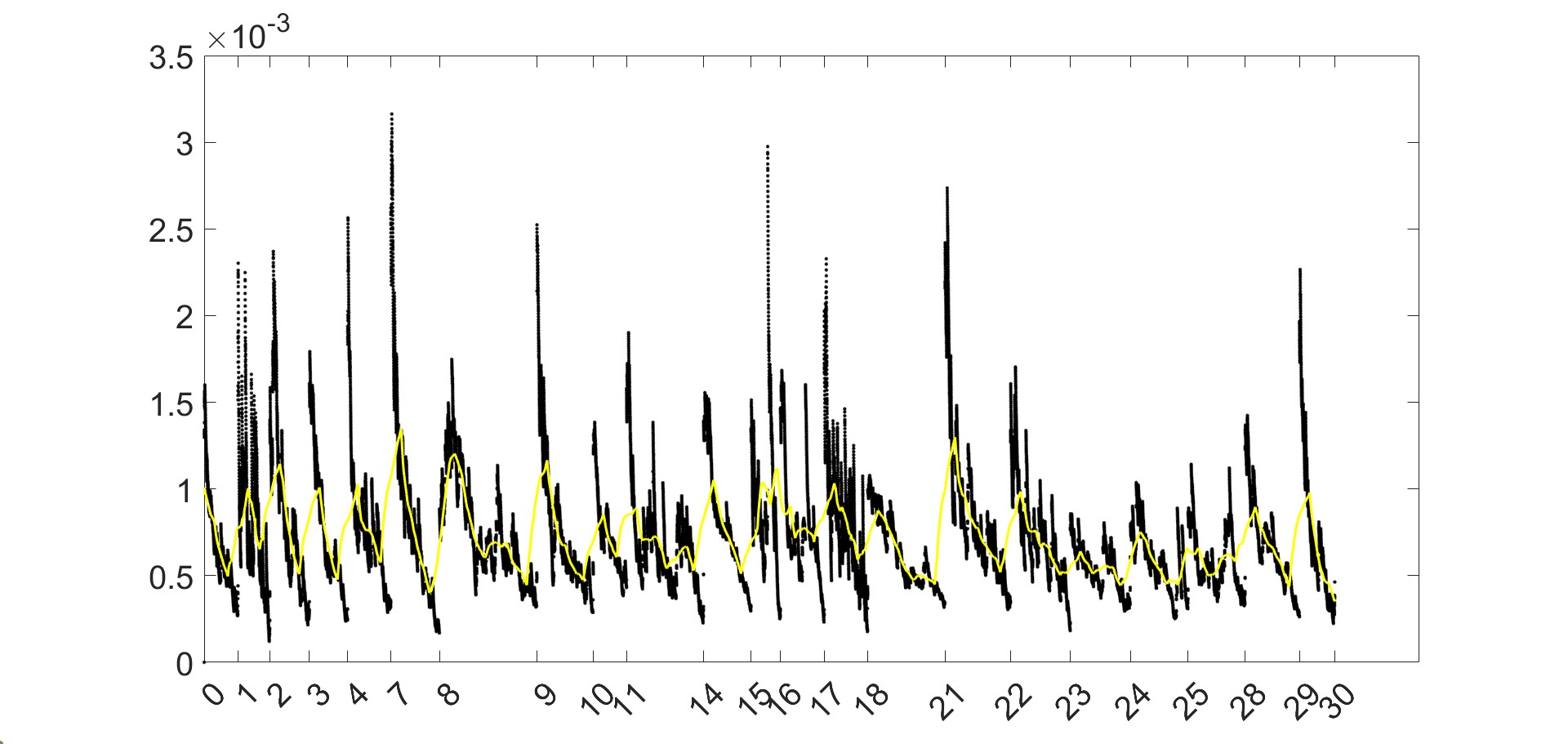}}
         \hspace{1mm}
	\subfigure[June, 2021 - FLGT]
	{\includegraphics[width=7cm]{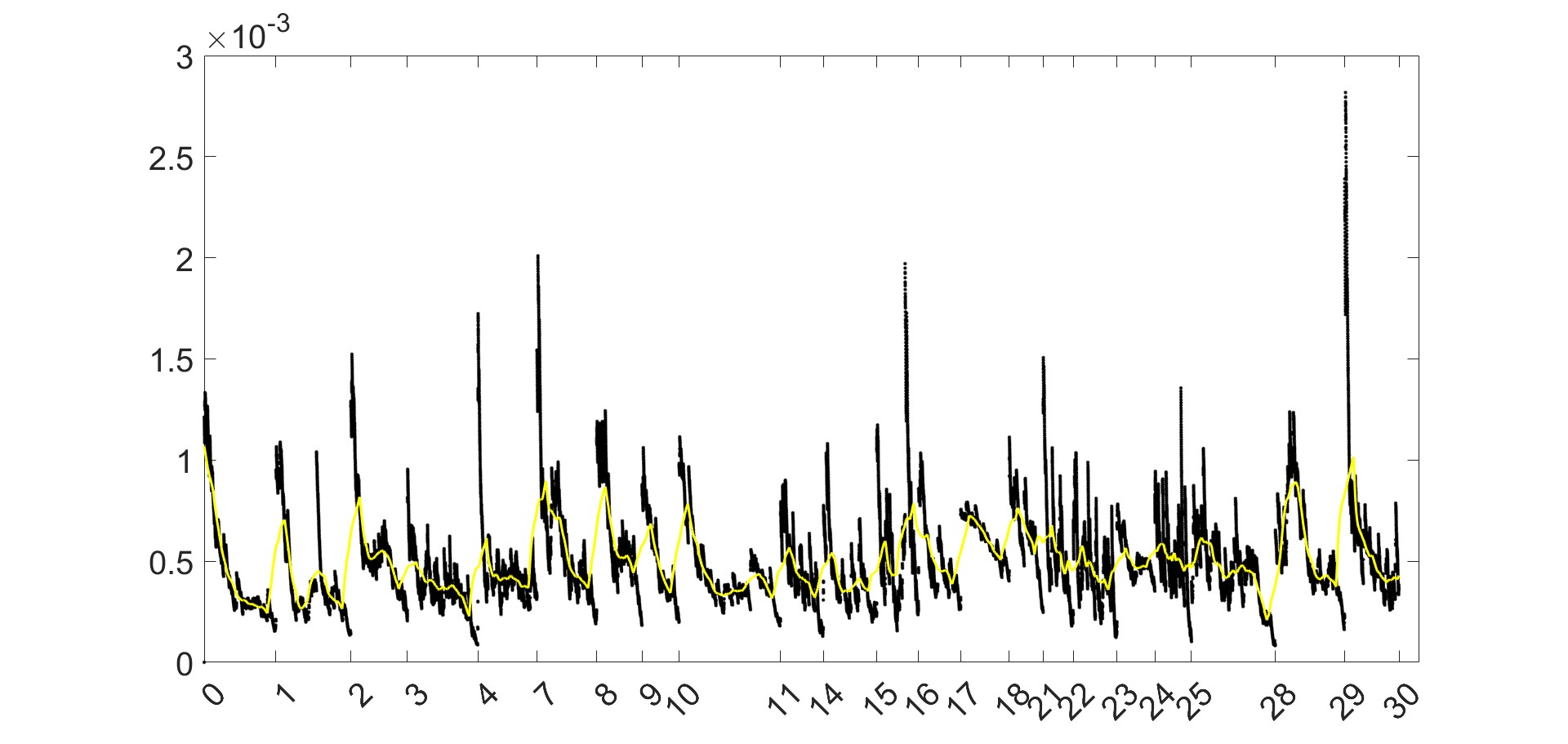}}
	\caption{\footnotesize 
    Estimates of the time-varying volatility of the SDAMH model (black lines) and its moving average (yellow solid lines). Overview of June, 2021 trading month.}
	\label{sigmatv}
\end{figure}

Figure~\ref{b0tv} shows the filtered $b_{0,t}$ (black dotted line) for the entire month of June 2021 for MSFT (top left panel), AMZN (top right panel), for CLSK (bottom left panel), and for FLGT (bottom right panel). We also report a smoothed path obtained from a moving average of the estimated parameter (green line). These panels show three main important characteristics. First, the estimated parameter seems to fluctuate around an unconditional monthly constant level. Second, the instantaneous impact shows a recurring path, corresponding to an intra-day pattern. Third and more importantly, the pattern is not constant throughout the days and the time series of $b_{0,t}$ shows occasional spikes and breaks away from its long-term value. An example of the latter occurred on June 16, 2021 for both stocks~\footnote{June 16, 2021 is the day of the FOMC announcement which will be investigated in detail in Section \ref{section:fomc} for the more liquid MSFT and AMZN stocks.}. These features are consistently captured by the integrated dynamics of the SDAMH model, which can adapt quickly in case of abrupt changes in market conditions and to describe an intra-day pattern, possibly different on different days. Figure~\ref{sigmatv} reports the time-varying volatility filtered for the same month for the four stocks (top left panel MSFT, top right panel AMZN, bottom left panel CLSK, and bottom right panel FLGT). The declining daily pattern is consistent with that observed for the return volatility in~\cite{buccheri2021score}, with the important difference that, in that case, the authors perform the analysis in physical time at a one-second frequency on trade price returns. The dynamics resemble $b_{0,t}$, showing a strong intra-day pattern and occasional spikes. Then, a natural question is whether the volatility drives the instantaneous impact or vice versa or whether both follow the dynamics of a common latent factor. We leave the answer to future investigation.
 
\begin{figure}[htt]
	\centering
           \textbf{Filtered time-varying parameters and LRCIRF$_{t}$}\par\medskip
	\subfigure[$b_{0,t}$ June 04, 2021 - MSFT]
	{\includegraphics[width=6.5cm]{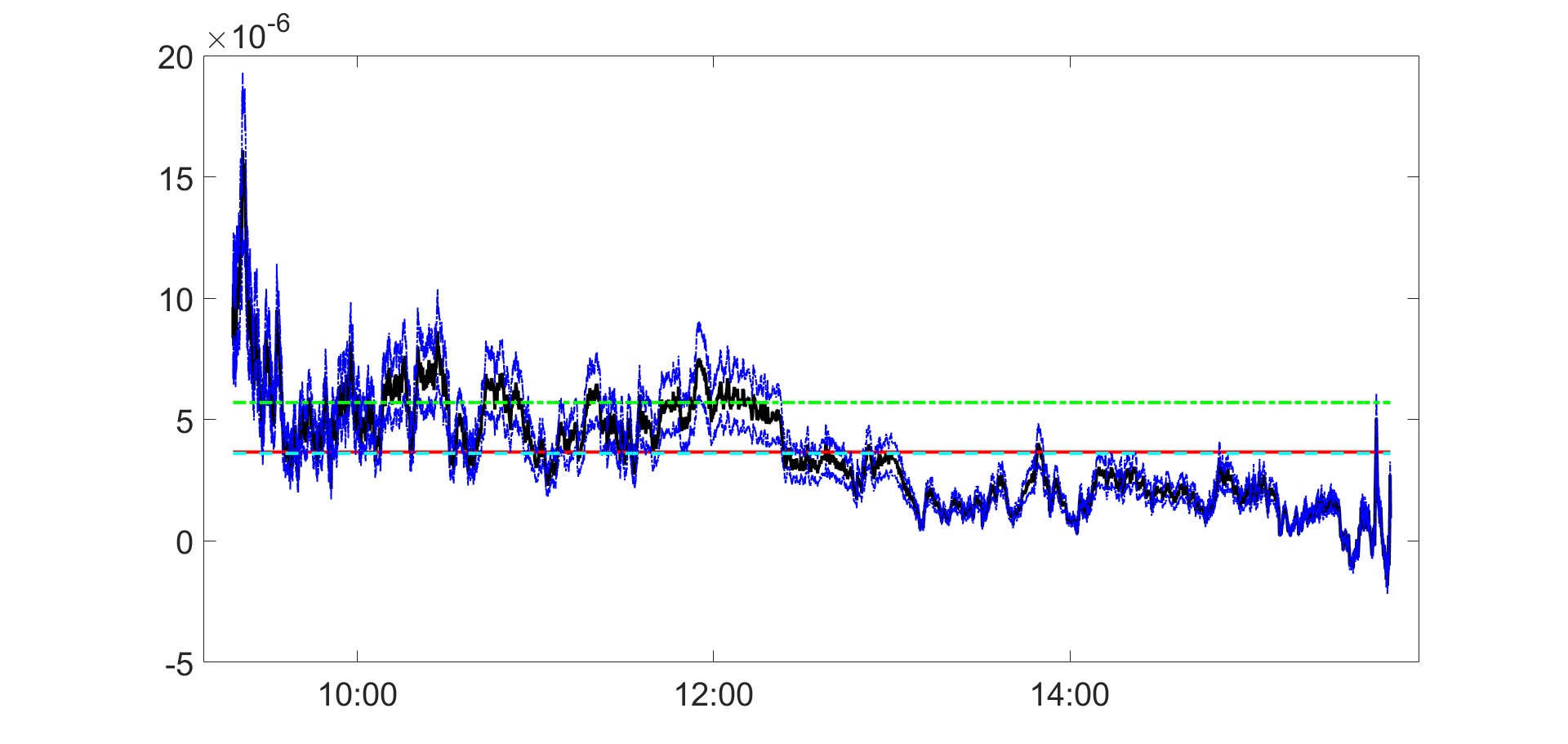}}
	\hspace{5mm}
	\subfigure[$b_{0,t}$ June 04, 2021 - AMZN]
	{\includegraphics[width=6.5cm]{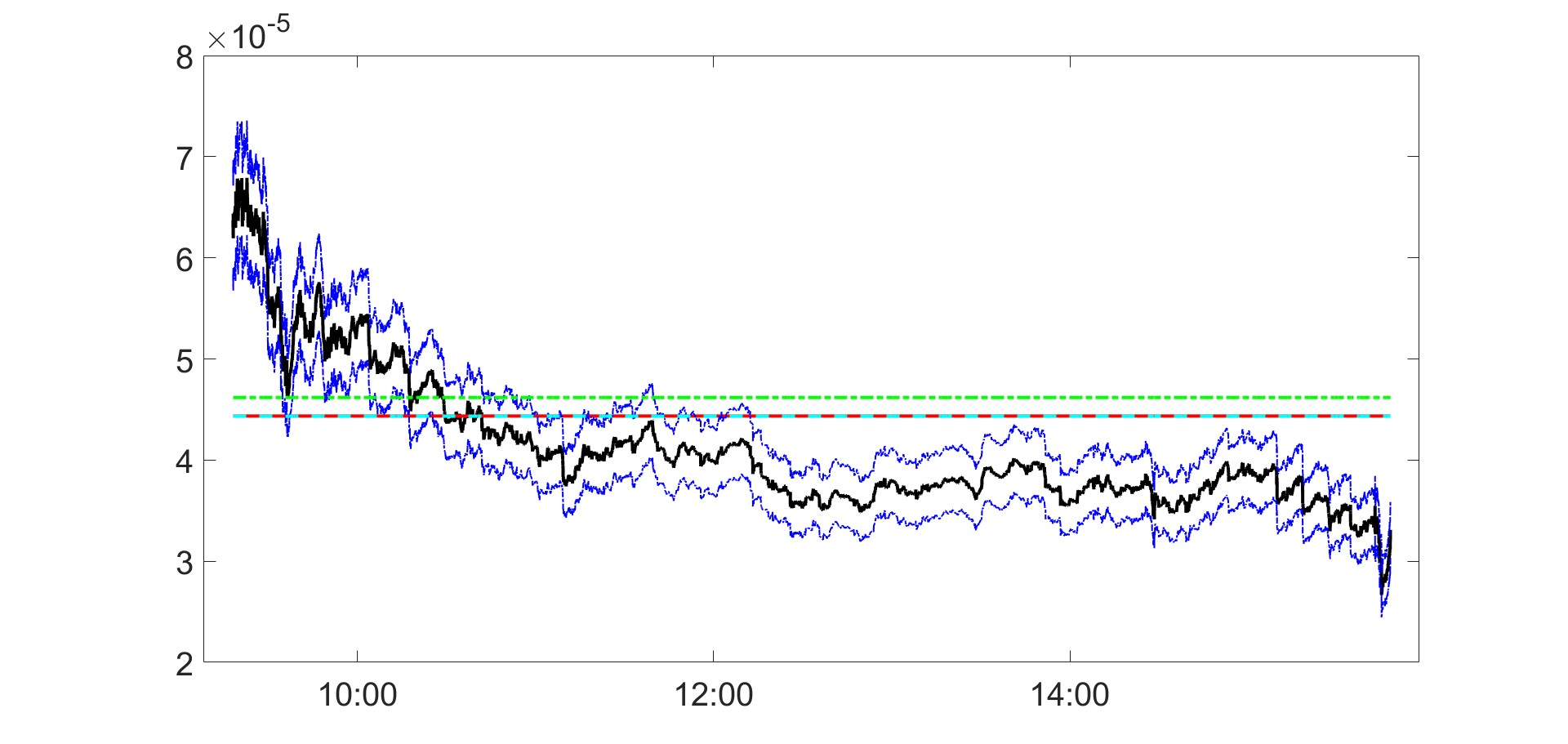}}
	\subfigure[$\text{LRCIRF}_{t}$ June 04, 2021 - MSFT]
	{\includegraphics[width=6.5cm]{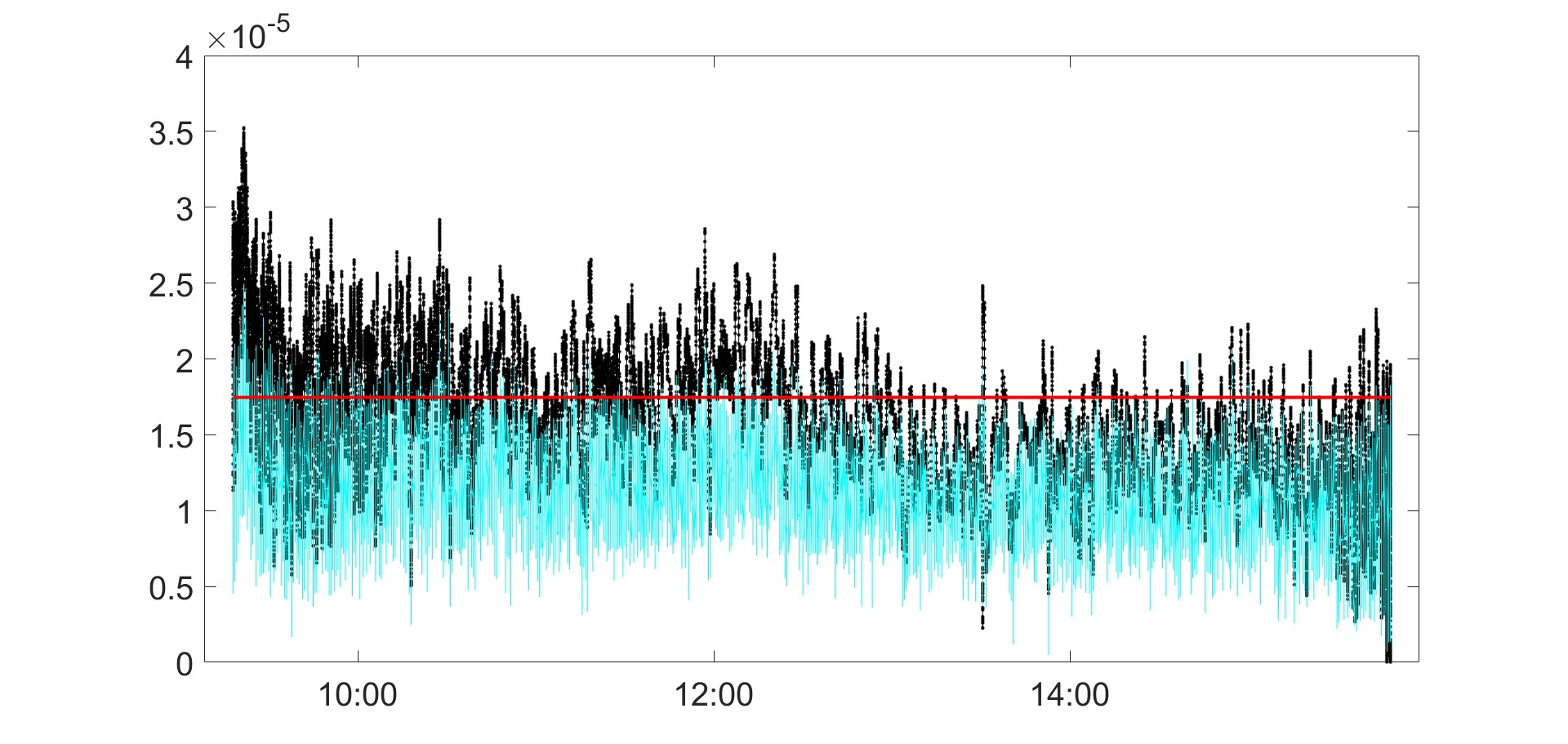}}
	\hspace{5mm}
	\subfigure[$\text{LRCIRF}_{t}$ June 04, 2021 - AMZN]
	{\includegraphics[width=6.5cm]{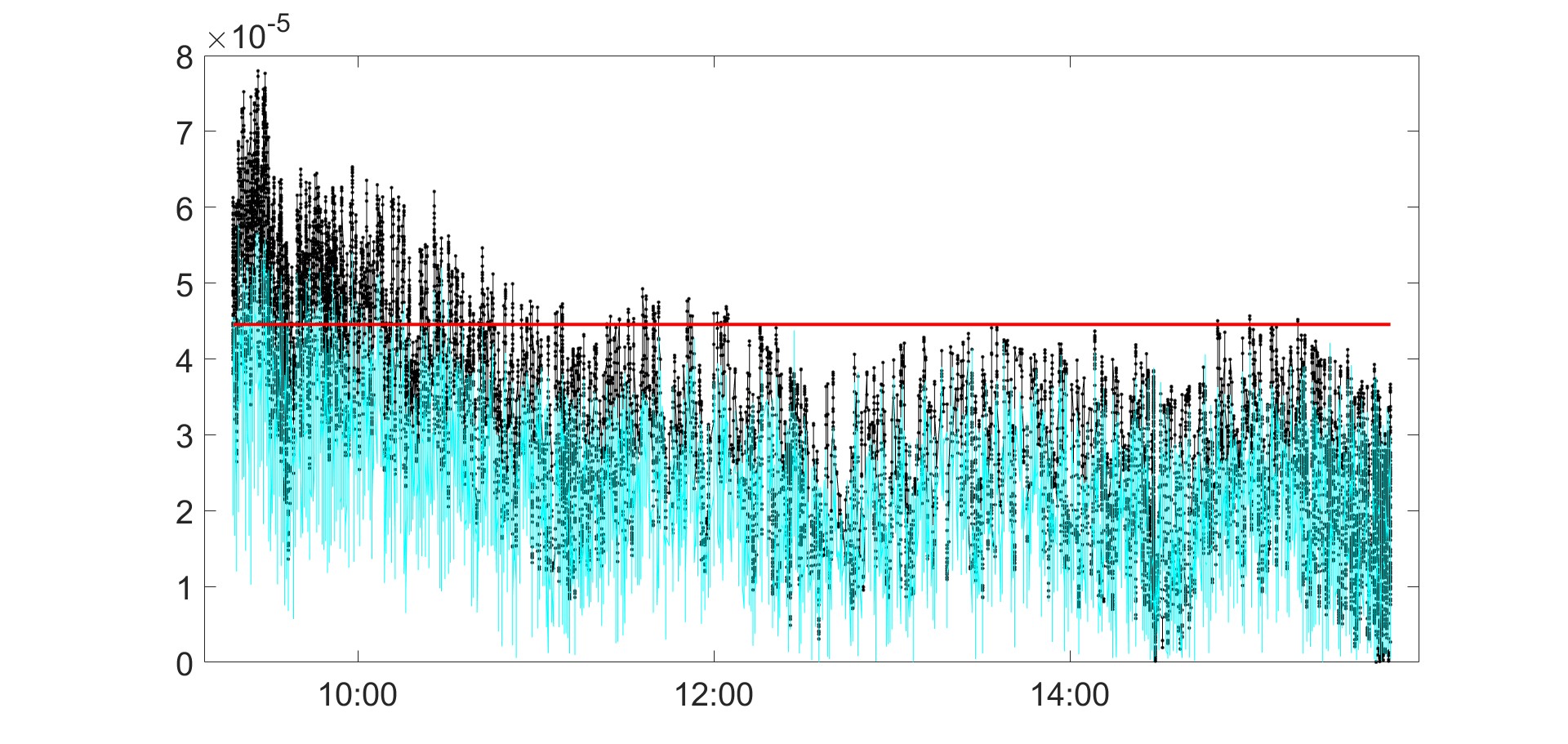}}
 	\subfigure[$\sigma_{t}$ June 04, 2021 - MSFT]
	{\includegraphics[width=6.5cm]{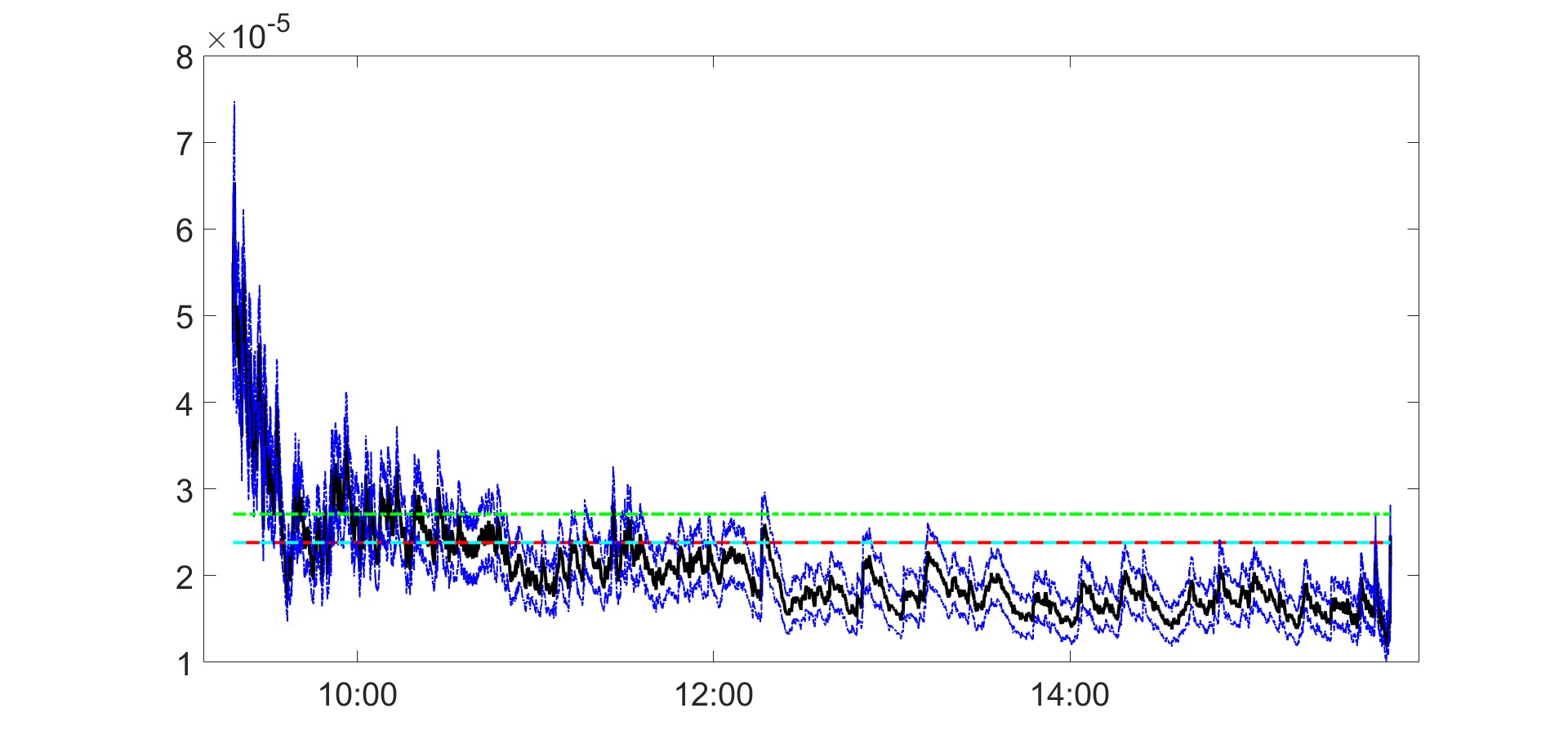}}
	\hspace{5mm}
	\subfigure[$\sigma_{t}$ June 04, 2021 - AMZN]
	{\includegraphics[width=6.5cm]{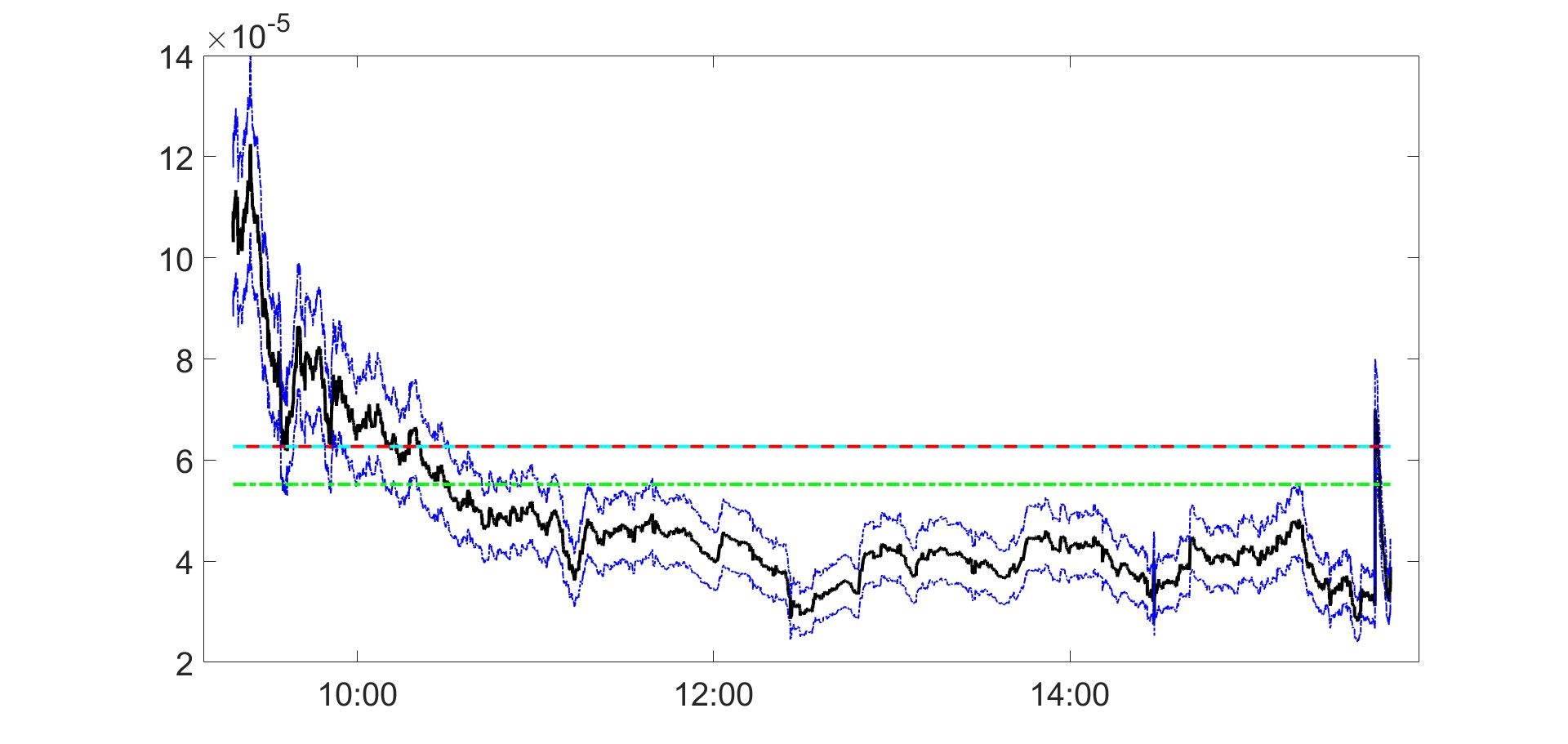}}
	\caption{\footnotesize Top panels: Estimates of the time-varying instantaneous impact parameter of the SDAMH model (black solid lines) and 95\% confidence bands (blue dashed lines), of the AMH model (cyan dashed lines), and of the AH model (red solid lines), and average intraday monthly variation (green dash-dotted line). Middle panels: Estimates of the long-term cumulative impulse response function ($\text{LRCIRF}_t$) of the SDAMH model (black dotted lines), of the AMH model (cyan dashed lines), and of the AH model (red solid lines). Bottom panels: Estimates of the time-varying volatility of the SDAMH model (black solid lines) and 95\% confidence bands (blue dashed lines). Results for MSFT and AMZN on June 04, 2021.}
	\label{b0tvsample}
\end{figure}

\begin{figure}[htt]
	\centering
    \textbf{Filtered time-varying parameters and LRCIRF$_{t}$}\par\medskip
	\subfigure[$b_{0,t}$ June 04, 2021 - CLSK]
	{\includegraphics[width=6.5cm]{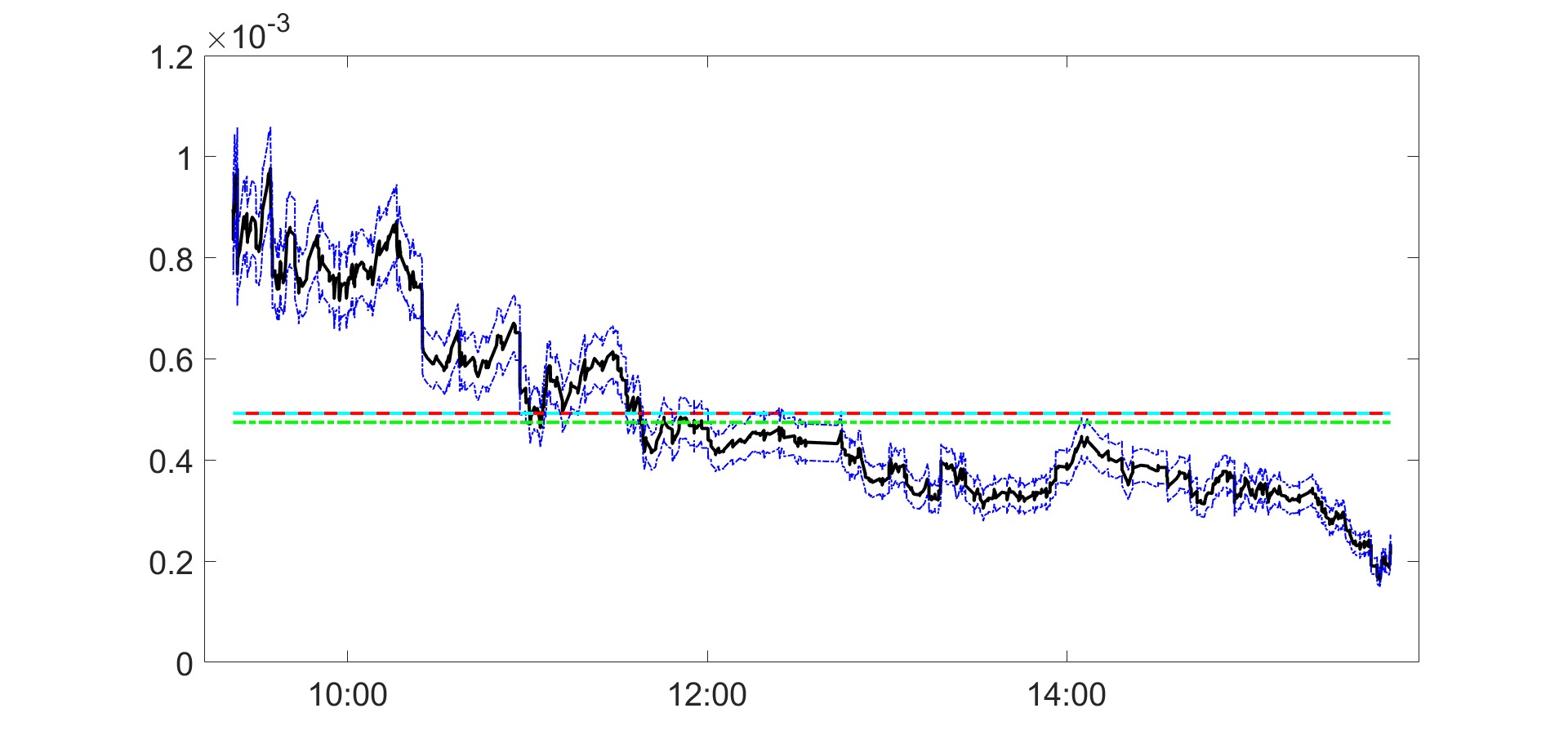}}
	\hspace{5mm}
	\subfigure[$b_{0,t}$ June 04, 2021 - FLGT]
	{\includegraphics[width=6.5cm]{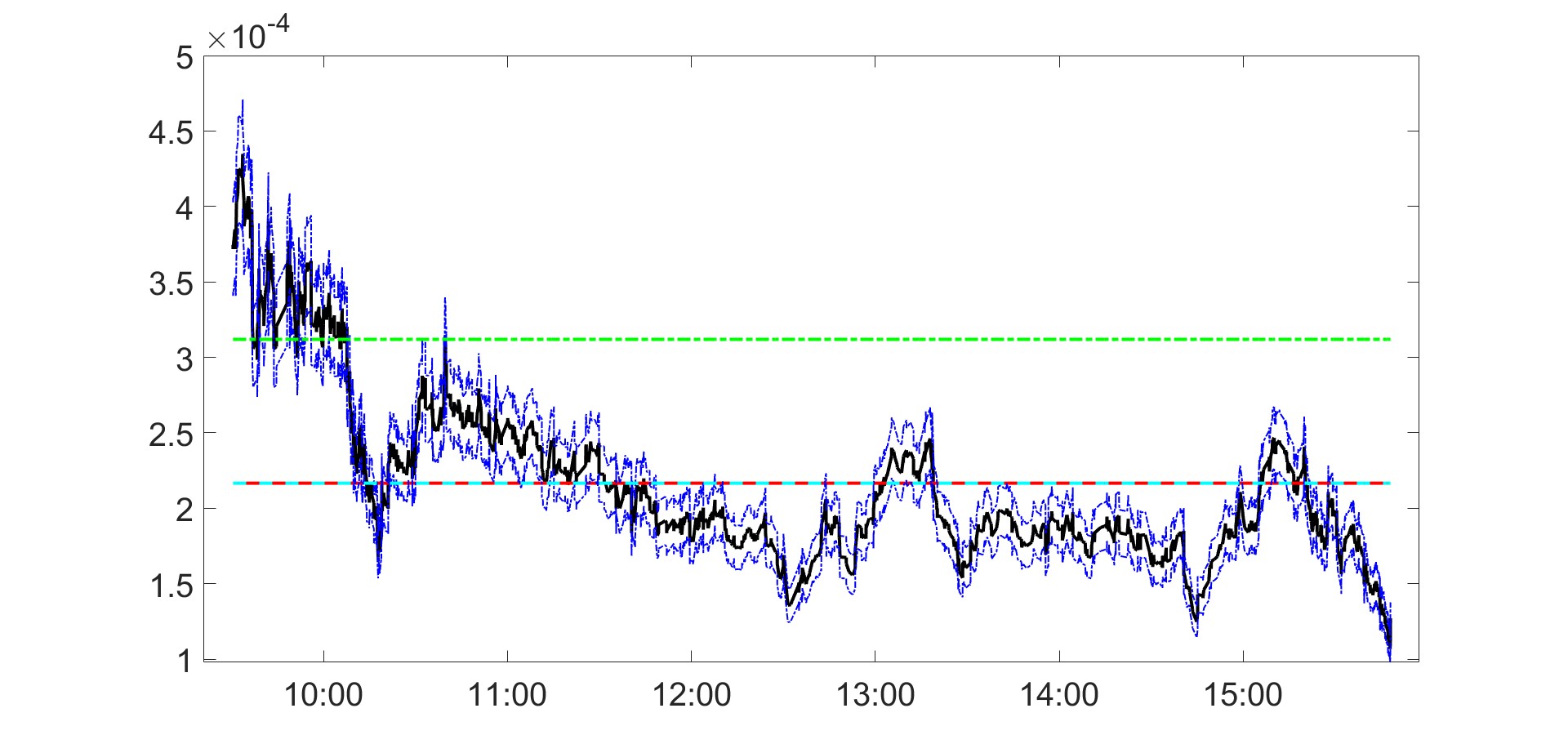}}
	\subfigure[$\text{LRCIRF}_{t}$ June 04, 2021 - CLSK]
	{\includegraphics[width=6.5cm]{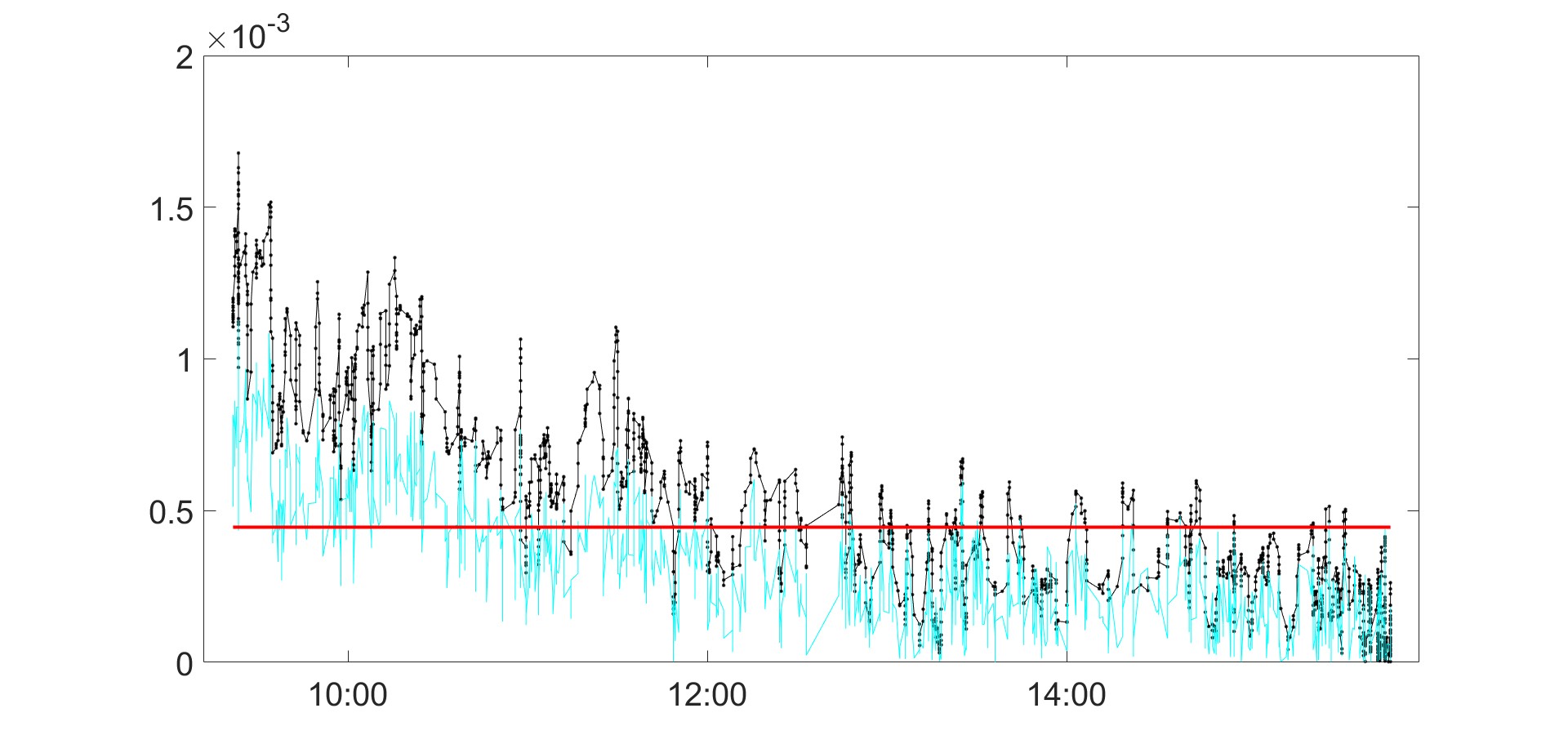}}
	\hspace{5mm}
	\subfigure[$\text{LRCIRF}_{t}$ June 04, 2021 - FLGT]
	{\includegraphics[width=6.5cm]{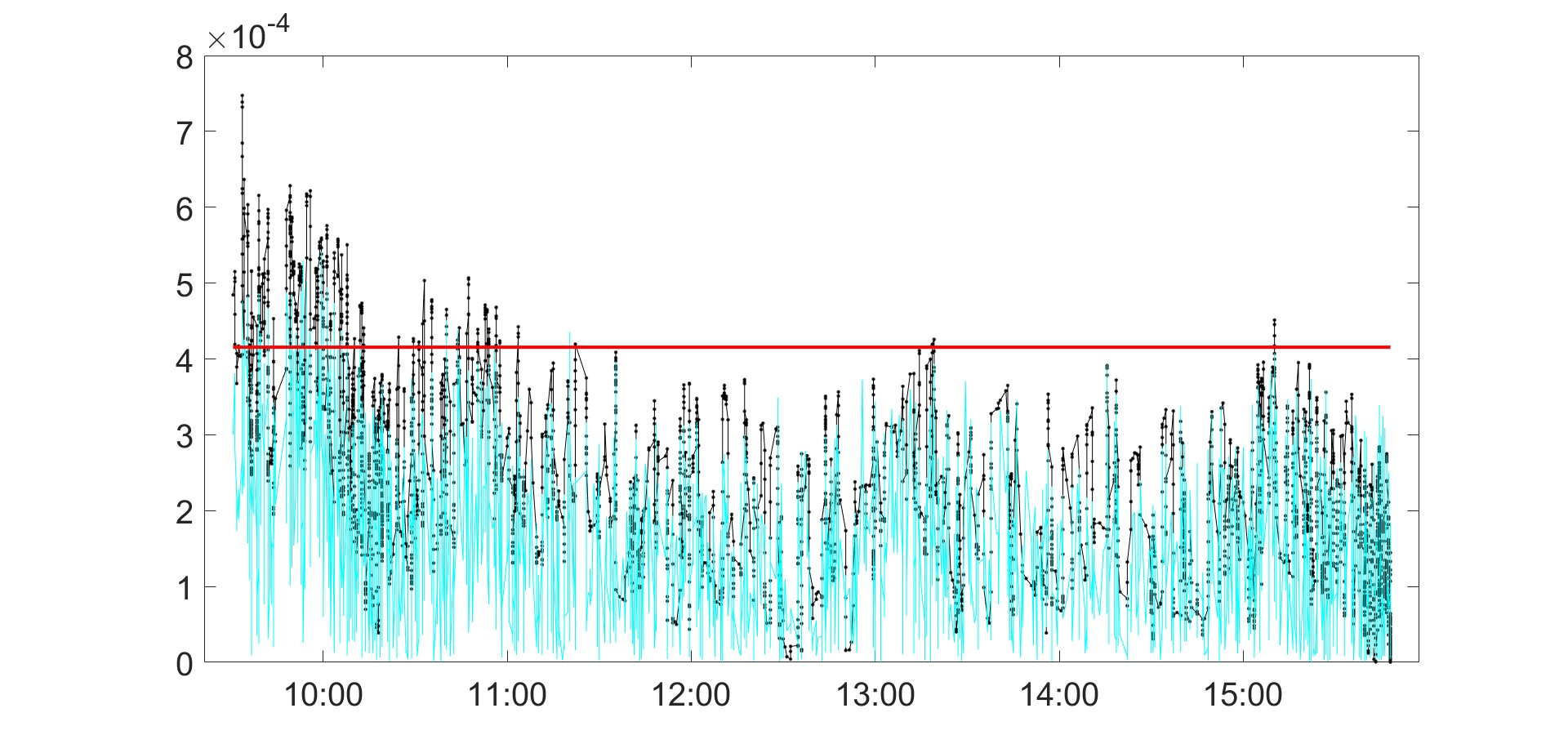}}
 	\subfigure[$\sigma_{t}$ June 04, 2021 - CLSK]
	{\includegraphics[width=6.5cm]{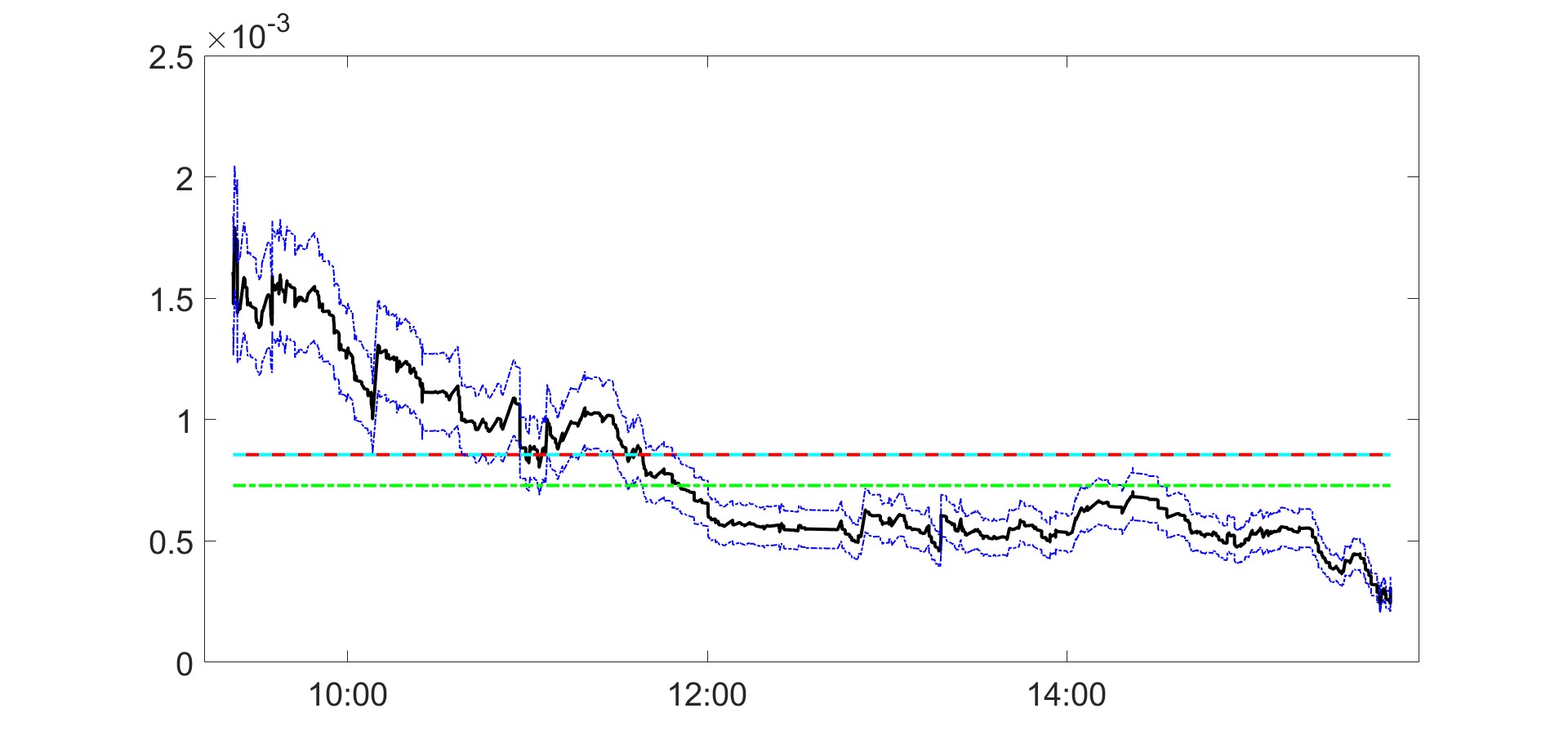}}
	\hspace{5mm}
	\subfigure[$\sigma_{t}$ June 04, 2021 - FLGT]
	{\includegraphics[width=6.5cm]{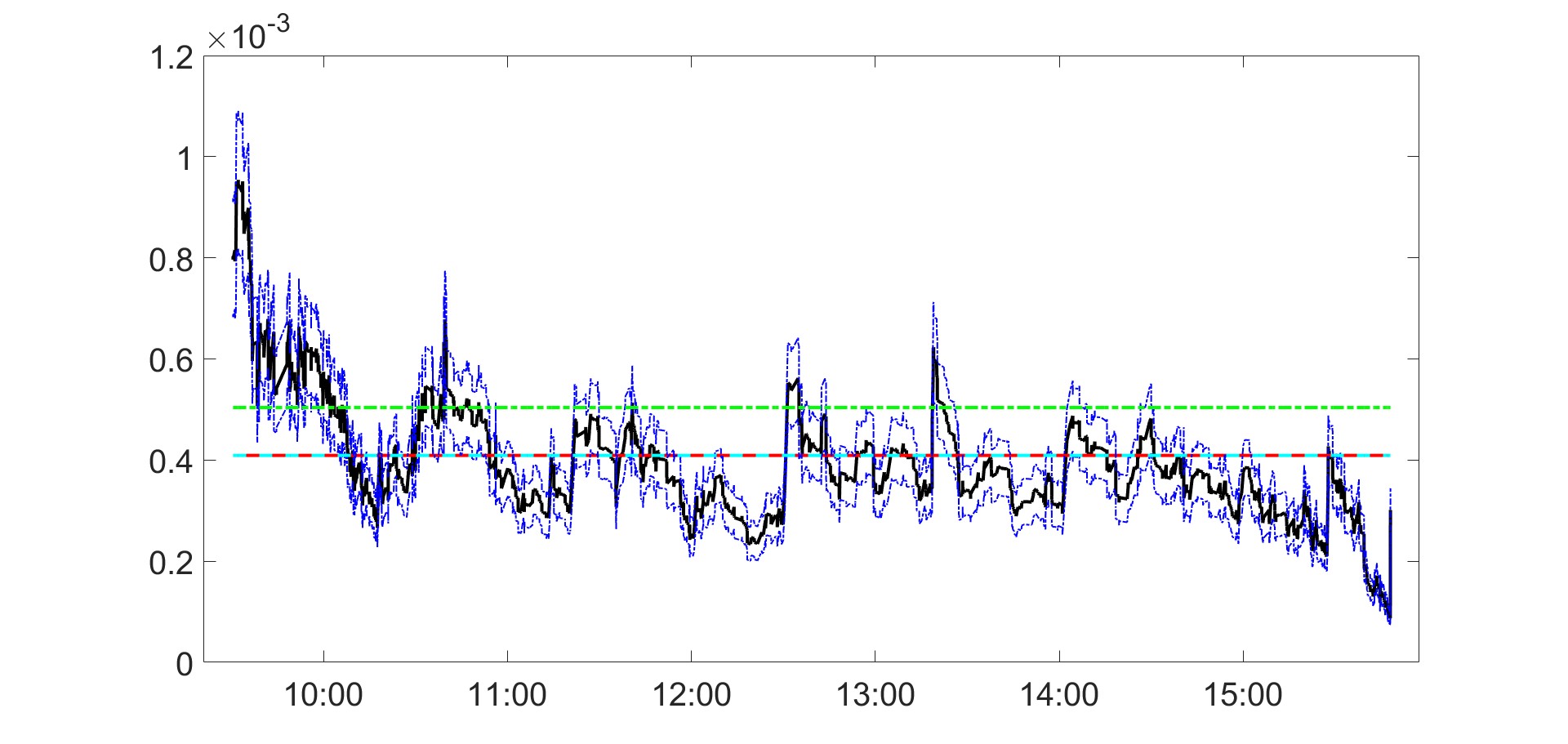}}
	\caption{\footnotesize 
    Top panels: Estimates of the time-varying instantaneous impact parameter of the SDAMH model (black solid lines) and 95\% confidence bands (blue dashed lines), of the AMH model (cyan dashed lines), and of the AH model (red solid lines), and average intraday monthly variation (green dash-dotted line). Middle panels: Estimates of the long-term cumulative impulse response function ($\text{LRCIRF}_t$) of the SDAMH model (black dotted lines), of the AMH model (cyan dashed lines), and of the AH model (red solid lines). Bottom panels: Estimates of the time-varying volatility of the SDAMH model (black solid lines) and 95\% confidence bands (blue dashed lines). Results for CLSK and FLGT on June 04, 2021.}
	\label{b0tvsample1}
\end{figure}

To better understand the role of the time-varying instantaneous impact in price discovery, we contrast the results with those from two constant parameter models, the AMH and a novel model that we name Aggregated Hasbrouck (AH). The latter follows by applying the lag aggregation to the Hasbrouck model. The AH model reads as 
\begin{align}\label{aggmhsbrouckmodel}
r_t & = \mu_{1} + a_1 r_{t-1} +  b_{0} x_t + b_1 x_{t-1} + \bar{b}_{10} x_t^{(10)} + \bar{b}_{100} x_t^{(100)} + v_{1,t}, & \nonumber \\ 
x_t & = \mu_{2}  + c_1 r_{t-1} +  d_1 x_{t-1} + \bar{d}_{10} x_t^{(10)} + \bar{d}_{100} x_t^{(100)} + v_{2,t}\,,
\end{align}
with constant volatility parameters and is a linear model with the same number of lags of the SDAMH model. A comparison with the AH model highlights the effect of the non-linearities associated with the logit and DCS specifications, while contrasting with the AMH allows to isolate the effect of the DCS specification.  We consider the dynamics for one representative day, namely June 04, 2021 for the sample stocks. The top panels of Figures \ref{b0tvsample} and \ref{b0tvsample1} show the temporal evolution of the time-varying parameter $b_{0,t}$ (black dots) of the SDAMH model compared with the corresponding constant parameter $b_0$ for the AMH (cyan dashed lines) and the AH (red solid lines) models. We notice how the time variation of $b_{0,t}$, modeled through integrated DCS dynamics, enlightens several unprecedented features concerning the impact of $x_t$ on $r_t$. The price impact is not constant throughout the day but is high during the early hours of the day and decreases as time approaches the end of the trading activity. As mentioned previously, the impact shows occasional intra-day spikes, which could be associated with abrupt liquidity crises. Relying on \cite{blasques2016sample}, we compute the time-varying confidence bands for $b_{0,t}$ using Monte Carlo simulations (dashed blue lines). The bands confirm the sizeable temporal variability of $b_{0,t}$. The instantaneous impact estimates associated with the constant parameter models (AMH and AH) are not suitable for capturing these dynamical features. Furthermore, we notice that the estimated values for $b_0$ for the AH and AMH models are numerically consistent with the daily average of the filtered time-varying parameter $b_{0,t}$. Please refer to Table~\ref{constant_bknot} for the numerical details. The bottom panels of Figures \ref{b0tvsample} and \ref{b0tvsample1} show the filtered dynamics of $\sigma_t$ for the same sample day and stocks.
\begin{table}[!ht]
  \centering
    \begin{tabular}{@{\extracolsep{1pt}} lcccc}
    & \textbf{MSFT}      
    & \textbf{AMZN}
    & \textbf{CLSK}
    & \textbf{FLGT}\\ 
    \hline
    AH &   4.141 $ \cdot 10^{-6}$   & 4.891$ \cdot 10^{-5}$ & 4.822$ \cdot 10^{-4}$ & 3.256$ \cdot 10^{-4}$ \\
    & (1.064 $\cdot 10^{-6}$) & (5.436 $\cdot 10^{-6}$) & (7.413 $\cdot 10^{-5}$) & (6.060 $\cdot 10^{-5}$) \\
    AMH &  4.140 $ \cdot 10^{-6}$ & 4.888$ \cdot 10^{-5}$ & 4.823$ \cdot 10^{-4}$ & 3.255$ \cdot 10^{-4}$ \\ 
    & (1.061 $ \cdot 10^{-6}$) & (5.439 $ \cdot 10^{-6}$) & (7.653 $ \cdot 10^{-5}$) & (6.817 $ \cdot 10^{-5}$) \\
    SDAMH (time average) & 4.702$ \cdot 10^{-6}$ &  4.618$ \cdot 10^{-5}$ & 4.746$ \cdot 10^{-4}$ & 3.119$ \cdot 10^{-4}$ \\
    \hline
    \end{tabular}
    \caption{\footnotesize Estimated values for $b_0$ and associated standard deviations (in brackets) for the constant parameter models, AH and AMH.}
    \label{constant_bknot}
\end{table}

\subsection{Conditional CIRF Analysis}
\label{section:cirf}

As stated in the original \cite{hasbrouck1991measuring} paper, one of the objectives of the model is to calculate the information content of a trade from the LRCIRF obtained with the impulse response analysis. While this is a straightforward computation for a constant parameter SVAR, the non-linearity of the logit specification and time variability of $b_{0,t}$ and $\sigma_t$ in our framework require a simulation approach, as described in Section \ref{section:cirf}, which provides a time $t$ conditional IRF. The crucial difference, and advantage, of our model with respect to competitors is to provide a time-varying LRCIRF that depends both on the return and trade history (because of the non-linearity from the logit and the DCS specifications) and the time variability of the instantaneous impact and return volatility. To decouple the role of the different components in $\text{LRCIRF}_t$, we define $\mu^{(*)}_{t-1} \doteq \mu^{(1)}_{t}-b_{0,t}x_t$, which is the mean of $r_t$ conditional on the past return $r_{t-1}$ and past trade signs. We refer to the quantity $\mu^{(*)}_{t-1}$ as the \textit{state} of the market at specific trade $t$. For the computation of $\text{LRCIRF}_t$ we use the CIRF at lag $H=20$ averaged across $S=1,000$ simulations. The choice $H=20$ is justified by the fact that for all stocks we observe that the CIRF is approximately constant at longer horizons~\footnote{To be precise, at this lag, the increase of the CIRF is smaller than 1\%. See also Figure~\ref{phasesMSFT} showing the IRFs for the two stocks on a specific day.}. It is worth stressing that, since for the linear models the IRF does not depend on the state of the market and it is linear in the shock, computing it adding an extra buy order to a buy/sell trade is equivalent to compute it by adding an extra sell order and changing the overall sign. In contrast, because of the highly non-linear specification, for the AMH and SDAMH models the computation of the IRF depends both on the nature of the shock -- a buy or a sell order -- and on the state of the market -- that is, whether one shocks an existing buy or sell trade. In the following analysis, we compute the LRCIRF shocking the market with a buy trade. We repeated the computation with a sell trade and verified that the overall message is not changing (while the long-term impact changes, as expected). These results are available upon request.

The middle panels of Figures \ref{b0tvsample} and \ref{b0tvsample1} show the temporal evolution of the time-varying LRCIRF, i.e., in Hasbrouck's jargon, the information content of a trade or its long-term impact. Apart from noisy fluctuations, we observe a time variation of the SDAMH LRCIRF (black dotted lines), which partially mirrors the dynamics of $b_{0,t}$ in the upper panels. The reason is, of course, the non-linearity of our model, whose LRCIRF$_t$ depends both on the time-varying $b_{0,t}$ and on the state $\mu^{(*)}_{t-1}$ of the market at time $t$. As expected, the LRCIRF implied by the AH model is constant through time. Finally, it is important to notice that the AMH LRCIRF (cyan dashed lines) varies with time and, as expected, changes conditionally on the state of the market, even though the instantaneous impact parameter $b_0$ is constant. The message from these panels is insightful. The comparison of Hasbrouck's trade information content -- a quantity that, by construction in the original model, is constant throughout the day -- with the time-varying proxies provided by the AMH and SDAMH cases makes evident the value added by our approach. 
The information content varies on a tick-time basis because the market environment changes, and the instantaneous impact is time-varying. This behavior is due to the combined effect of dropping the linear specification of the VAR model and including a DCS specification for the impact dynamics. Performing the original Hasbrouck procedure on a running window during the day to capture possible time variation significantly differs from what we do here. Two are the main reasons. First, the model remains linear; then, by construction, the LRCIRF can not depend on the current state of the market. Second, the level of the model parameters will change over different estimation windows, but the parameters will remain fixed when computing the impulse response functions. Our approach marks an important difference because, during the LRCIRF Monte Carlo computation, the instantaneous time-varying impact reacts dynamically to different drawings of future returns and trade signs. Then, we pay a reasonable price for computational efficiency, but we gain a more realistic description of return and sign behavior, with impact and volatility fast reacting to unexpected market events.

To disentangle and quantify the role of the time-varying impact parameter $b_{0,t}$ and of the state $\mu^{(*)}_{t-1}$ on the LRCIRF, we consider the regression model:
\begin{equation}
    \text{LRCIRF}_{t} = \gamma_{0} + \gamma_{1} b_{0,t} + \gamma_{2} |\mu^{(*)}_{t-1}| + \eta_{t}. \label{eqreg1}
\end{equation}
We estimate the regression coefficients for each of the days in our sample for all tickers and we report in Table \ref{regtab} the average of the estimated coefficients~\footnote{The average parameter estimates are computed as $\gamma_i := \sum_{\tau = 1}^{22}\gamma_{i,\tau}/22, \ i \in \{0,1,2,3\}$, where $\gamma_{i,\tau}$ represents the $i$-th coefficient estimated on day $\tau$. The same holds for $R^2$.}. Furthermore, we count the number of times that we do not reject the null hypothesis at the significance level 5\% ($\# \{\text{Pval} > 5\% \}$) and report $R^2$. Most of the time, the point estimates are statistically significant (see $\# \{\text{Pval} > 5\% \}$ in Table \ref{regtab}). Moreover, the high $R^2$ shows that most of the variation of the long-run CIRF is captured by the two explanatory variables. Specifically, the $\gamma_0$ coefficient is statistically significant, as well as the other parameters $\gamma_1$ and $\gamma_2$. The $\gamma_{1}$ is positive, while the $\gamma_{2}$ is negative~\footnote{Their results on the coefficients are valid also when performing a regression with only one independent variable at the time.}. The sign of these coefficients validates the finding that the information content of stock trade positively depends on the instantaneous impact and negatively on the state of the market. According to the regression, the information content of stock trade consists of three parts: the constant ($\gamma_0$), the instantaneous impact ($\gamma_1 b_{0,t}$), and the \textit{state} of the market ($\gamma_{2} |\mu^{(*)}_{t-1}|$). We quantify the weight of each part by computing the monthly (absolute) average of each component divided by the sum across the components. The analysis indicates that the most important term explaining the information content of stock trades is the instantaneous impact component, followed by the \textit{state} component. The least important part is the constant. In fact, for MSFT, the term $\gamma_1 b_{0,t}$ accounts for 58\%, the term $\gamma_{2} |\mu^{(*)}_{t-1}|$ for 37\%, and the remaining $\gamma_0$ for 5\% of the weight. For AMZN, the weights are 69\%,  27\%, and 4\%, respectively. For CLSK, the weights are 65\%,  30\%, and 5\%. Similarly, for FLGT, the weights are 72\%,  24\%, and 4\%. We quantify the variability of $\text{LRCIRF}_t$ explained by the two explanatory variables using the decomposition of variance of the linear model in Eq.~\eqref{eqreg1}. Table \ref{anvaperc} shows that on average the variance portion explained by the time-varying impact parameter $b_{0,t}$ is greater than that explained by the (absolute) value of the conditional information of the state. This indicates the importance of having a market impact that varies over time to adequately account for the variability of the information content of the trade. To better understand the relation between $b_{0,t}$ and $\mu^{(*)}_{t-1}$, we consider the linear model
\begin{equation}
 b_{0,t} = \gamma_{0} + \gamma_{3} |\mu^{(*)}_{t-1}| + \eta_{t}\,. \label{eqreg4}
\end{equation}
The estimated coefficient $\gamma_{3}$ is negative and statistically significant, indicating an inverse relation between the variables. When the absolute state $|\mu^{(*)}_{t-1}|$ is large (small), that is, the past history of price and trades predicts a large (small) price change, the instantaneous impact of a trade $b_{0,t}$ is small (large). This behavior is consistent with the fluctuating liquidity hypothesis first introduced in \cite{lillo2004long} and further investigated in \cite{taranto2014adaptive}. As a robustness check of this analysis, in Online Appendix \ref{section:valresmhm} we extend the regression study in three directions. First, to partially control for the high-variability of the impact and volatility during the opening and closing phases of the market, we restrict the regression to observations in the time interval from 10:30 to 16:00. The estimated coefficients align with those estimated by using the whole trading day. As a second robustness check, we consider the AMH model, which inherits non-linearity and state dependence from the logit specification but has constant instantaneous impact and volatility. By regressing the LRCIRF against $|\mu^{(*)}_{t-1}|$, we find, again, that the state affects the LRCIRF. These findings further support the importance of the state in determining the information content of stock trades. As a final robustness check, we examine the role of the trade size on the long-run CIRF and in order not to suffer from the possible attenuation bias caused by pooling buy and sell orders, we run separate regressions conditioned on the trade direction. As shown by the detailed results presented in Online Appendix D, the trade size has no systematic incremental explanatory power over the time-varying instantaneous impact $b_{0,t}$ and the market state $|\mu_{t-1}^{(*)}|$. The baseline specification without explicit volume regressors is thus fully sufficient.
\begin{table}[!ht]
  \centering
   \textbf{Regression results}\par\medskip
   		\resizebox{150mm}{40mm}{
    \begin{tabular}{@{\extracolsep{1pt}} lllllllll}
			& \multicolumn{4}{c}{\textbf{MSFT}}      & \multicolumn{4}{c}{\textbf{AMZN}}               \\ 
			\hline
			          & $\gamma_0$ & $\gamma_1$ & $\gamma_2$  & $\gamma_{3}$ & $\gamma_0$ & $\gamma_1$ & $\gamma_2$ & $\gamma_{3}$    \\ 
	Est &   1.464$\cdot 10^{-5}$   & 0.602   &  -5.501$\cdot 10^{-3}$  & - &  2.283$\cdot 10^{-5}$  & 0.841 &  -9.164$\cdot 10^{-3}$  & -   \\
			$ \# \{\text{p} > 5 \% \}$ & 5 &  0        & 3 & -   &      4    &    1   &  3  &-          \\ 
			$R^2$& 0.551& & & &  0.423 & &  \\\hline
	Est   &  4.628$\cdot 10^{-6}$     & -   &  -  &  -3.738$\cdot 10^{-3}$  &     1.565$\cdot 10^{-5}$  & -& -  &  -6.328$\cdot 10^{-3}$  \\
			$ \# \{\text{p} > 5 \% \}$ & 3 &  -  & - & 4     &    1 & -  &  -  &  4         \\ 
			$R^2$& 0.373 & & & & 0.384 & &  \\
   			& \multicolumn{4}{c}{\textbf{CLSK}}      & \multicolumn{4}{c}{\textbf{FLGT}}               \\ 
			\hline
			          & $\gamma_0$ & $\gamma_1$ & $\gamma_2$  & $\gamma_{3}$ & $\gamma_0$ & $\gamma_1$ & $\gamma_2$ & $\gamma_{3}$    \\ 
	Est &   9.919$\cdot 10^{-5}$   & 1.048   &  -1.394$\cdot 10^{-3}$  & - &  2.813$\cdot 10^{-5}$  & 1.011 &  -6.636$\cdot 10^{-3}$  & -   \\
			$ \# \{\text{p} > 5 \% \}$ & 6 &  3        & 5 & -   &      5    &    3  &  5  &-          \\ 
			$R^2$& 0.447& & & &  0.363 & &  \\\hline
	Est   &  4.625$\cdot 10^{-5}$     & -   &  -  &  -1.050$\cdot 10^{-3}$  &     4.043$\cdot 10^{-5}$  & -& -  &  -5.500$\cdot 10^{-3}$  \\
			$ \# \{\text{p} > 5 \% \}$ & 5 &  -  & - & 4     &    4& -  &  -  &  5         \\ 
			$R^2$& 0.331 & & & & 0.304 & &  \\
			\hline 
    \end{tabular}}
   \caption{\footnotesize Results of regressions of Eqs. \eqref{eqreg1} (top) and \eqref{eqreg4} (bottom). Temporal average of the point estimates (Est), number of times we do not reject the null hypothesis at 5\% level $ \# \{\text{p} > 5 \% \}$, and average $R^2$.}
   	\label{regtab}
\end{table}
\begin{table}[!ht]
  \centering
       \textbf{Variance decomposition}\par\medskip
    \begin{tabular}{@{\extracolsep{1pt}} lllllllll}
		&  & \textbf{MSFT} &  &  & \textbf{AMZN} &  \\ \hline
		& $\text{Var} \ b_{0,t} $ & $\text{Var} \ |\mu^{(*)}_{t-1}| $ & $\text{Var} \ \text{Err} $ & $\text{Var} \ b_{0,t} $ & $\text{Var} \ |\mu^{(*)}_{t-1}| $ & $\text{Var} \ \text{Err}$ \\ \hline
        Est (\%) & 34.5 &   26.3 &   39.2 &   31.4 &   21.3 &  47.3 \\
        Std (\%) & (8.6) &   (7.4) &  (13.0) &  (10.3)   & (6.4) &  (20.1) \\		 
        		&  & \textbf{CLSK} &  &  & \textbf{FLGT} &  \\ \hline
		& $\text{Var} \ b_{0,t} $ & $\text{Var} \ |\mu^{(*)}_{t-1}| $ & $\text{Var} \ \text{Err} $ & $\text{Var} \ b_{0,t} $ & $\text{Var} \ |\mu^{(*)}_{t-1}| $ & $\text{Var} \ \text{Err}$ \\ \hline
        Est (\%) & 32.7 &   28.1 &   39.2 &   39.2 &   25.6 &  45.2 \\
        Std (\%) & (9.1) &   (6.3) &  (12.2) &  (8.4)   & (5.9) &  (14.7) \\		\hline 
    \end{tabular}
  \caption{\footnotesize Temporal average (Est) of the portions of variance decompositions of $b_{0,t}$ ($\text{Var} \ b_{0,t}$), of $|\mu^{(1)}_{t}|$ ($\text{Var} \ |\mu^{(*)}_{t-1}|$) and of the error part (Var Err) and daily standard deviation (Std) for MSFT, AMZN, CLSK, and FLGT.} 
	\label{anvaperc}
\end{table}

\section{Event study: FOMC announcement}
\label{section:fomc}

To highlight the ability of our model to capture in real-time the dynamics of market impact and of the information content of trade, as described by the LRCIRF, we focus the analysis on a day with pre-announced macroeconomic news. Specifically, we consider June 16, 2021, when, at 14:00, the Federal Open Market Committee released the announcement communicating decisions to implement the monetary policy stance. The most eagerly awaited one was the choice of the interest rate on reserve balances (IORB). The Board of Governors of the Federal Reserve System voted to raise the IORB from 0.10 \% to 0.15 \%. In the following, we study the evolution of the CIRF and the behavior of its long-term value to this day~\footnote{The FOMC analysis is conducted for the liquid stocks in our sample. Results for the other stocks are available upon request.}. 
 
\subsection{Long run cumulative impulse response function}

 Figure~\ref{lfomcall} reports the filtered dynamics of the long-term cumulative impulse response function for MSFT and AMZN on June 16, 2021. Panels (a) and (d) show the $\text{LRCIRF}_t$ for the three competing models. The estimates of the SDAMH model (black dotted lines) are jointly affected by the conditional information at time $t$ and by the level of the instantaneous impact parameter. We observe that the long-term impact is higher at the beginning of the day when the instantaneous impact $b_{0,t}$ is higher (see panels (b) and (e)) and when $\mu^*_{t-1}$ is lower (see panels (c) and (f)). In the first phase of the day, the variation of $\text{LRCIRF}_t$ resembles the dynamics of $b_{0,t}$. At 14:00 the FOMC releases the news about the IORB. Consequently, the $\text{LRCIRF}_t$ increases sharply, reaching its maximum at the announcement time, and then it declines with several upward shocks as the time approaches the end of the day.  For the AMH model (cyan dashed lines), the $\text{LRCIRF}_t$ exhibits a less pronounced but similar behavior. On the one hand, the constant impact parameter keeps the $\text{LRCIRF}_t$ of the AMH model lower than the one of the corresponding DCS version. On the other hand, the relevance of the state (please refer to Online Appendix \ref{section:valresmhm}) makes the two dynamics similar and very reactive to announcements. Finally, the constant $\text{LRCIRF}_t$ of the AH model  (red lines) is much higher than the average LRCIRF of the other two models. Clearly, on days when one expects a large variability of the trade information content (before and after the announcement)
 the standard Hasbrouck's model does adapt to changing market conditions. When peaks are present (as in the FOMC day) the AH model tends to overestimate the information content. The non-linearity and the time variation of the instantaneous impact of the SDAMH model are better suited to capture the dynamic nature of the trade information content. 

\subsection{The shape of the CIRF}

\begin{figure}[htt]
    \centering
    \textbf{Estimates of LRCIRF$_t$, $b_{0,t}$ and $|\mu_{t-1}^*|$ during an FOMC day}\par\medskip
    \begin{minipage}{0.45\textwidth}
        \centering
        \subfigure[]
        {\includegraphics[width=\textwidth,height=4cm]{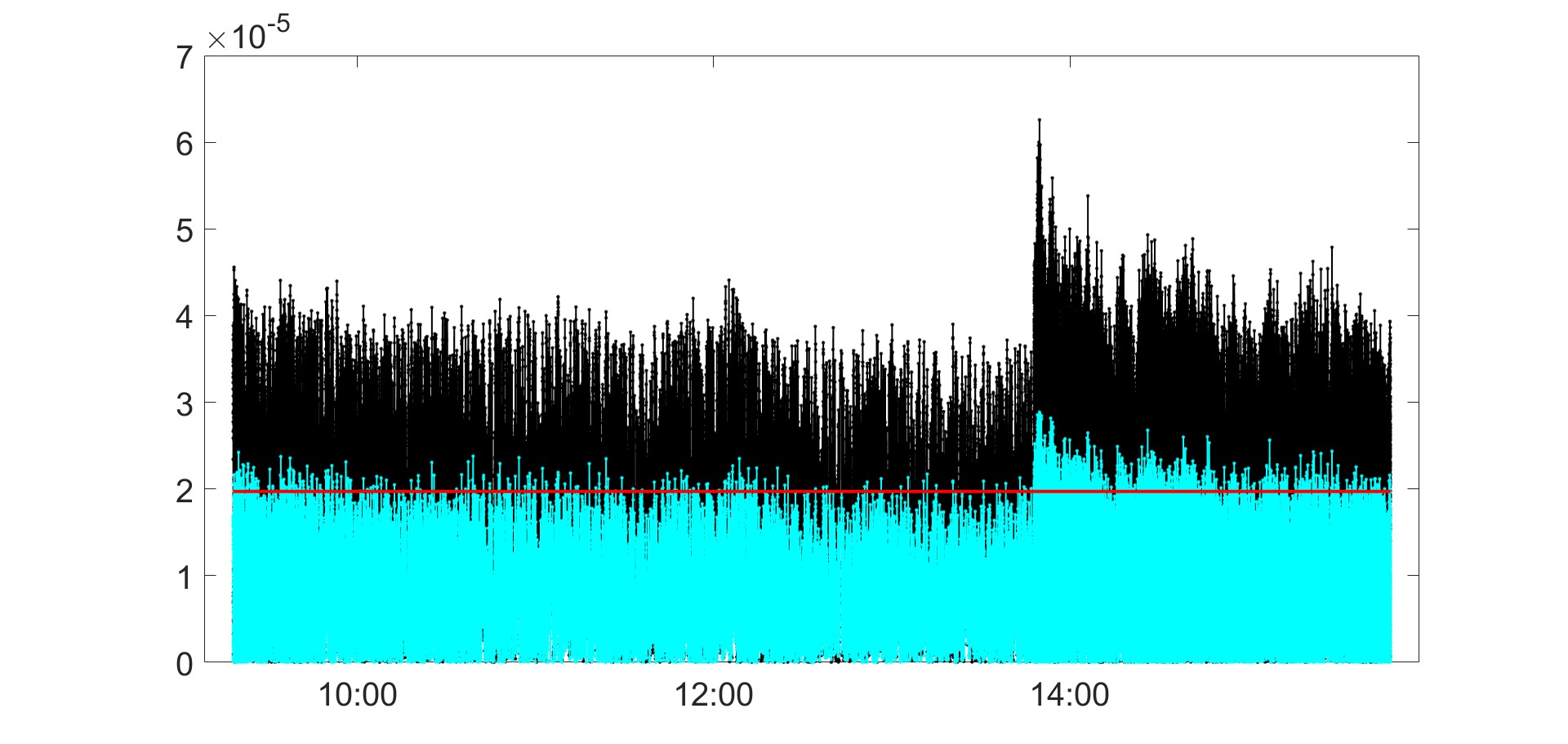}}\par\medskip
        \subfigure[]
        {\includegraphics[width=\textwidth,height=2.5cm]{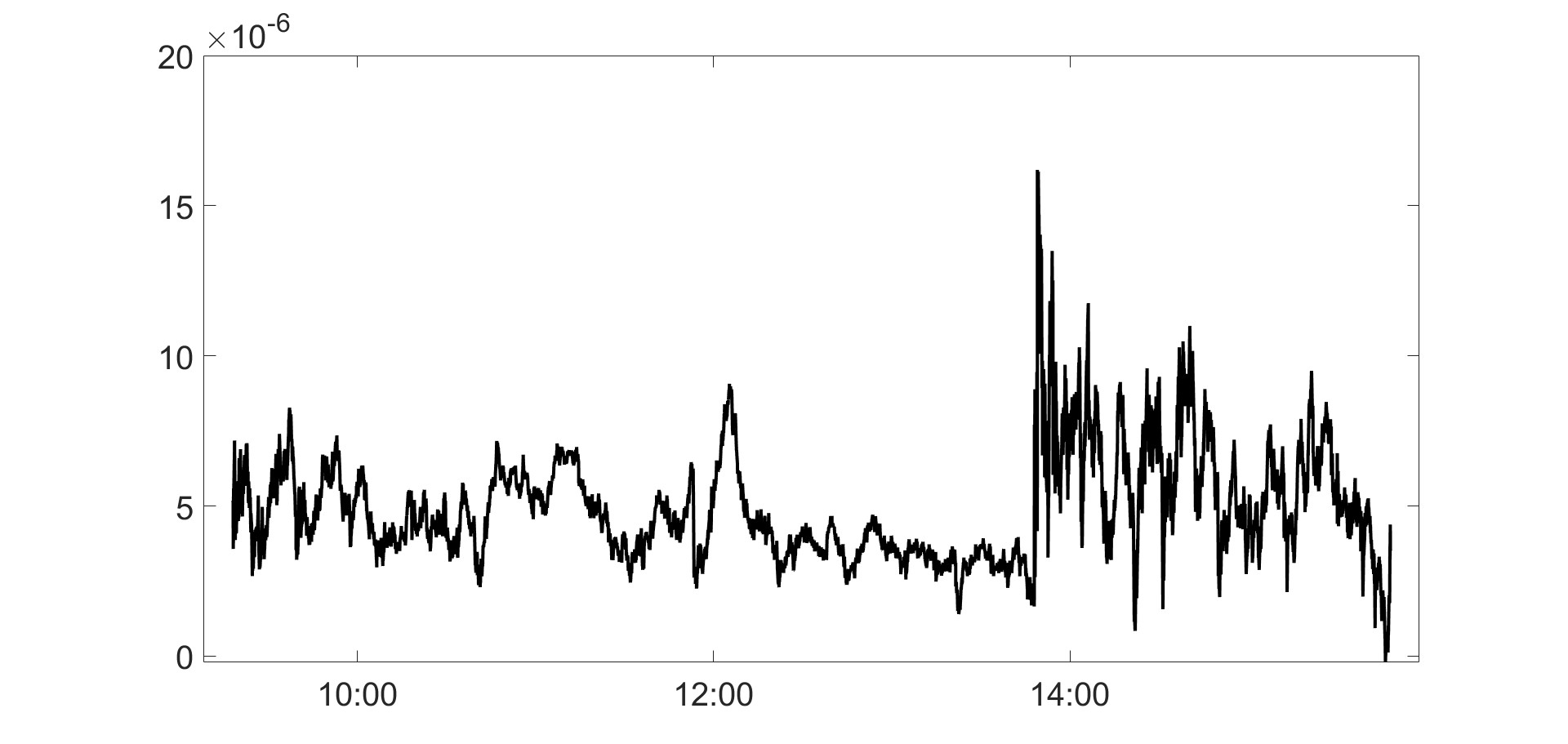}}\par\medskip
        \subfigure[]
        {\includegraphics[width=\textwidth,height=2.5cm]{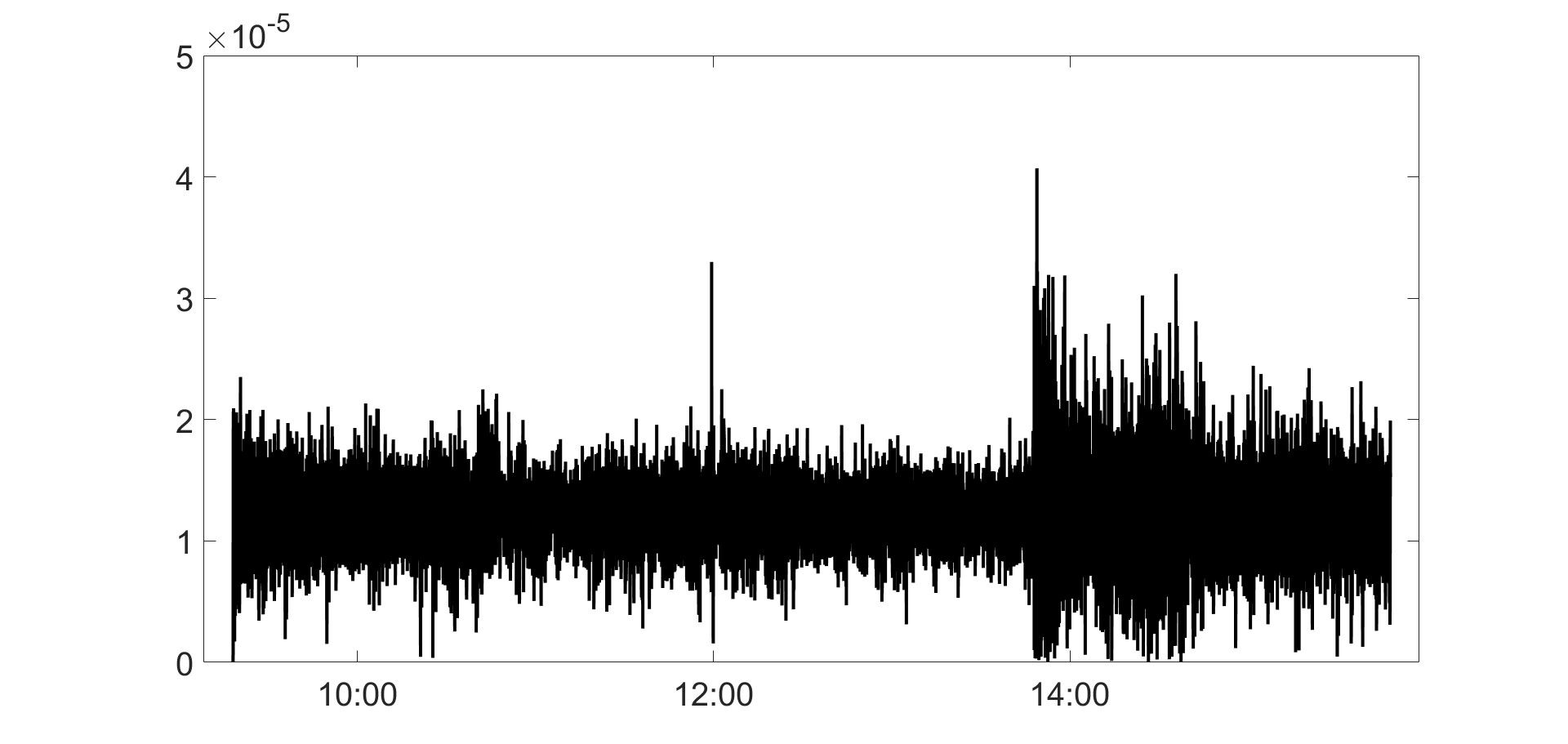}}
    \end{minipage}
    \begin{minipage}{0.45\textwidth}
        \centering
        \subfigure[]
        {\includegraphics[width=\textwidth,height=4cm]{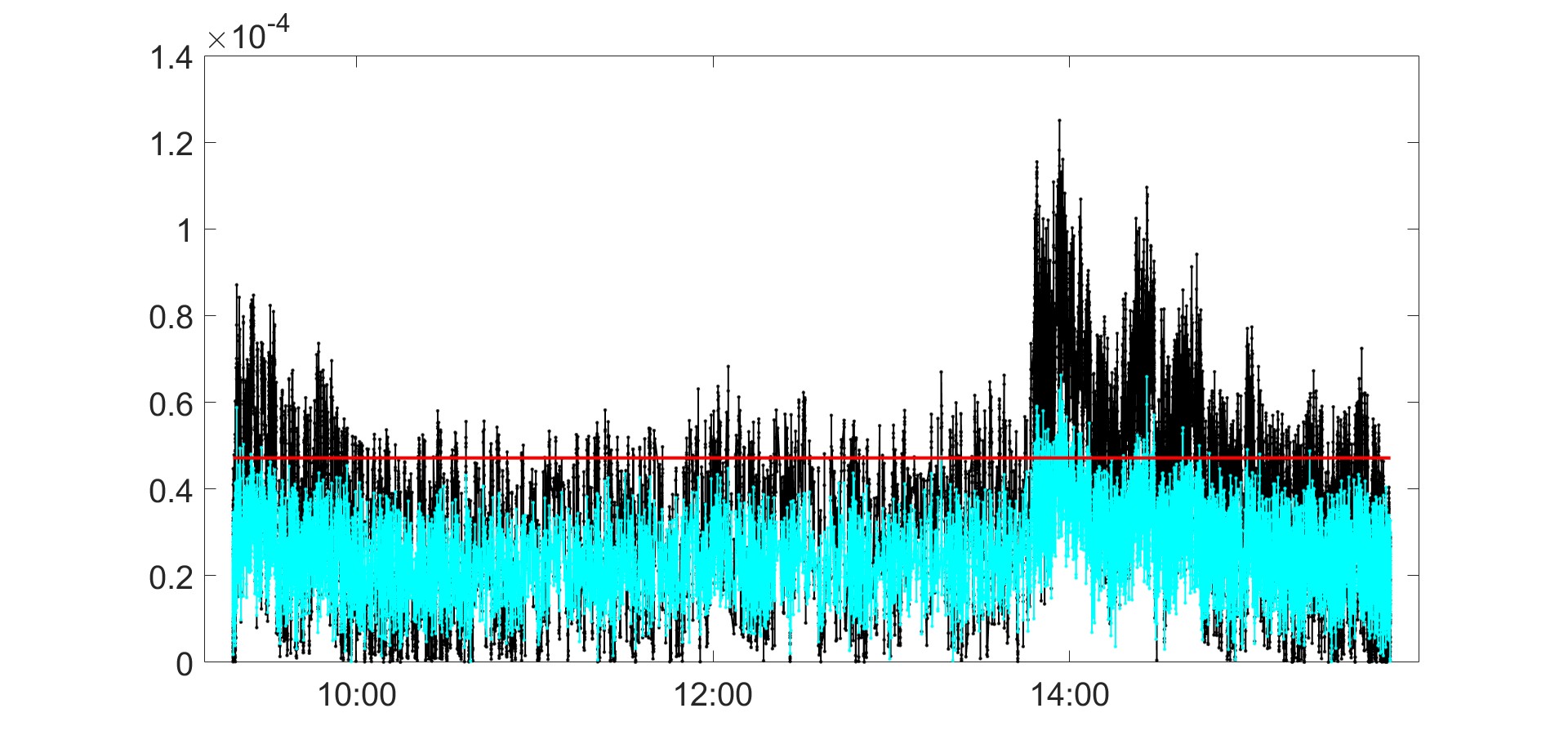}}\par\medskip
        \subfigure[]
        {\includegraphics[width=\textwidth,height=2.5cm]{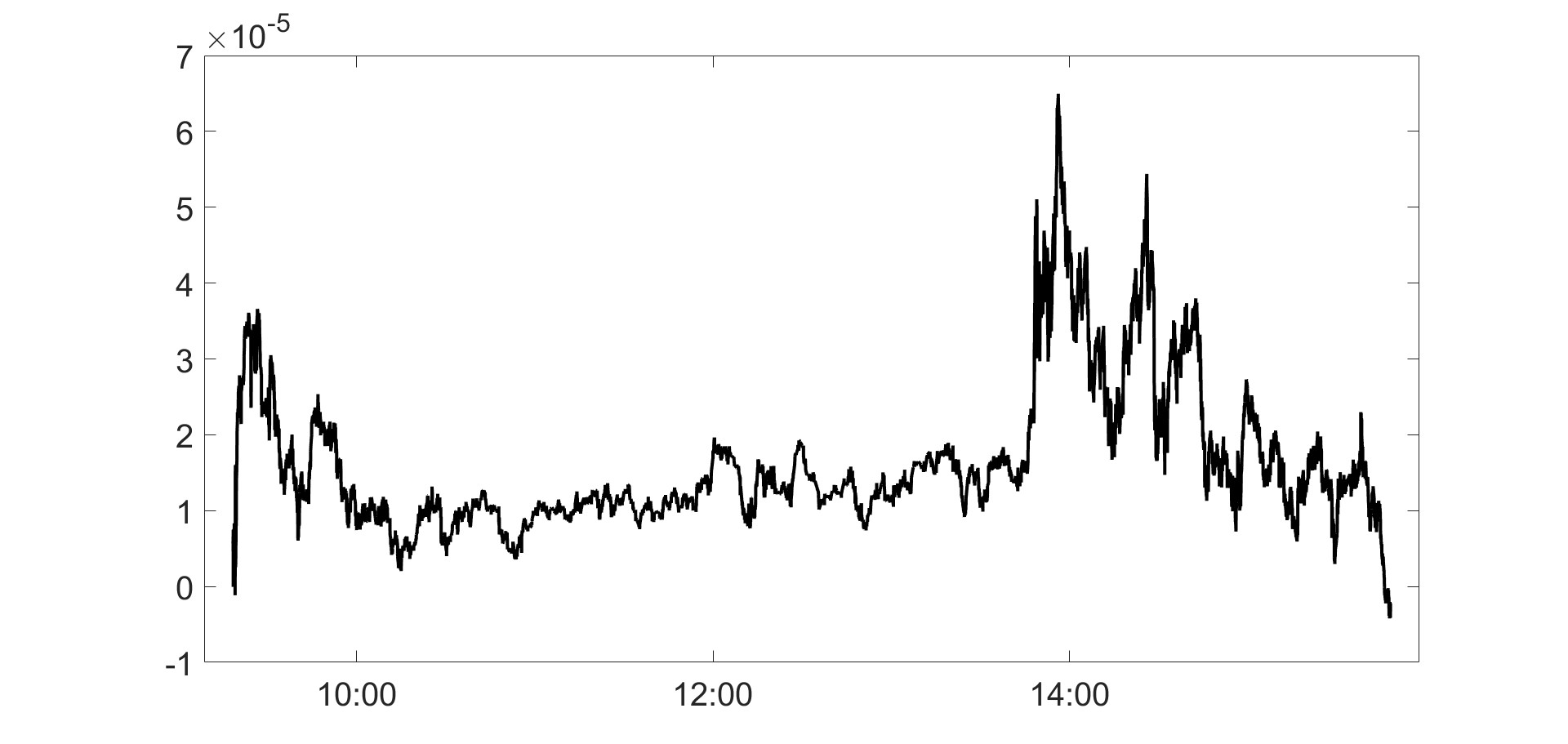}}\par\medskip
        \subfigure[]
        {\includegraphics[width=\textwidth,height=2.5cm]{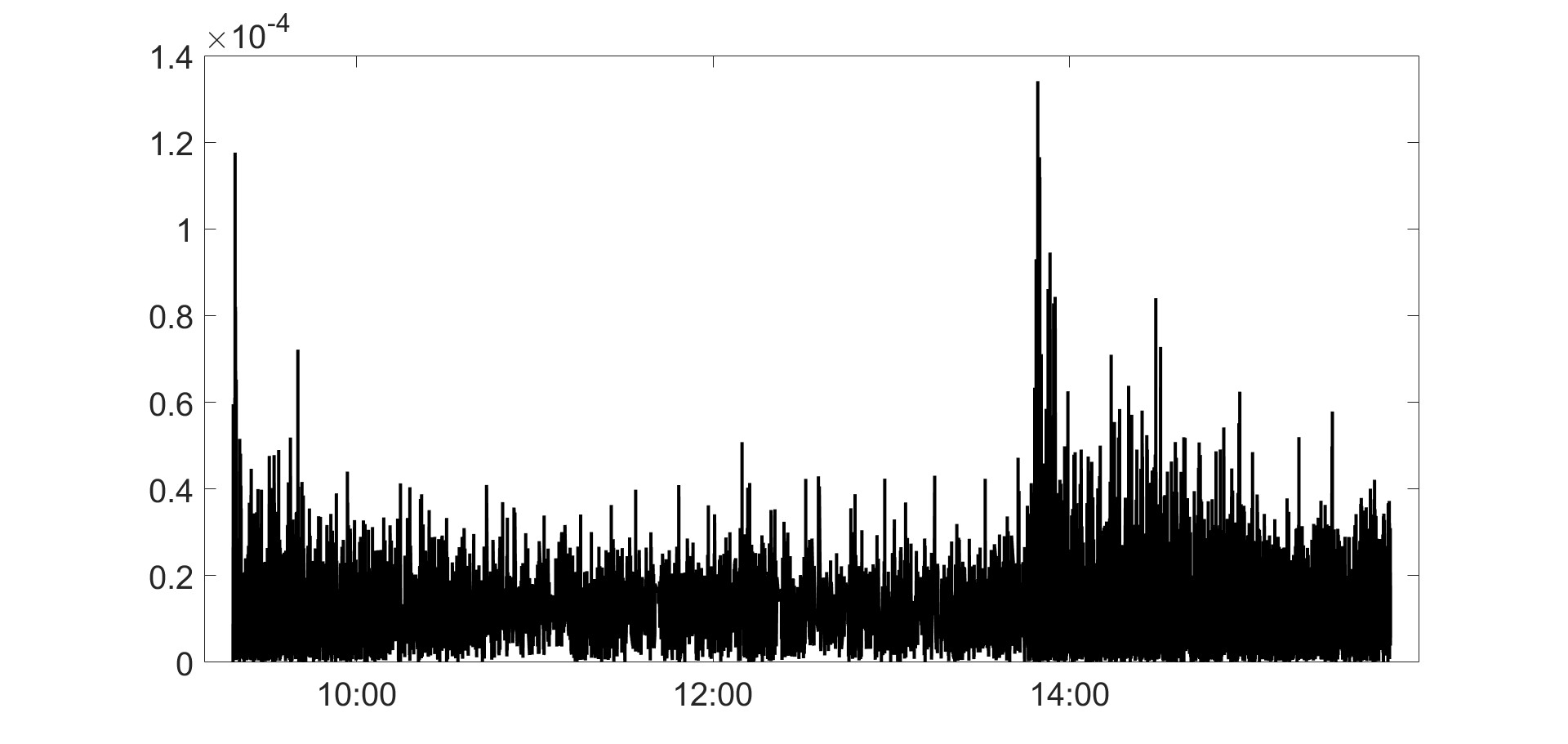}}
    \end{minipage}
    \caption{\footnotesize Panel (a) and (d): long-run cumulative impulse response ($\text{LRCIRF}_t$, y-axis) for the SDAMH model (black dotted lines), for the AMH model (cyan dashed lines), and for the AH model (red solid lines). Time in hours and minutes (hh:mm, x-axis). Panel (b) and (e): instantaneous impact $b_{0,t}$. Panel (c) and (f): absolute state $|\mu^*_{t-1}|$. Results for MSFT (left panels) and AMZN (right panels) on June 16, 2021.}
    \label{lfomcall}
\end{figure}

\begin{figure}[htt] 
	\centering
           \textbf{CIRF$_{t,h}$ estimates during FOMC - I}\par\medskip
 \subfigure[Time 09:45-09:50]{%
		\includegraphics[width=0.3\textwidth]{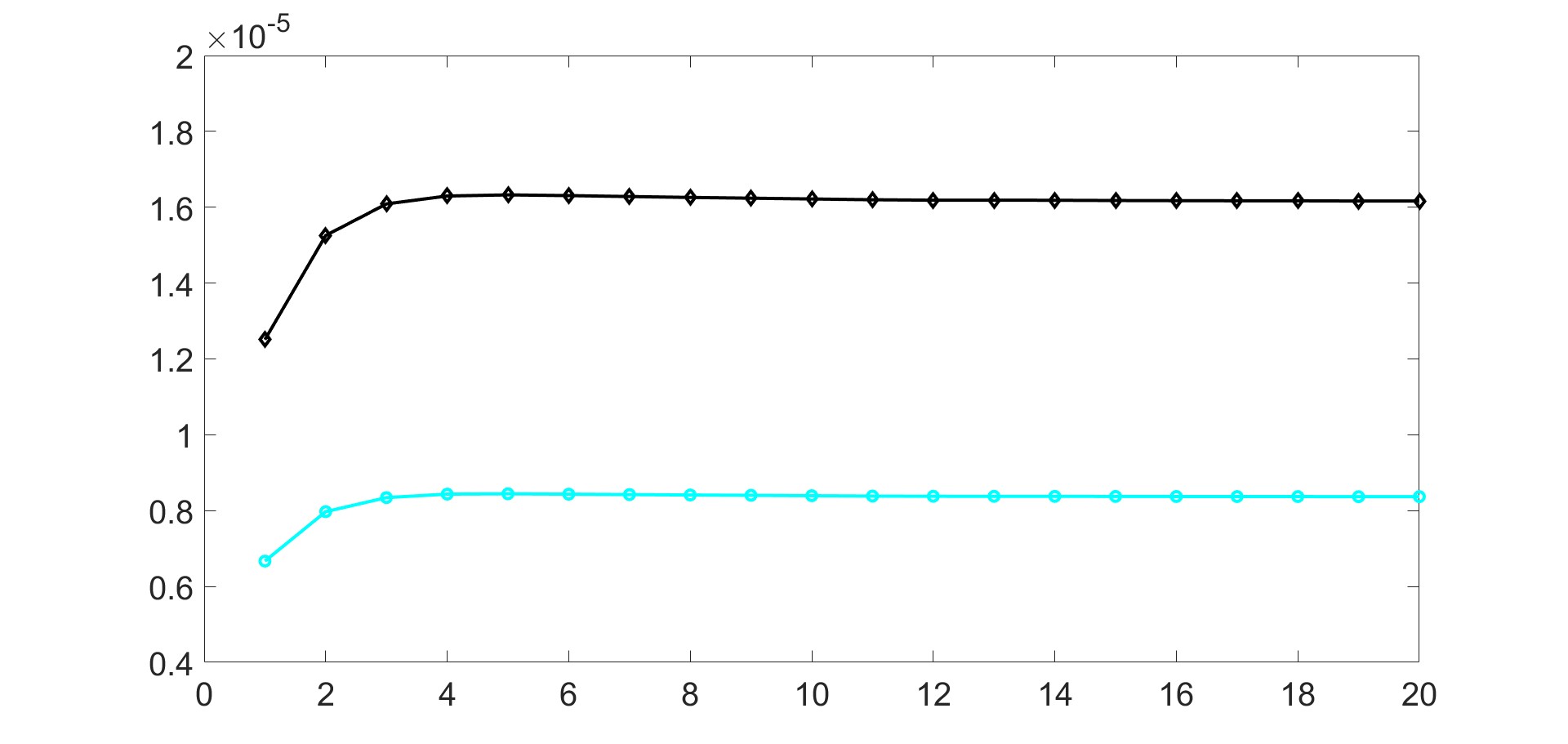}%
	} 
	\subfigure[Time 11:00-11:05]{%
		\includegraphics[width=0.3\textwidth]{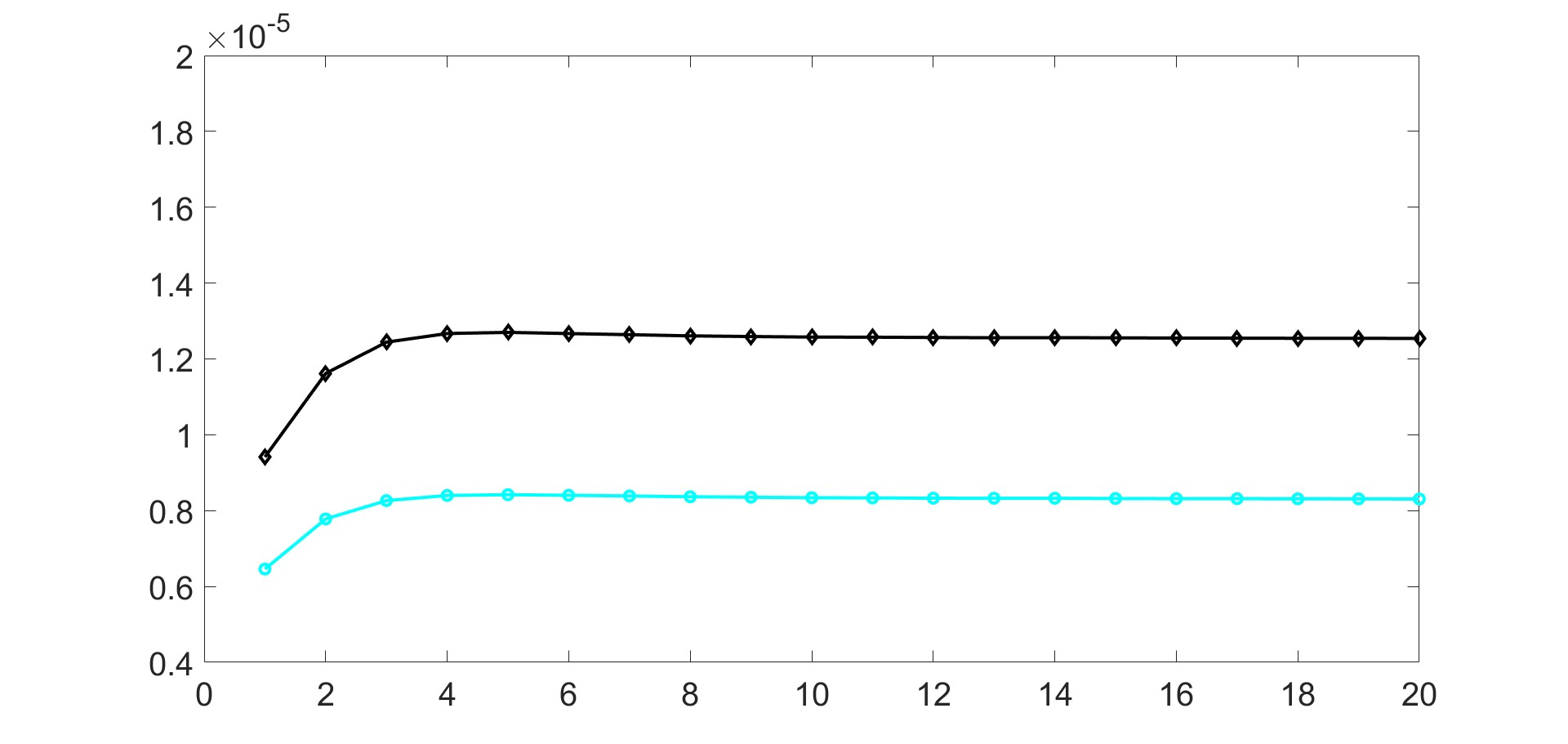}%
	} 
	\subfigure[Time 15:50-15:55]{
		\includegraphics[width=0.3\textwidth]{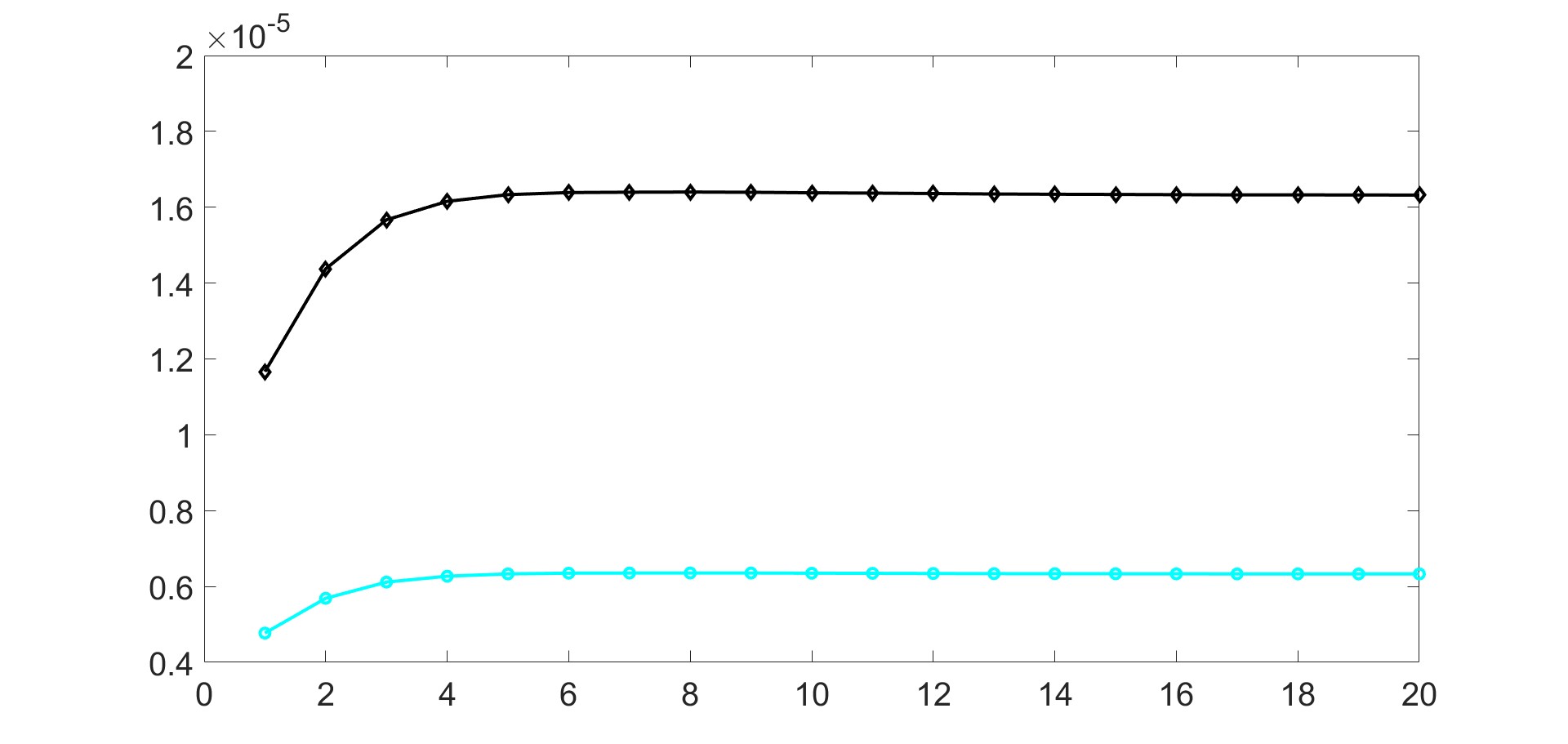}
	} \\
	\subfigure[Time 09:45-09:50]{
		\includegraphics[width=0.3\textwidth]{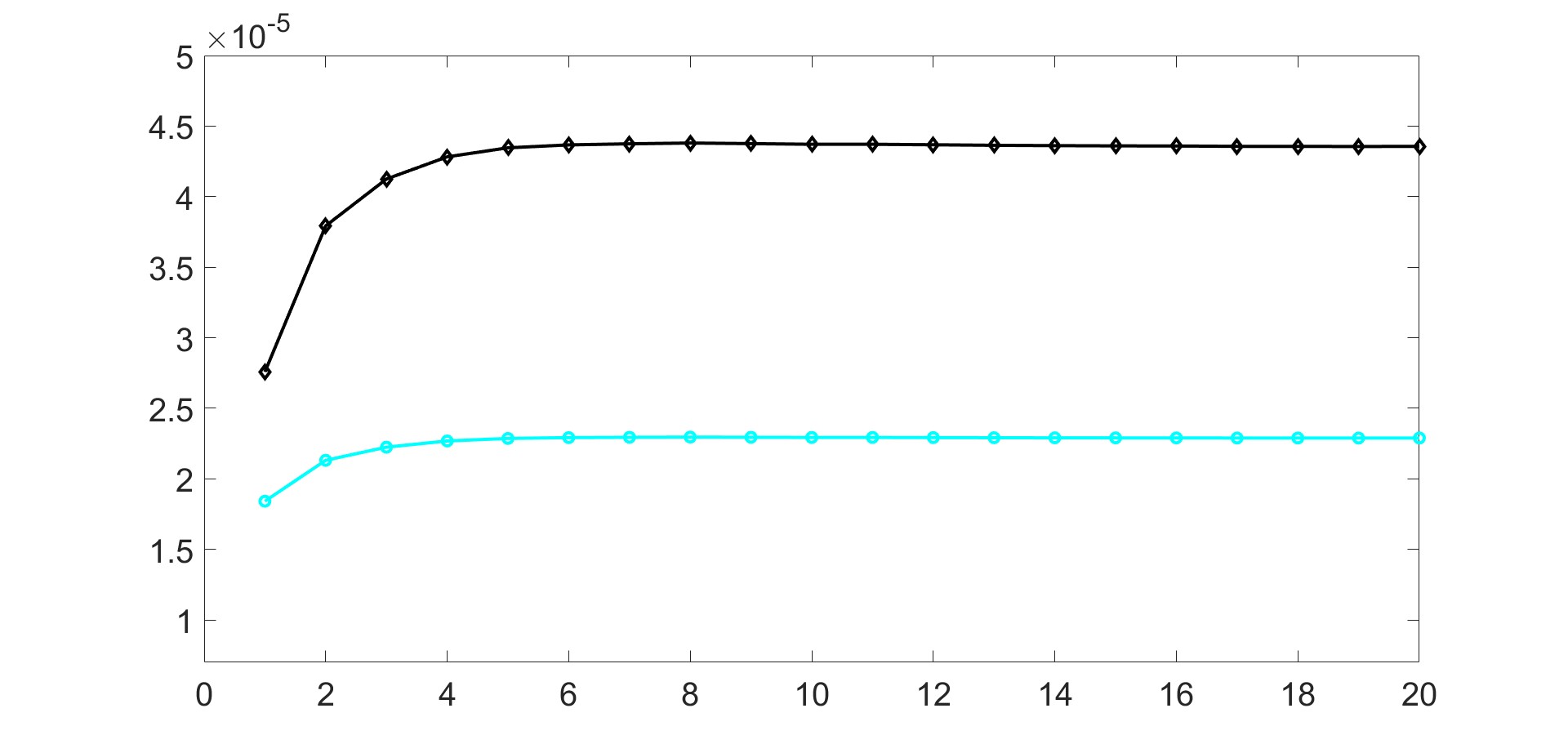}%
	} 
	\subfigure[Time 11:00-11:05]{
		\includegraphics[width=0.3\textwidth]{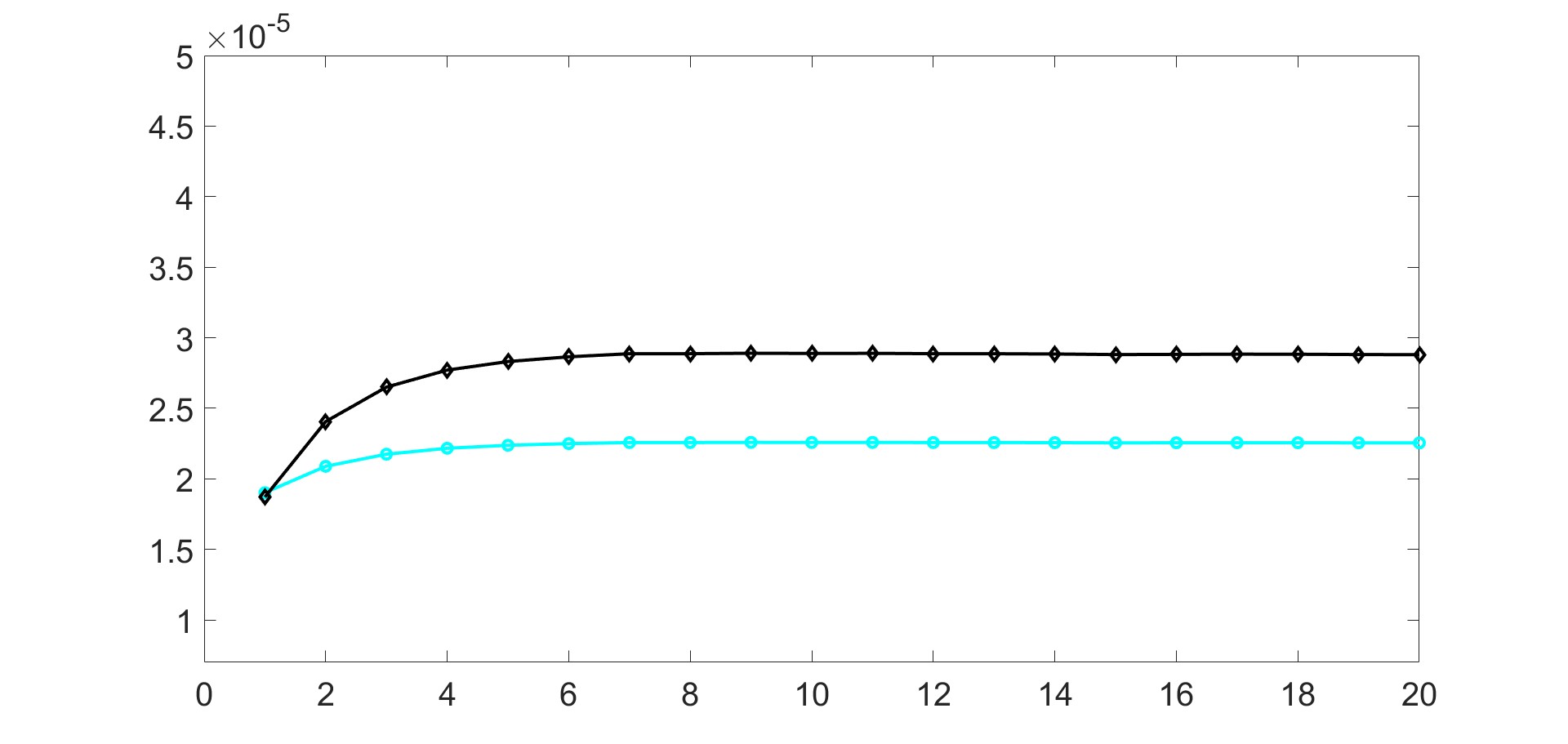}
	} 
	\subfigure[Time 15:50-15:55]{
	\includegraphics[width=0.3\textwidth]{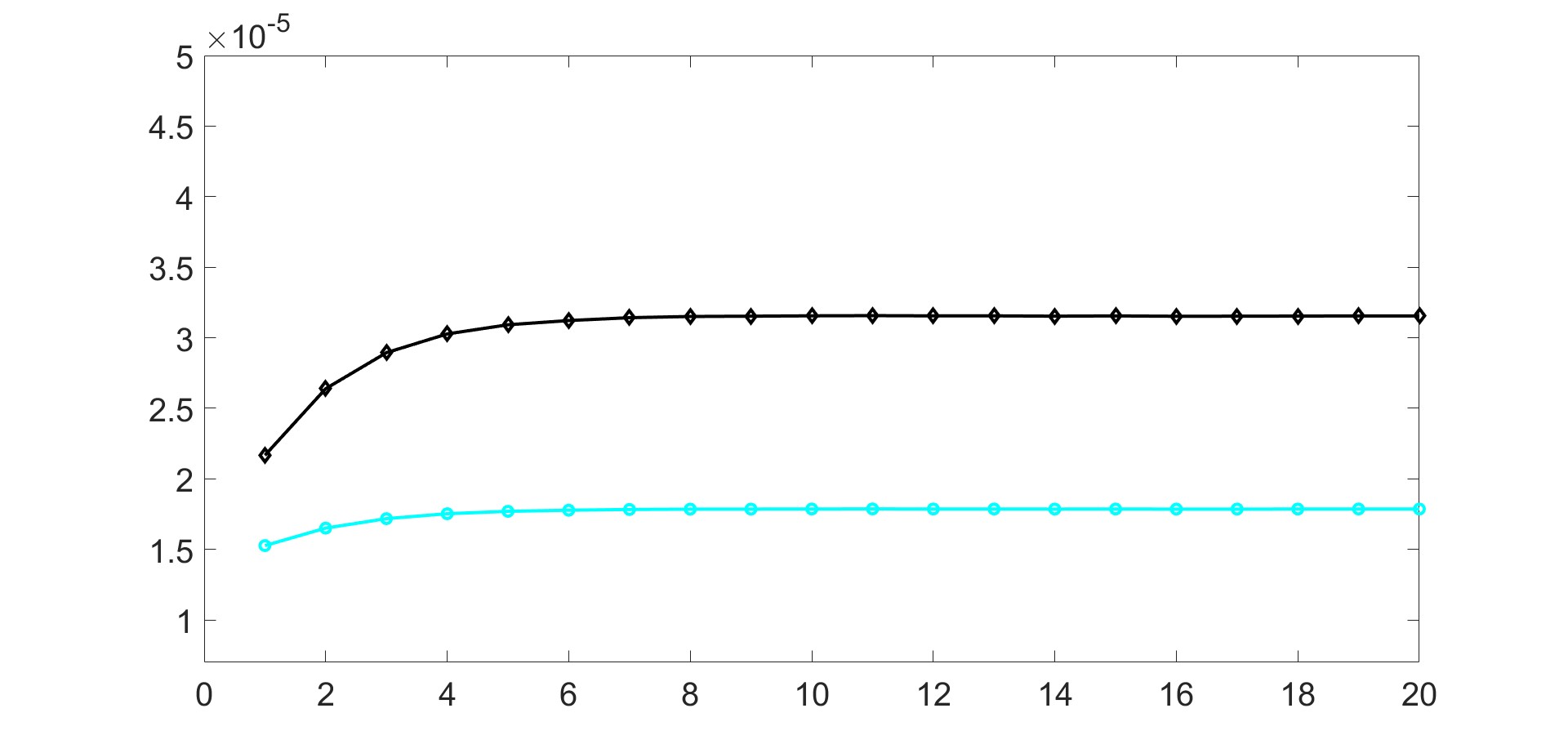}%
}
	\caption{\footnotesize Cumulative response $\text{CIRF}_h$ for $h = 1,\dots, H$ of SDAMH (black diamonded lines), and for AMH (cyan circled lines) model. The lines are average over trades in the specified intervals corresponding to different market phases, namely the opening (left), calm period (middle), and closing (right). Results for MSFT (top) and AMZN (bottom) on June 16, 2021.}\label{phasesMSFT}
\end{figure}

 The previous analysis focused on the long-term impulse response function. This section illustrates the evolution of the CIRF during the day of the FOMC announcement as a function of the time horizon and the time of day. To highlight the variability of the cumulative impulse response at each time step $t$ and at each horizon $h$, we select three different phases of the day, namely from 09:45 to 09:50 as an ``opening market phase", from 11:00 to 11:05 as a ``calm market phase", and trades from 15:50 to 15:55 as the "closing market phase", and we compute the averaged $\text{CIRF}_{t,h}$ estimated at each trade inside the periods considered above. Figure~ \ref{phasesMSFT} reports the estimated $\text{CIRF}_{t,h}$ in the three phases for the SDAMH model (black diamond lines) and the AMH model (cyan circle lines). To quantify the speed of convergence and the long-term limit of the CIRF, we fit the model
\begin{equation}\label{expmdl}
	\text{CIRF}_{t,h} =  c_t+ \kappa_t e^{-\phi_t h} 
\end{equation}
for fixed $t$. The convergence rate, that is, the speed with which the information of a trade is impounded on the price, is measured by the parameter $\phi_t$. We saw above that the opening market phase is characterized by high levels of instantaneous impact, low values of the state, and thus high levels of the information content of the stock trades. The left panels of Figure~\ref{phasesMSFT} show that the permanent level of SDAMH tends to be higher than the level of the AMH model, and achieves faster convergence. Table \ref{phi} reports the estimated value of $\phi_t$  and we notice that the estimate $\phi_t$ for the score-driven model is larger than that for the static parameter model. During the calm phase, the instantaneous impact parameter is small and fluctuates around its unconditional value and the state is higher (see Figure~\ref{lfomcall}). In this market situation, the structure of the CIRF is mainly affected by the state ($\mu_{t-1}^*$), as the instantaneous impact can be considered constant. From the middle panels of Figure~\ref{phasesMSFT}, we notice that the speed of convergence and the permanent level in the two models are comparable. The estimates of $\phi_t$ in Table \ref{phi} are similar. The closing phase of the market is similar to the opening one. The right panels of Figure~\ref{phasesMSFT} show that the SDAMH estimates for the long-term CIRF are higher than those of the AMH model and the speed of convergence is faster for the score-driven model (see Table \ref{phi}). It appears that the SDAMH cumulative responses adapt quickly in case of different market conditions, due to the dynamics of the time-varying parameter.

\begin{table}[htt]
	\centering
           \textbf{Speed of convergence estimates - I}\par\medskip
	\begin{tabular}{llllllll}
		&         & \textbf{MSFT} &         &  &       & \textbf{AMZN} &         \\ \hline
		& Opening & Calm & Closing &   & Opening & Calm & Closing \\ \hline
	  SDAMH &     0.373     &  0.311   &  0.335   &    &   0.676       &  0.492    &   0.366      \\
	   		 & (0.072)    &  (0.081)   &  (0.077)   &    &  (0.096)  &  (0.222)   &   (0.078)      \\ 
	     AMH &     0.314     &  0.299   &  0.312  &     &   0.635       &  0.498   &   0.353 \\ 
	      		 & (0.077)    &  (0.083)   &  (0.092)   &    &  (0.113)  &  (0.200)   &   (0.064)      \\ \hline
	\end{tabular}
	\caption{\footnotesize Speed of convergence parameter $\phi_t$ from Eq.\eqref{expmdl} during the opening, the calm and the closing phase of the market and length of the 95\% confidence bands in brackets.} 
\label{phi}
\end{table}

\begin{figure}[htt] 
 	\centering
      \textbf{CIRF$_{t,h}$ estimates during FOMC - II}\par\medskip
 	\subfigure[Time 13:55 to 13:59]{%
 		\includegraphics[width=0.28\textwidth]{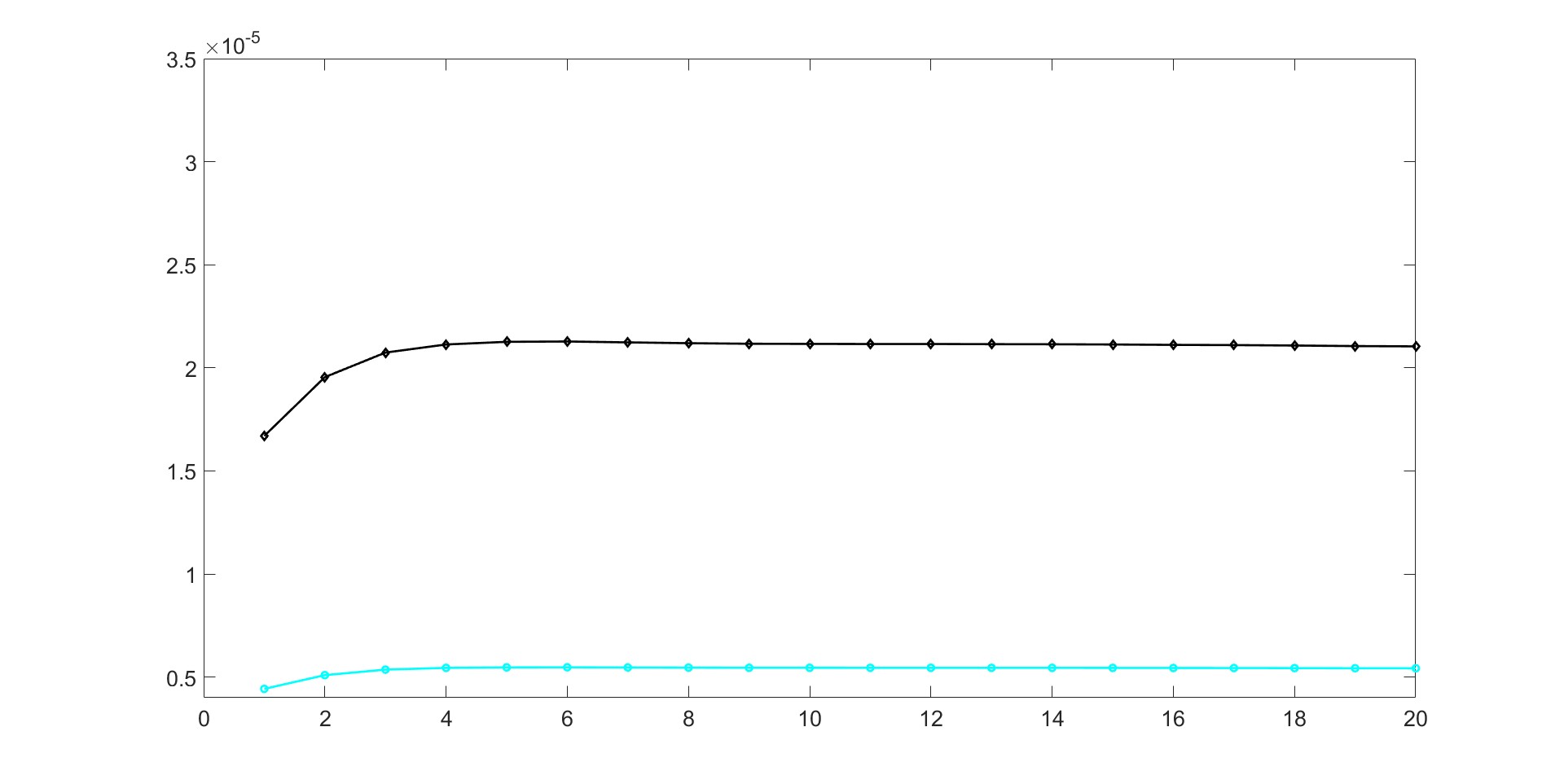}%
 	} 
 	\subfigure[Time 14:00]{%
 		\includegraphics[width=0.3\textwidth]{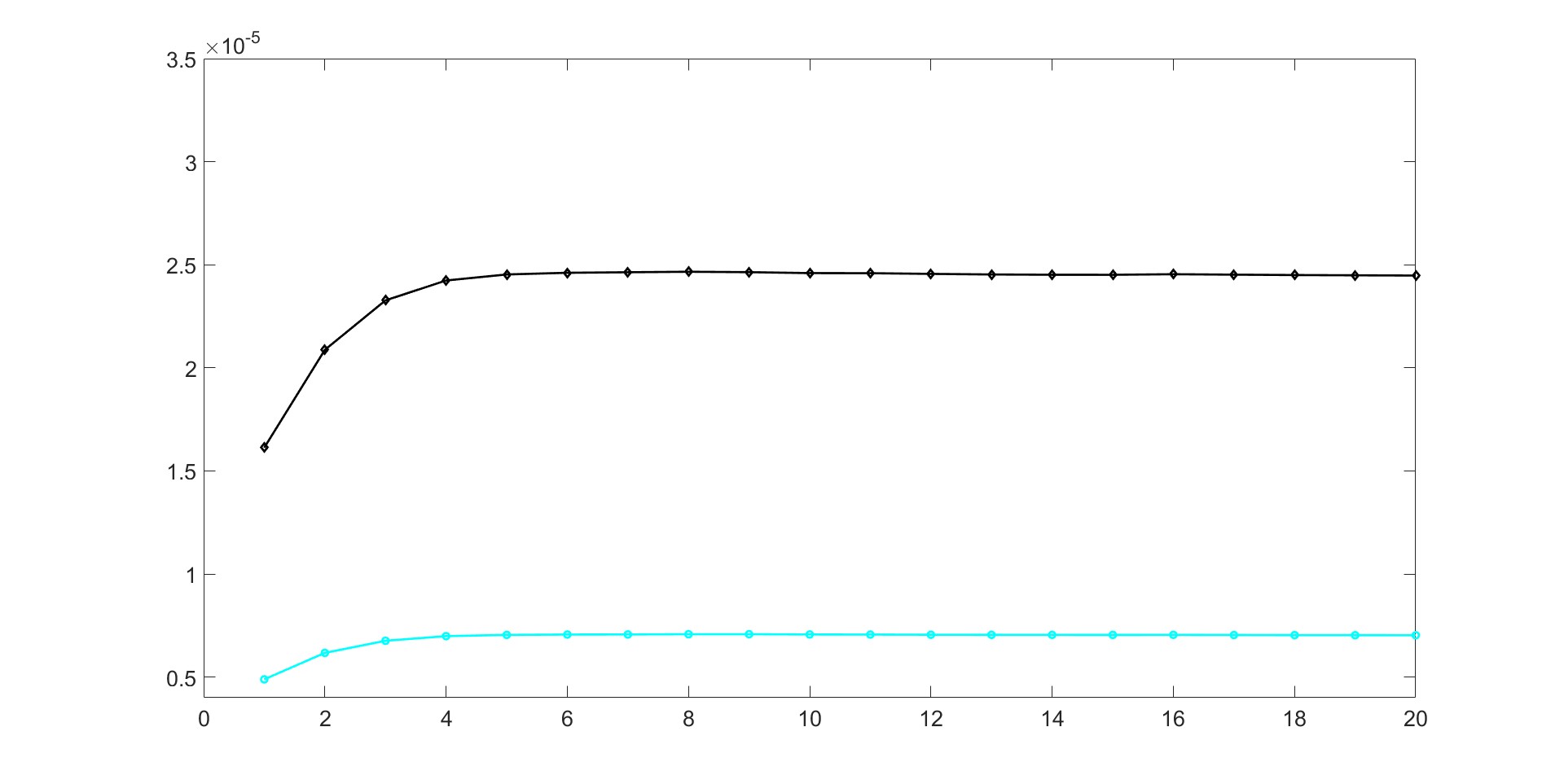}%
 	} 
 	\subfigure[Time 14:01-14:05]{
 		\includegraphics[width=0.3\textwidth]{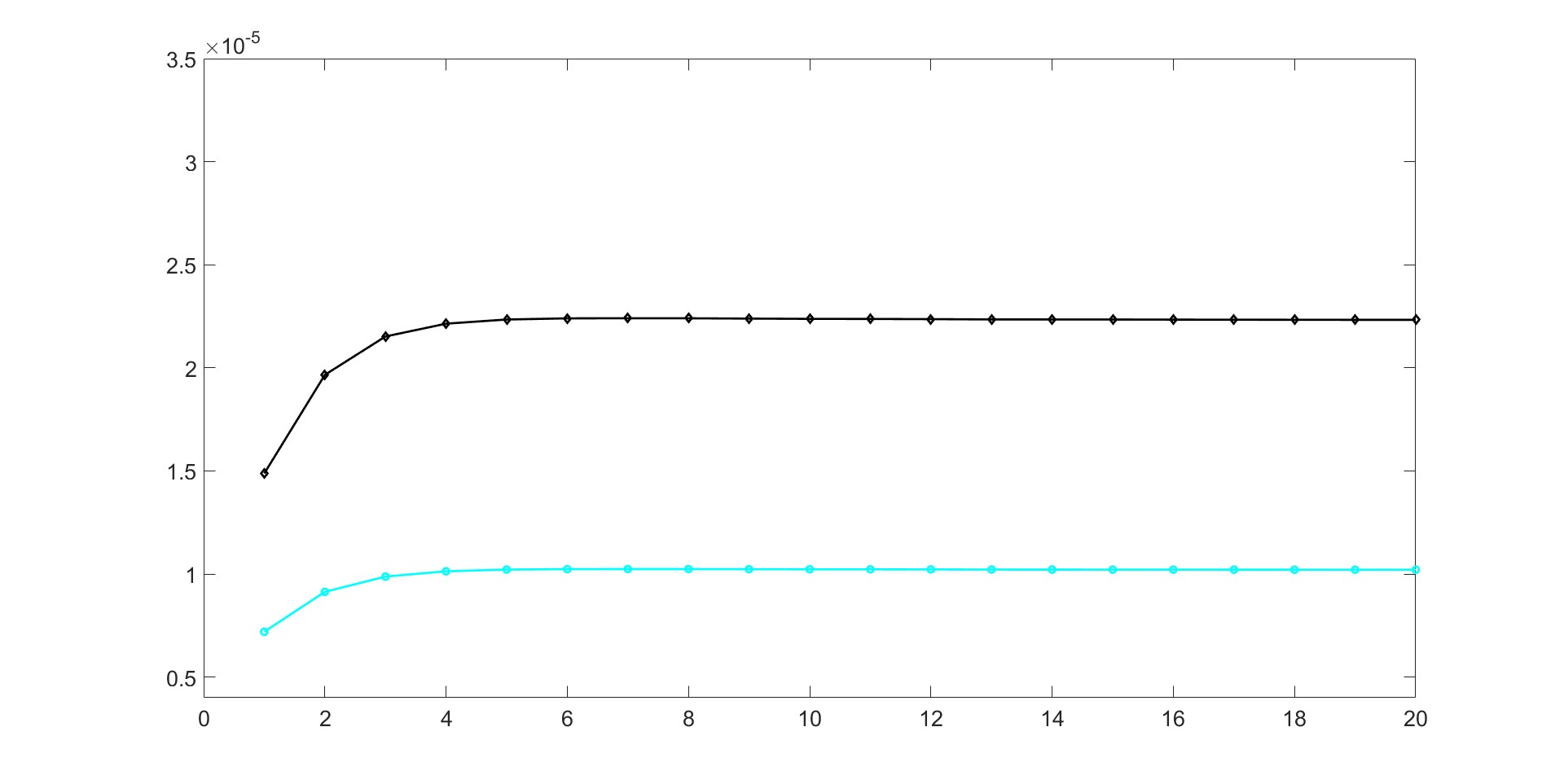}
 	} \\
 	\subfigure[Time 13:55 to 13:59]{
 		\includegraphics[width=0.3\textwidth]{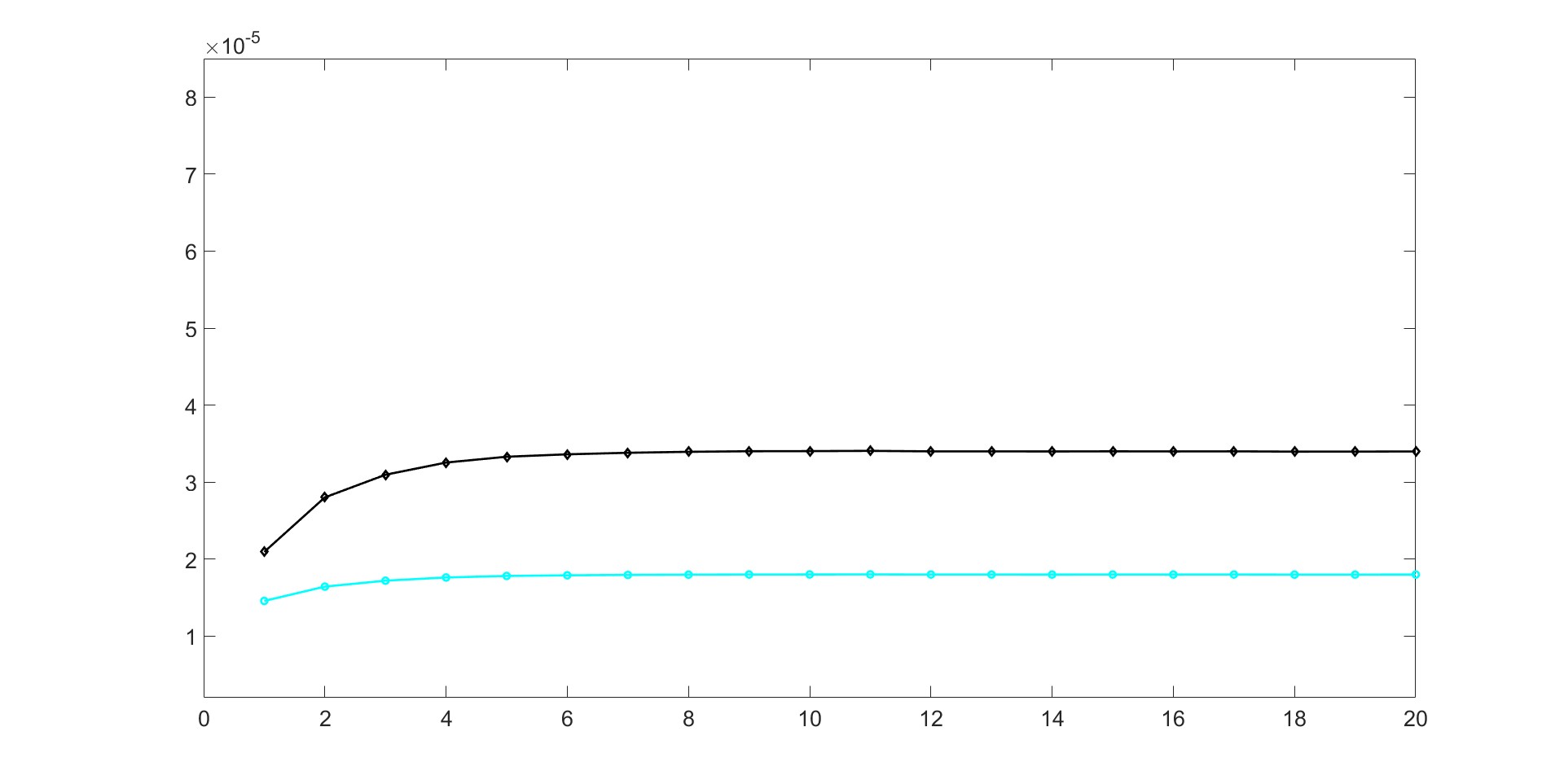}%
 	} 
 	\subfigure[Time 14:00]{
 		\includegraphics[width=0.3\textwidth]{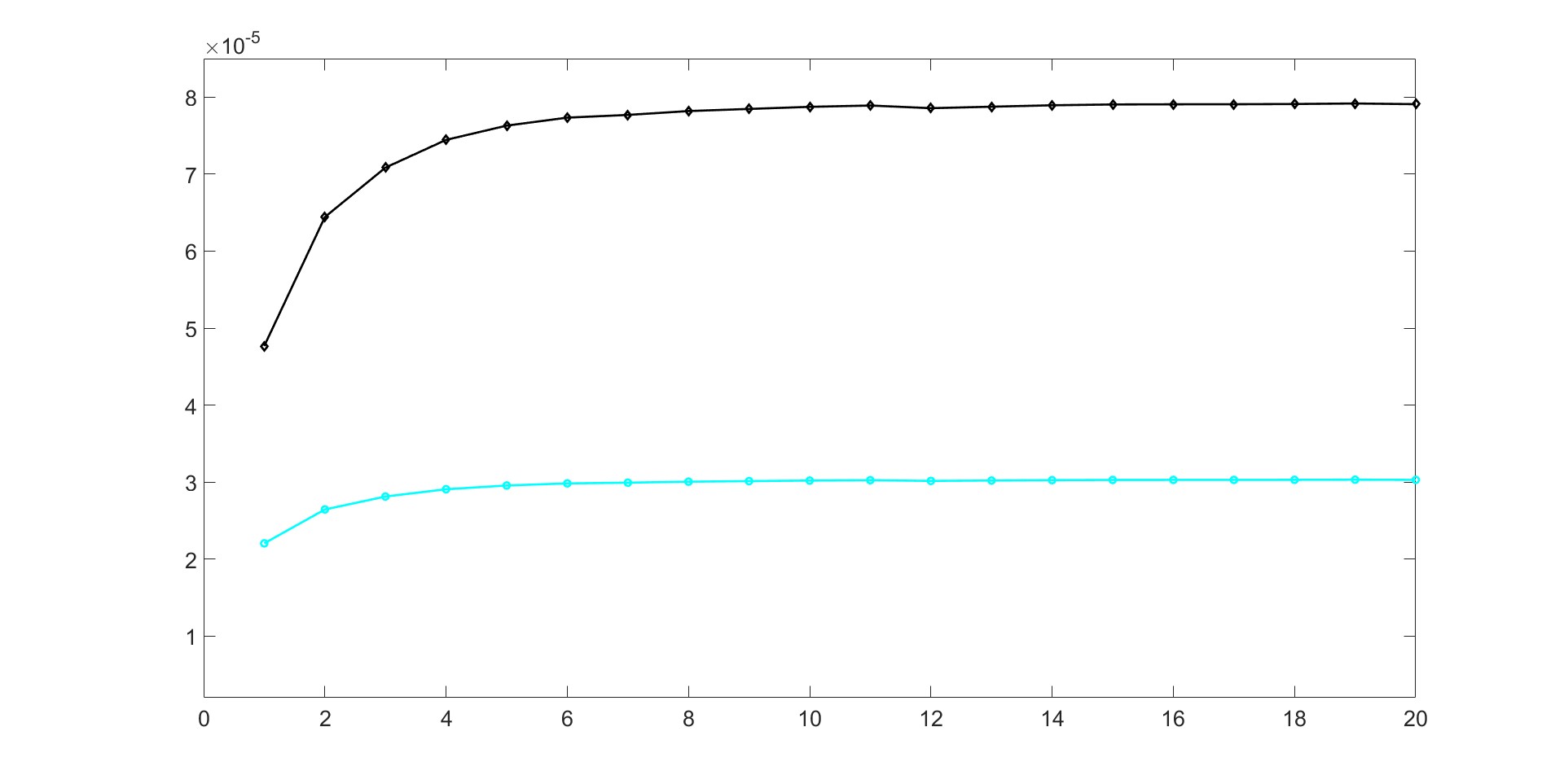}
 	} 
 	\subfigure[Time 14:01-14:05]{
 		\includegraphics[width=0.3\textwidth]{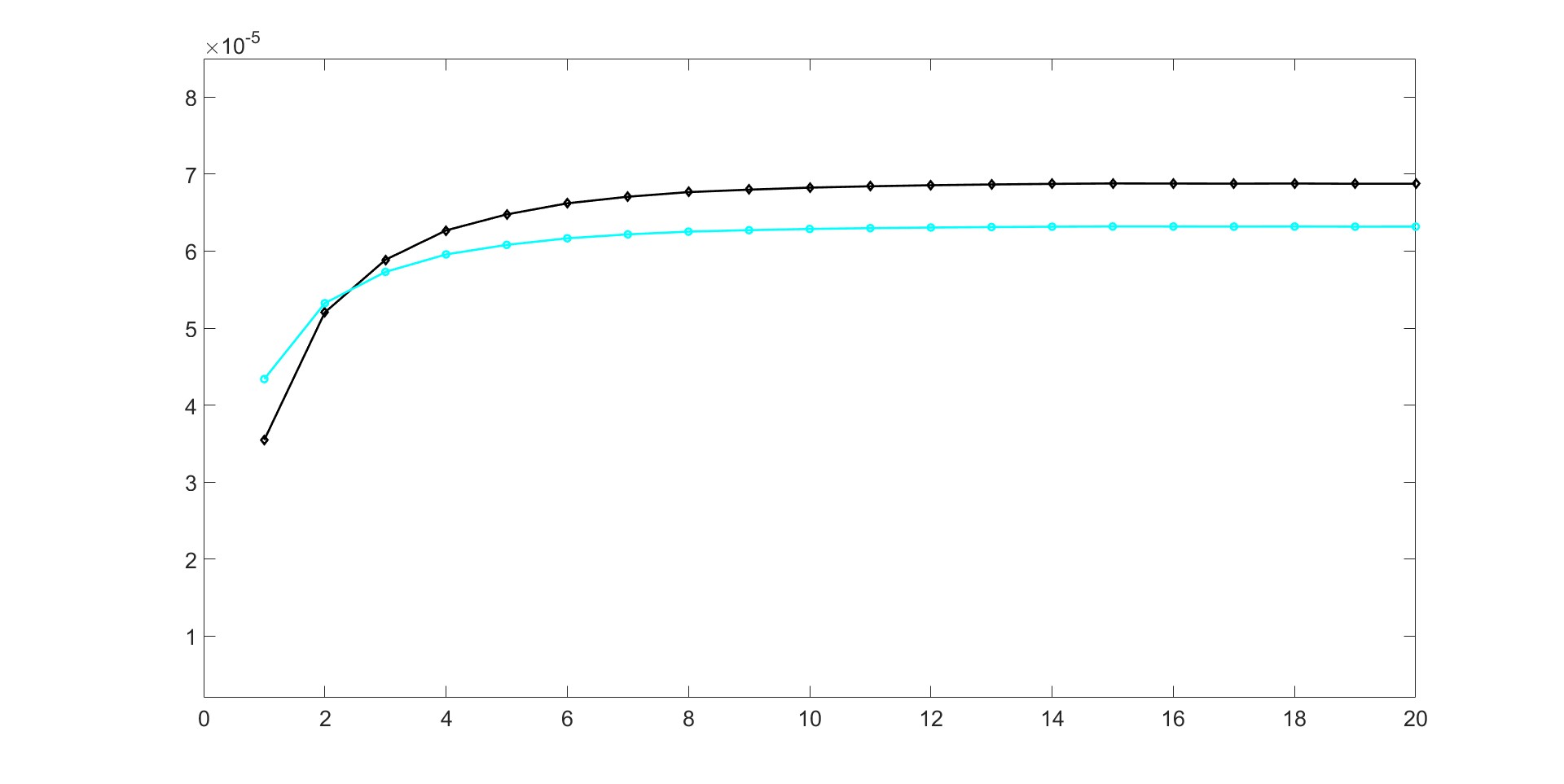}%
 	}
 	\caption{\footnotesize Cumulative response $\text{CIRF}_h$ for $h = 1,\dots, H$ of SDAMH (black diamonded lines), and for AMH (cyan circled lines) model. The lines are averages over trades in the specified intervals corresponding to different periods around the FOMC announcement, namely before (left), during (middle), and after (after) the event. Results for MSFT (top) and AMZN (bottom) on June 16, 2021.}
 	\label{aroundfomcMSFT}
 \end{figure}

We now focus our attention on the informativeness of trades around the time of the FOMC announcement (14:00) by computing the average of $\text{CIRF}_{t,h}$ over three-time intervals: from 13:55 to 13:59,  at 14:00, and from 14:01 to 14:05. Figure~\ref{aroundfomcMSFT} shows the results for MSFT (top panels) and AMZN (bottom panels), and Table \ref{phi0} reports the estimated parameter $\phi_t$ for all curves. Before the FOMC announcement, the LRCIRF of the SDAMH model was higher than the one of the AMH model. The long-term value of SDAMH at each $h$ tends to be high already before the time of the announcement and has a faster adjustment speed than that of the AMH model. The same holds during the FOMC announcement (middle panels). After the announcement (right panels), the level of the SDAMH $\text{CIRF}_h$ is still higher than the one of the AMH model, as well as its speed of adjustment. This is mostly evident for MSFT, while for AMZN the CIRF are similar. From the comparison, we conclude that the effect of the time-varying parameters $b_{0,t}$ and $\sigma_t$ make the score-driven model more sensitive to changes in market conditions. In fact, this is evident when comparing how the level and the speed of adjustment change during the three phases. Compared to the AMH model, the SDAMH model seems to better perceive changes in market conditions before the event, and it seems to be more sensitive to the release of the FOMC announcement during the event. After the announcement, the SDAMH estimated average value tends to adjust (in terms of levels and speed). In a less evident way, Table \ref{phi0} shows that the "speed of convergence" of the AMH model is anyway sensitive 5 minutes after the announcement when the peak of informativeness has passed. 
  \begin{table}[htt]
 	\centering
             \textbf{Speed of convergence estimates - II}\par\medskip
 	\begin{tabular}{llllllll}
 		&         & \textbf{MSFT} &         &  &       & \textbf{AMZN} &         \\ \hline
 		& Before & During & After &   & Before & During & After \\ \hline
 		SDAMH & 0.734    &  0.948      &  0.745      &    &  0.689  &  0.850   &   0.701      \\
 	      	 & (0.074)  &  (0.234)    &  (0.183)    &    &  (0.152)  &  (0.141)   &   (0.194)      \\
 	    AMH   & 0.672    &  0.763      &  0.711      &    &  0.666  &  0.689  &   0.673  \\  
 		       & (0.081)  &  (0.055)    &  (0.135)    &    &  (0.143)  &  (0.216)   &   (0.148)      \\ \hline
 	\end{tabular}
 	\caption{\footnotesize Speed of convergence parameter $\phi_t$ from Eq.\eqref{expmdl} before, during, and after the announcement and length of the 95\% confidence bands in brackets.} 
 	\label{phi0}
 \end{table}
 
\section{Time-varying impact for transaction cost analysis}
\label{section:trancost}

In this section, we present a financial application, namely, the real-time estimation of transaction costs due to market impact. Consider a trader who wants to sell a large amount $Q$ of shares over a time interval $[0,T]$. As typical in the optimal execution literature, we assume the trader partitions the interval in $N$ equal size subintervals of length $\tau=T/N$ and the number of shares traded in interval $n$ ($n=1,...,N)$ is denoted with $v_n \tau$ where $v_n$ is the trading velocity\footnote{As customary a negative (positive) value of $v_n$ means that the trader sells (buys) in the interval.}. Clearly it is $\sum_{n=1}^N v_n \tau=-Q$. We assume that market impact is described by an \cite{almgren2001optimal} model, which implies that the efficient stock price at the end of the $n$-th interval is modeled by
$$
S_{n+1}=S_n+\sigma\sqrt{\tau}\epsilon_{n+1}+\beta v_{n+1} \tau\,,
$$
where $\beta$ is the permanent (linear) impact coefficient and $\epsilon_i$ are standard Gaussian variables i.i.d. Let $X_n$ be the cash account of the trader at the end of the interval $n$. The Almgren-Chriss model postulates that its dynamics is
$$
X_{n+1}=X_n-v_{n+1}S_n\tau
-kv^2_{n+1}\tau.
$$
In this expression, $k$ is the (linear) temporary coefficient and represents the cost incurred by the trader because of the price concession needed to attract counterparts in a short time interval. It is straightforward to prove (see \cite{gueant2016financial}) that the expected terminal cash account for a generic execution is 
$$
{\mathbb E}[X_N-X_0]=QS_0-\frac{\beta}{2}Q^2+\frac{\beta}{2}\sum_{n=1}^N v^2_{n}\tau^2-k\sum_{n=1}^N v^2_n\tau\,.
$$
Suppose for simplicity that the trader executes a TWAP (which is also optimal if the agent is risk neutral), i.e. $v_n=-\frac{Q}{N\tau}$. The expected implementation shortfall 
$$
{\mathbb E}[X_N-X_0]-QS_0=-Q^2\left(\frac{\beta}{2}\left(1-\frac{1}{M}\right)+\frac{k}{T}\right)\simeq -Q^2\left(\frac{\beta}{2}+\frac{k}{T}\right)\,.
$$
From this expression, it is evident that the expected terminal cash and the expected implementation shortfall depend on the impact coefficients $\beta$ and $k$ which must be estimated from the data to assess the cost of a large order. The estimation of $\beta$ is typically performed via linear regression of the mid-price change over a macroscopic time interval (e.g. 5 min) against the net order flow, defined as the difference between the volume of buy and sell market orders during the time interval (see, for example, \cite{cartea2016incorporating}). This estimation method has two major pitfalls: (i) $\beta$ is assumed to be constant during the period used in the regression and (ii) the method requires long estimation intervals (typically one day) to obtain reliable estimates. In the following, we show that the SDAMH model provides time-varying estimates of $\beta$, which can be used successfully to predict permanent impact without the need to aggregate data in physical time. To this end, consider a time interval in which $M$ trades are executed in the market. A way of quantifying permanent impact is to estimate the following ratio $\hat \beta_t = \frac{\sum_{j=t-M}^{t-1}\Delta q_j}{\sum_{j=t-M}^{t-1}v_j}\,$, where $\Delta q_t=q_t-q_{t-1}$ is the midpoint price change and $v_t$ the signed volume of the $t-$th trade. To use our model that considers trade signs, we notice that $\hat {\bar{\beta}}_t \simeq \frac{\bar{q}_t}{\bar{v}_t} \frac{\sum_{j=t-M}^{t-1} r_j}{\sum_{j=t-M}^{t-1}x_j} = \frac{\bar{q}_t}{\bar{v}_t} \hat \beta^{(s)}_t$, where $\bar q_t$ and $\bar v_t$ are, respectively, the average midpoint price and the average absolute trading volume in the interval. In the above equation, we have introduced the quantity (the superscript $(s)$ stands for trade sign)
\begin{equation}\label{sumrsumx}
	\begin{split}
		\hat \beta^{(s)}_t := \frac{\sum_{j=t-M}^{t-1} r_j}{\sum_{j=t-M}^{t-1} x_j}\,.
	\end{split}
\end{equation}
The estimator requires the aggregation of $M$ events and might be highly fluctuating especially when the denominator is close to zero, i.e. when the number of buy and sell trades is approximately the same. For these reasons, it is customary to estimate the permanent impact by performing a linear regression of the aggregate price change versus the aggregate order flow in a longer time window, such as a day (as in \cite{cartea2016incorporating}). The SDAMH model, instead, provides an estimation that is updated at each trade, does not require event aggregation, and avoids the high fluctuations for approximately balanced markets. In Online Appendix \ref{section:derivpif} we show that if the data follows an SDAMH model, the estimator of the time-varying impact coefficient is approximated by
\begin{equation}\label{piamh}
\hat\beta^{(s)}_t \simeq \frac{\bar{b}_{0,t} + b_1 \omega_t^{(1)} + \bar{b}_{10} \omega_t^{(L_1)} + \bar{b}_{100} \omega_t^{(L_2)} }{1-a_1}\,,
\end{equation}
where $\bar{b}_{0,t}$ is the local average of the instantaneous impact estimated from trade $t-M$ to $t-1$ and $\omega_t^{(1)}, \omega_t^{(L_1)}$, and $\omega_t^{(L_2)}$ are local averages of the trade signs in the same interval (please refer to Online Appendix \ref{section:derivpif} for details). Finally, the estimator of the time-varying permanent impact coefficient reads
\begin{equation}\label{piamhqtvt}
\hat {\bar{\beta}}_t \simeq \frac{\bar{q}_t}{\bar{v}_t}  \frac{\bar{b}_{0,t} + b_1\omega_t^{(1)} + \bar{b}_{10} \omega_t^{(L_1)} + \bar{b}_{100} \omega_t^{(L_2)}}{1-a_1}. 
\end{equation}
Clearly, $\bar q_t$ and $\bar v_t$ can be also computed with a local estimation using, for example, an EWMA~\footnote{Alternatively, one could develop a score-driven Hasbrouck model where the first variable is the mid-point price change $\Delta q_t$ (instead of the return $r_t$) and the second variable is the signed volume $v_t$ (instead of the trade sign $x_t$). We leave this extension for future work.}. Eq.~\eqref{piamhqtvt} depends both on instantaneous impact and on the \textit{state} of the market. The final expression for $\hat \beta^{(s)}_t$ is consistent with the findings of this work: The permanent impact positively depends on the instantaneous impact parameter, through $\bar{b}_{0,t}$, and negatively on the state, through the $\omega_t$'s coefficients. In the absence of the \textit{state} coefficients, the lagged impacts of the order flow on the permanent impact are directly measured by the $b_1$, $\bar{b}_{10}$ and $\bar{b}_{100}$. The coefficients of $\omega_t$' s can be interpreted as factors that weigh the effect of the change in the \textit{state} of the market on the permanent impact. For example, consider $\omega_t^{(1)} := \bar{x}_{t-1} / \bar{x}_t$, where $\bar{x}_{t}$ is the average of the trade signs in the interval $[t-M, t-1]$. We have that if the number of buy orders is higher than that in the previous interval (i.e. $\bar{x}_{t} > \bar{x}_{t-1}$), the order flow tends to be unbalanced by the increasing pressure of buy orders. The coefficient $\omega_t^{(1)}$ describes the change in the \textit{state} of the market and the effect on the permanent impact coefficient: $b_1$ is weighed by $\omega_t^{(1)} < 1$, which produces a reduction in the permanent impact coefficient. In contrast, the order flow tends to be unbalanced by the increasing pressure of sell orders. The coefficient $b_1$ is weighed by $\omega_t^{(1)} > 1$, which produces an increase in the permanent impact coefficient. The same holds for $\omega_t^{(L_1)}$ and $\omega_t^{(L_2)}$.

To validate our estimator, in Online Appendix \ref{section:mcpif} we present a Monte Carlo experiment showing that the SDAMH approach provides estimates of the permanent impact coefficient consistent with those of a regression approach on aggregated data. This result supports the reliability of the SDAMH approach to estimate the permanent impact. Moreover, the full exploitation of tick-by-tick returns and trade-sign data allows us to obtain estimates with low error bands of the permanent impact coefficient in simulations.

\begin{figure}[htt]
	\centering
    \textbf{Time-varying estimates of permanent impact}\par\medskip
	\subfigure[June 04, 2021 - MSFT]
	{\includegraphics[width=7.5cm]{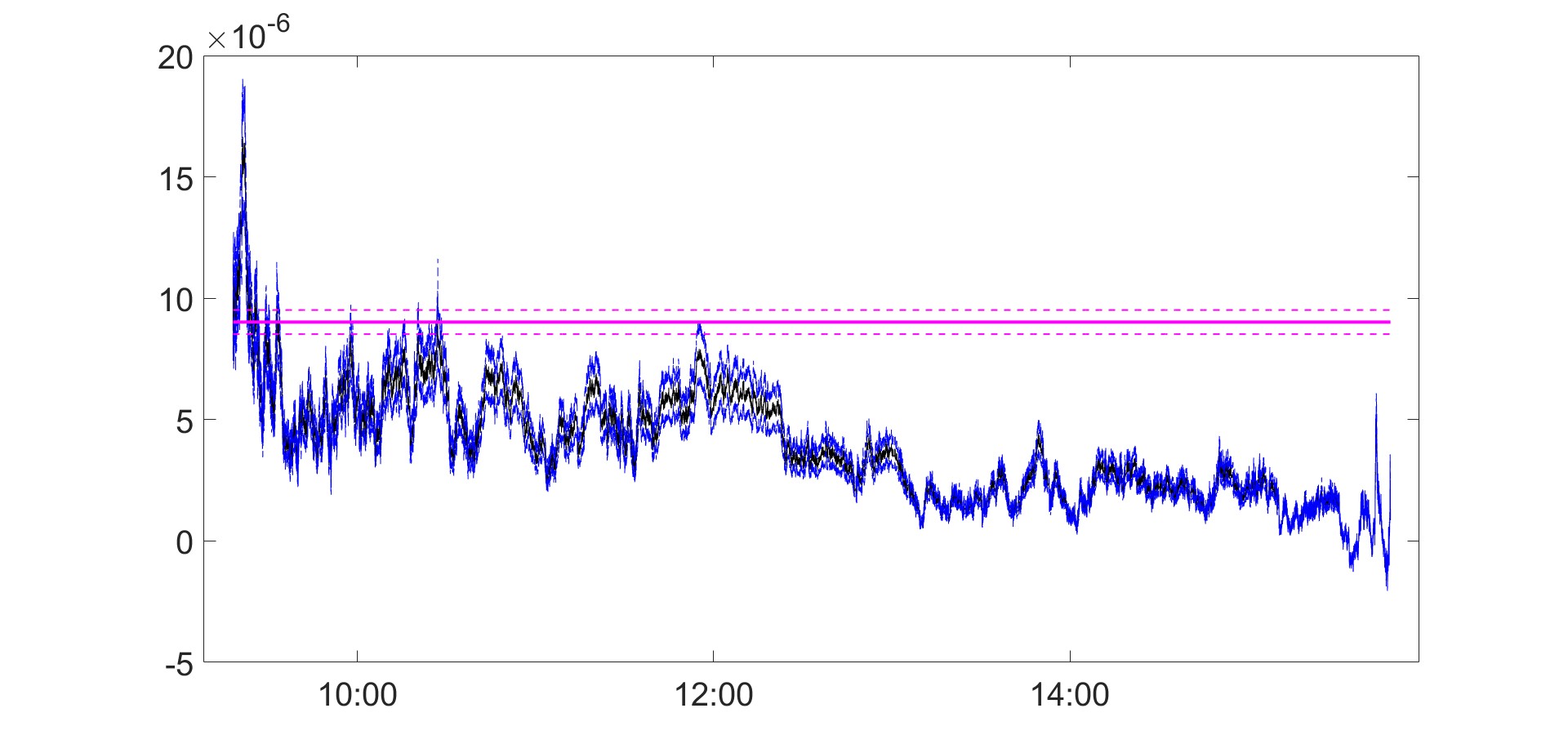}}
	\hspace{5mm}
	\subfigure[June 04, 2021 - AMZN]
	{\includegraphics[width=7.5cm]{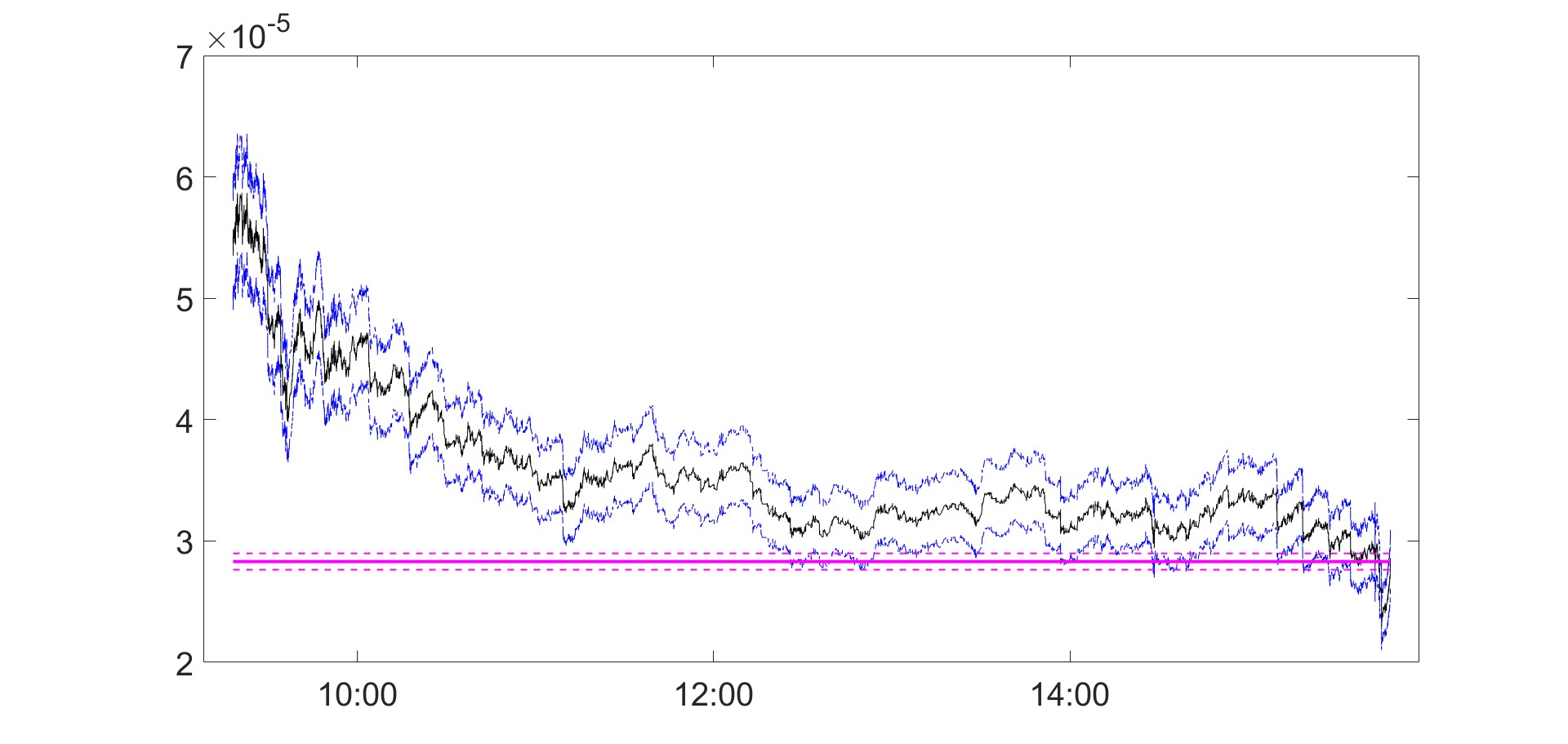}}
 	\subfigure[June 04, 2021 - CLSK]
	{\includegraphics[width=7.5cm]{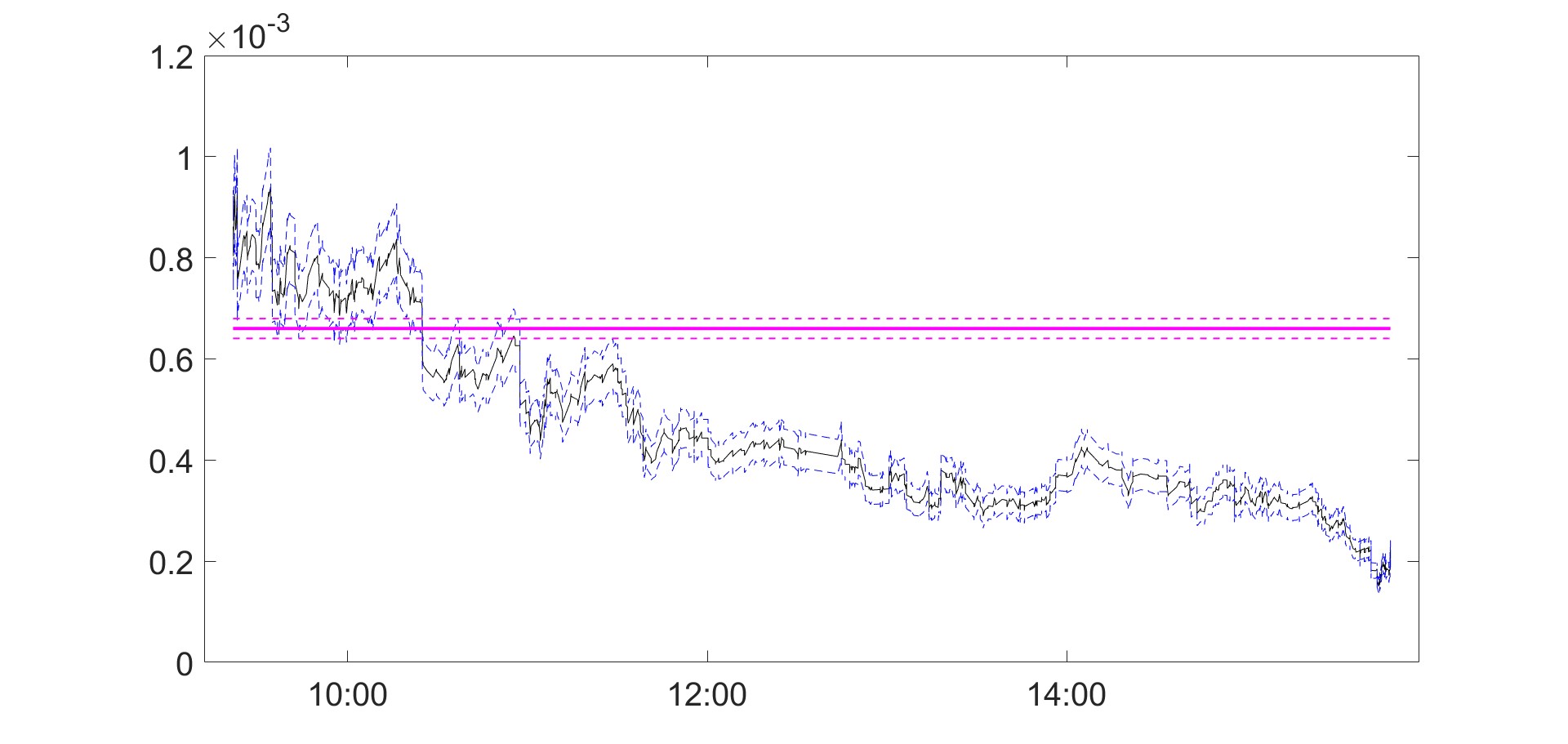}}
	\hspace{5mm}
	\subfigure[June 04, 2021 - FLGT]
	{\includegraphics[width=7.5cm]{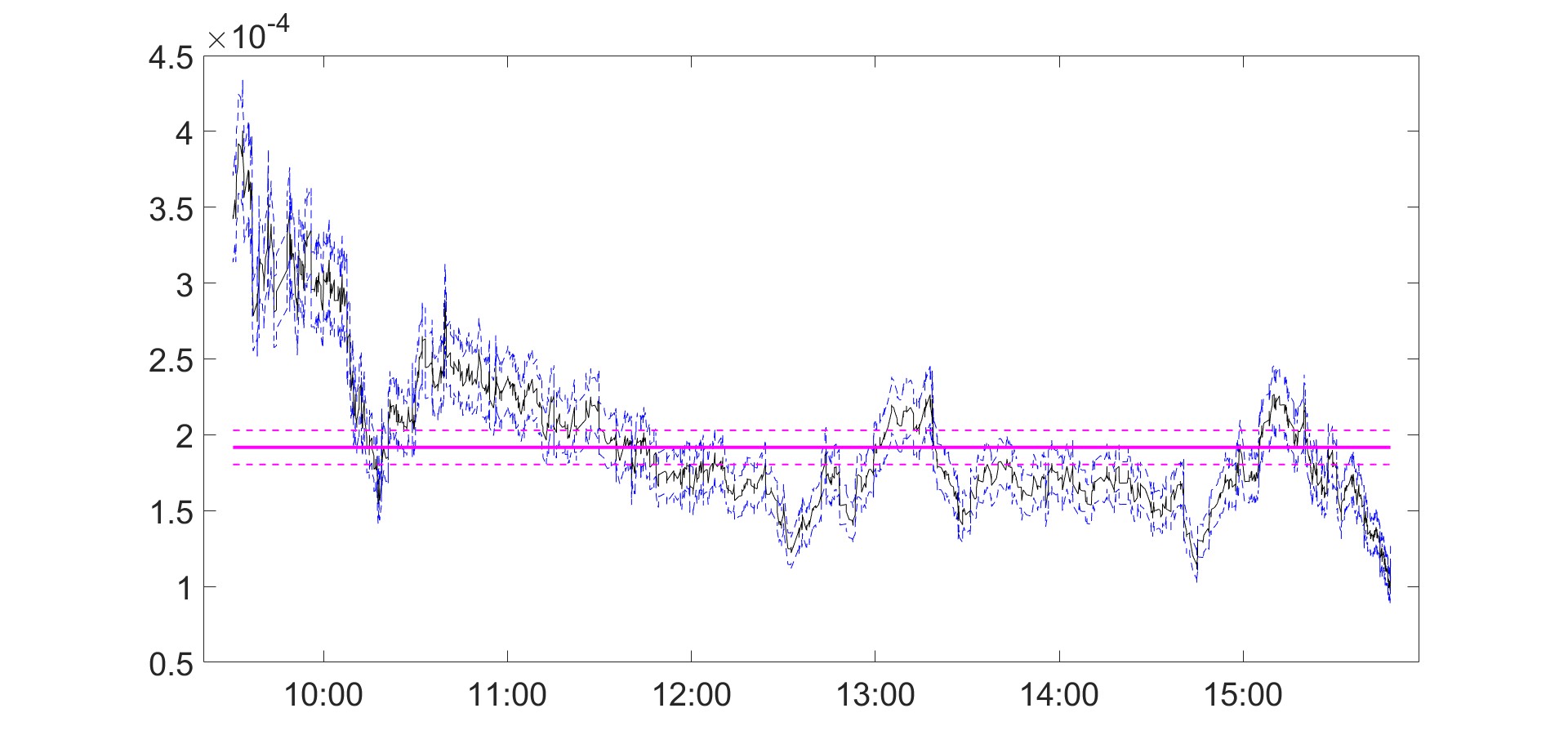}}
	\caption{\footnotesize 
    Estimates of $\hat{\beta}_{t}^{(s)}$ (black lines) and time-varying 95\% confidence bands (blue dashed lines), with a window of $M = 101$. Estimated regression $\hat{\beta}_1^{(reg)}$ (dash-dotted magenta lines) with 1 minute aggregated data and bands (dashed magenta lines). Results for MSFT (a), AMZN (b), CLSK (c), and FLGT (d) on June 04, 2021.}
	\label{betatvsample}
\end{figure}

We now present an application to real data. We compare the out-of-sample performance in estimating the permanent impact when using the two approaches. For this, we compute today's estimates of the permanent impact using coefficients estimated on the previous day. Specifically, we first obtain parameter estimates from the previous day for the DCS model. Then, we filter the time-varying instantaneous impact dynamics using today's observation in an online mode, and we calculate the impact using the equation~\eqref{piamhqtvt} with an aggregation window of $M$ = 101 trades. This window is wide enough for the initialization method to be effective. In this approach, the permanent impact estimation is conditional on the information flow of the day of the forecast. For the regression approach, we compute the estimate on the previous day's data, by regressing the one minute returns versus the contemporaneous net order flow (in this case the difference between the number of buyer and seller-initiated trades). In the end, we use the estimate as an unconditional prediction for the actual day. Figure~\ref{betatvsample} shows the estimated $\hat {\beta}^{(s)}_t$ for the investigated days. For all the stockstocksas in our sample, we observe clear dynamics of the permanent impact coefficient. We notice that the estimated time-varying $\hat \beta_t^{(s)}$ tends to fluctuate around $\hat \beta^{(reg)}$ but many observations lie outside the confidence bands estimated using OLS. Comparing the temporal average of $\hat \beta_t^{(s)}$ and $\hat \beta^{(reg)}$, we notice that the two approaches provide quite different point estimates: we have that their relative absolute difference is roughly 30\% for all the ticker of our sample. To statistically verify the difference between the two, we performed a t-test on the 22 days of the sample. The rejection of the null hypothesis of the t-test performed on the series of permanent impacts for both stocks makes us better realize the out-of-sample gap between the two approaches. 

The significance of the difference between the time-varying $\hat {\beta}^{(s)}_t$ and the constant $\hat \beta^{(reg)}$ can be better appreciated by quantifying the error on the first. To this end, we rely on the method of \cite{blasques2016sample} to estimate the uncertainty of the parameters in score-driven models by computing confidence bands of a parameter that varies over time ($b_{0,t}$ in our case, as in Figure~\ref{b0tvsample}) using Monte Carlo simulations. Then, by error propagation, we compute confidence bands of the $\hat \beta^{(s)}_t$ for the SDAMH model. Our calculations show that the width of the confidence bands is, on average, $1.144  \cdot 10^{-6}$ for MSFT, $4.398  \cdot 10^{-6}$ for AMZN, $7.775 \cdot 10^{-5}$ for CLSK, and $3.638 \cdot 10^{-5}$ for FLGT, that is, significantly smaller than the variability range of $\hat \beta^{(s)}_t$. In conclusion, it is evident that the estimated permanent impact varies considerably throughout the day and a regression estimation based on the entire trading day would not be able to capture this variability, leading to severe transaction cost mis-estimation.

\section{Conclusions}
\label{section:conclusion}

The seminal work of \cite{hasbrouck1991measuring} pioneered research on trade information content using an SVAR model to describe mid-quote price changes and trade signs. However, subsequent research has highlighted several drawbacks: misspecification of trade signs, estimated parameter values may lead to contradictory conclusions when the impact function is non-linear, and the model's lack of parsimony. Most notably, the assumption of a constant instantaneous impact—tied to liquidity—fails to account for the highly dynamic and time-varying nature of liquidity \cite{cont2014price}, \cite{mertens2022liquidity}. This paper introduces the Score-Driven Aggregated Modified Hasbrouck (SDAMH) model to address these limitations. First, we model mid-quote returns jointly with the probability of buy trades using a logistic function, introducing non-linearity. This allows the cumulative impulse response function (CIRF) to depend on the local market state, unlike Hasbrouck’s linear specification. Second, we ensure a parsimonious model of order-flow dynamics by adapting the heterogeneous specification of \cite{corsi2009simple} to capture long persistence with few parameters. Finally, we incorporate time-varying liquidity and volatility using observation-driven dynamics inspired by \cite{creal2013generalized} and \cite{harvey2013dynamic}. We test the SDAMH model on trades and quotes for ten selected NASDAQ stocks (AAPL, AMZN, BNFT, CLSK, EXTR, FLGT, GOOG, INTC, MO$^*$, and MSFT) in June, 2021, reporting results in the main text for MSFT, AMZN, CLSK, and FLGT. The model performs well, capturing intra-day liquidity and volatility variations, with the instantaneous impact adapting quickly to market changes, particularly around FOMC announcements. The CIRF depends on both the time-varying impact and market state, which Hasbrouck’s model cannot capture. We also link the SDAMH model to the regression approach of \cite{cartea2016incorporating} for estimating the permanent impact parameter in transaction cost analysis, showing that our approach provides consistent, more accurate estimates without the need for data aggregation. Future extensions of the current approach will consider modeling the interplay between price returns, trade signed volumes, as a proxy for market activity, and the liquidity outstanding on the opposite side of the limit order book. Furthermore, we have only considered the instantaneous impact and the volatility as time-varying quantities. One can extend the SDAMH to the case where the remaining parameters, especially those describing the role of past quotes and trades on market impact, and thus the ``memory" of the market, are time-varying.

\newpage
\singlespacing

\newpage

\doublespacing
\begin{center}
    {\LARGE Figures}
\end{center}

\newpage

\begin{figure}[htt]
	\centering
         \textbf{All sample time-varying estimates of the instantaneous impact}\par\medskip
	\subfigure[June, 2021 - MSFT]
	{\includegraphics[width=7cm]{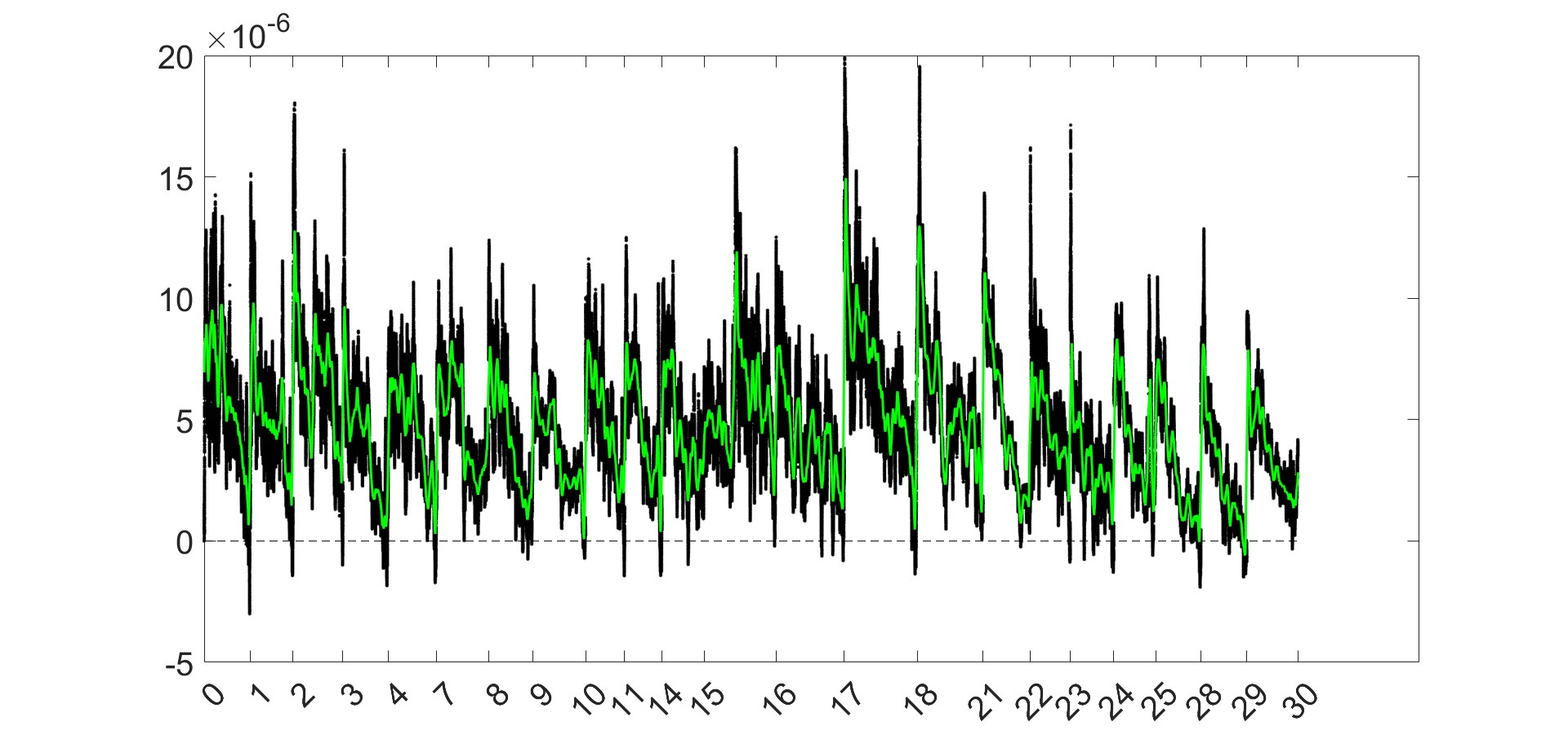}} 
        \hspace{1mm}
	\subfigure[June, 2021 - AMZN]
	{\includegraphics[width=7cm]{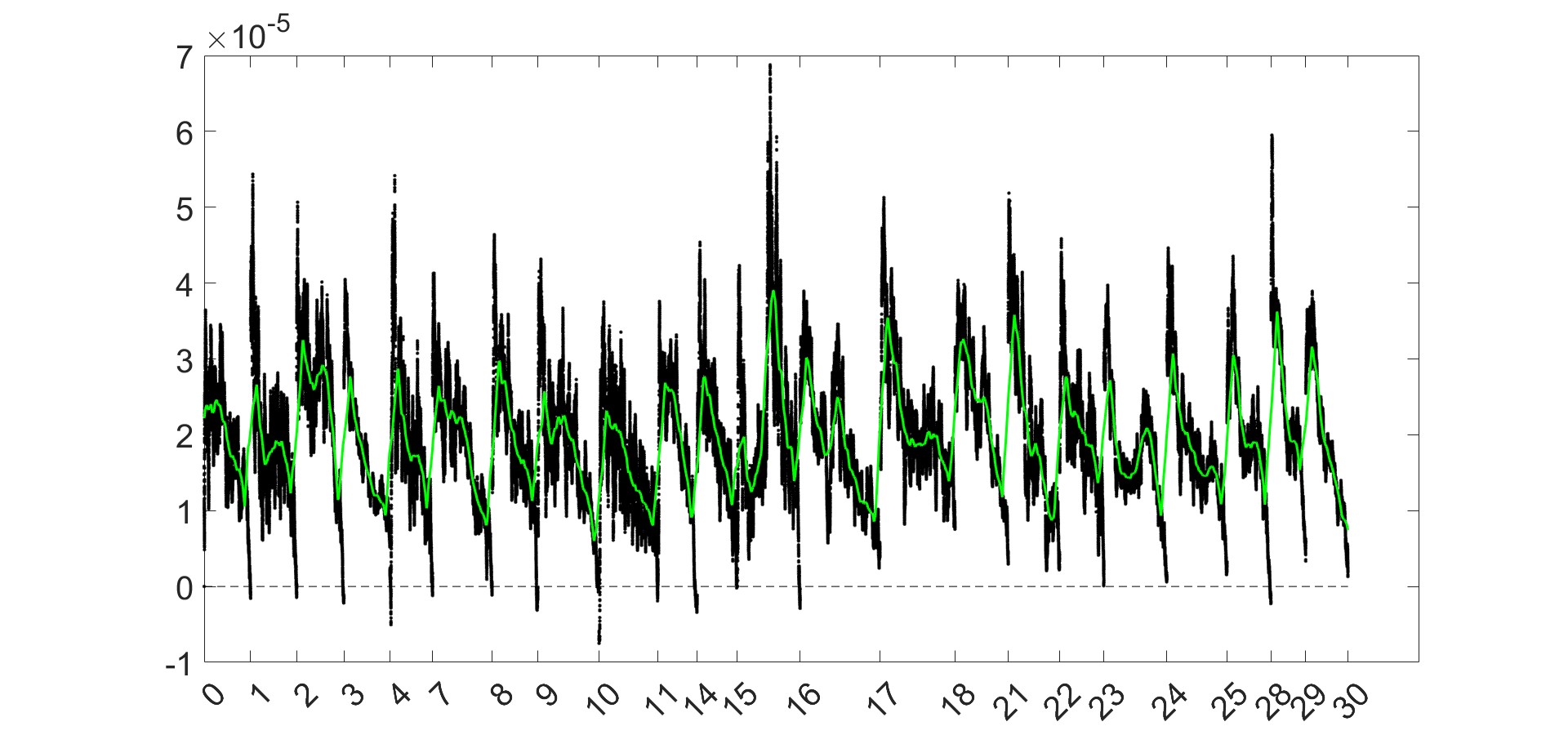}}
         \hspace{1mm}
	\subfigure[June, 2021 - CLSK]
	{\includegraphics[width=7cm]{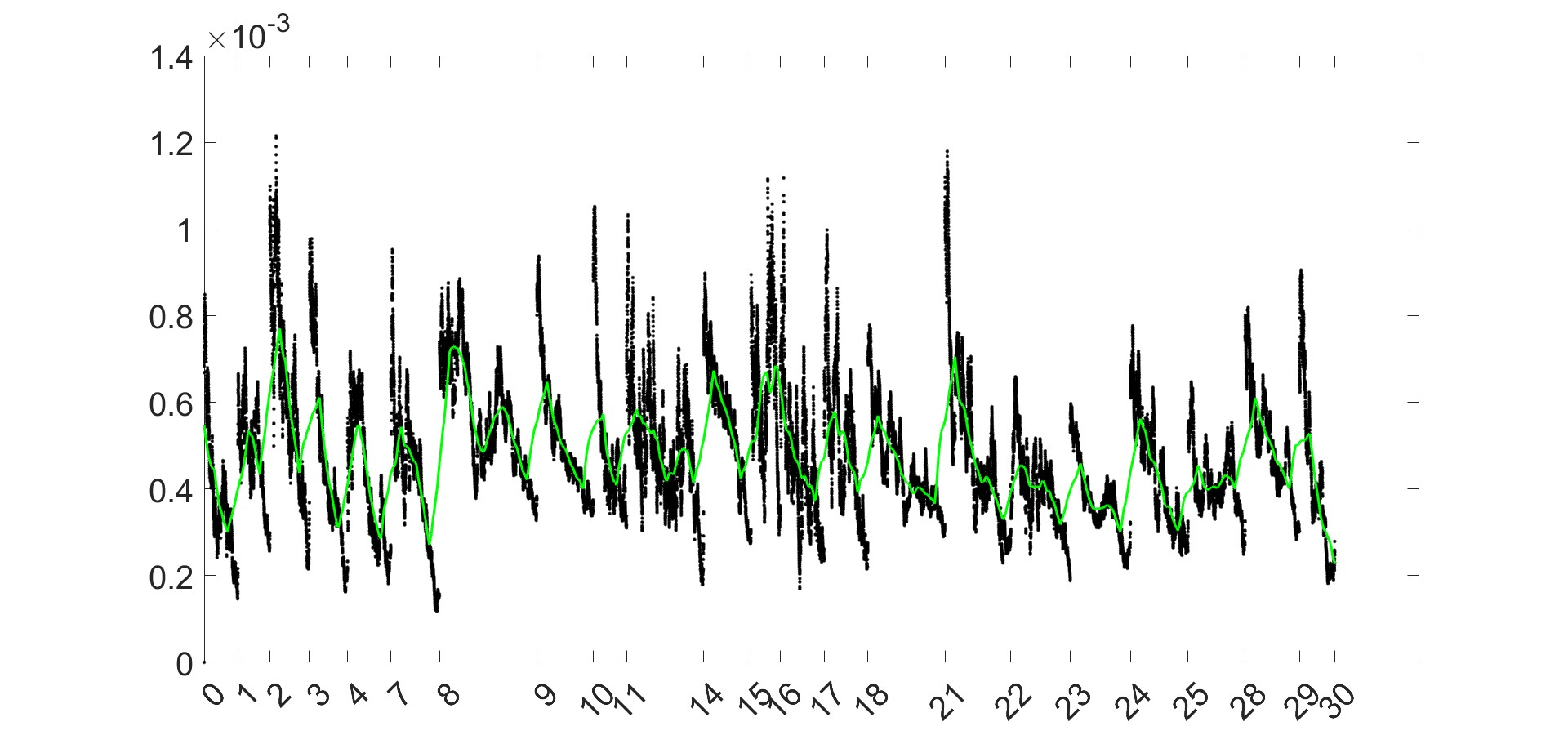}}
         \hspace{1mm}
	\subfigure[June, 2021 - FLGT]
	{\includegraphics[width=7cm]{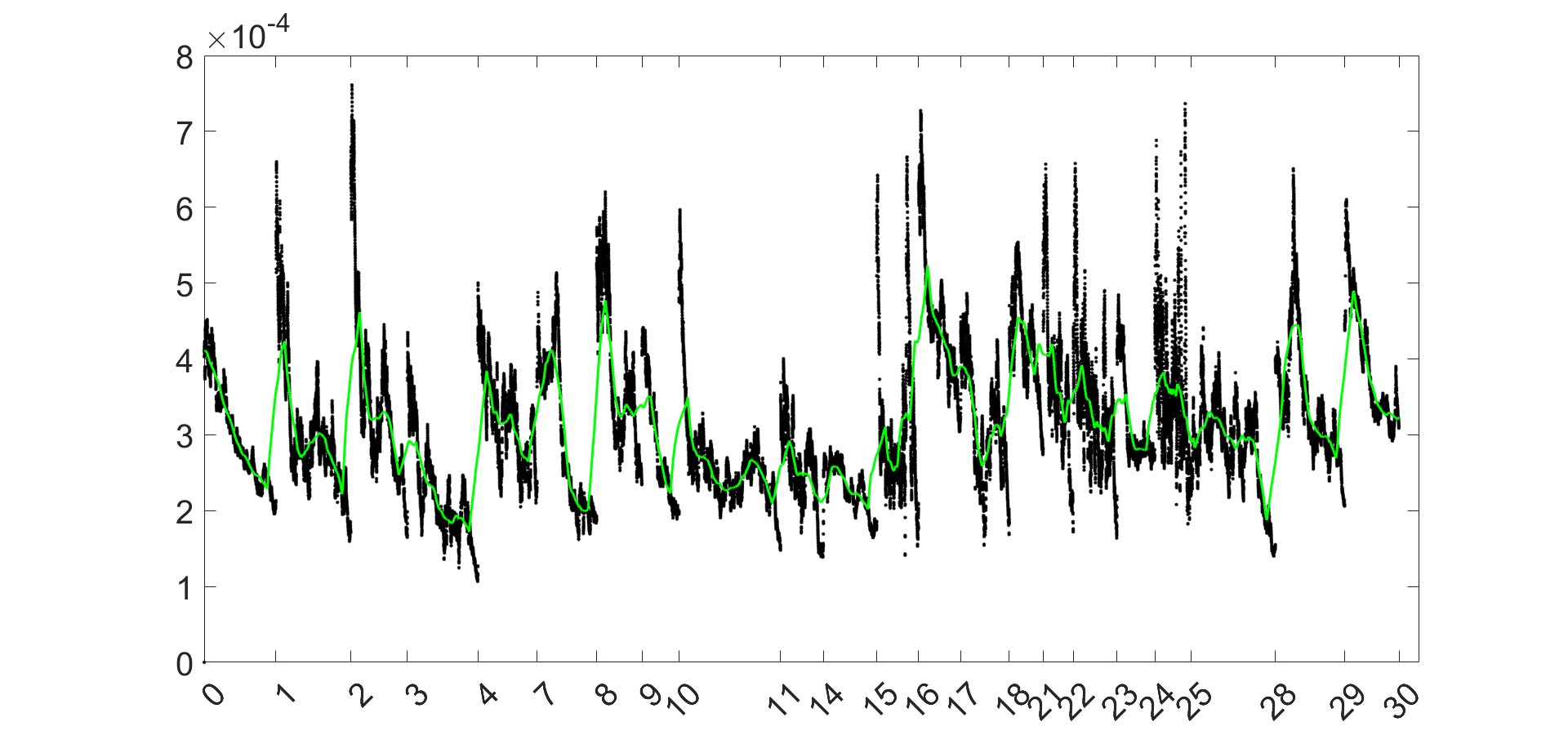}}
	\caption{\footnotesize 
    Estimates of the time-varying instantaneous impact parameter of the SDAMH model (black lines) and its moving average (green solid lines). Overview of June, 2021 trading month.}
	\label{b0tv}
\end{figure}

\newpage

\newpage

\newpage

\newpage

\newpage

\newpage

\newpage

\newpage

\doublespacing
\begin{center}
    {\LARGE Online supplementary material}
\end{center}

\newpage

\begin{appendices}

\section{Monte Carlo study}
\label{app:mc}

In this appendix, we present the finite-sample performance of MLE on data generated according to the SDAMH model and the results obtained when it is used as a filter of misspecified (and non-stationary) dynamics.

\subsection{Finite sample performance of MLE on simulations}
\label{app:finitesampleperforancemc}

 We first perform a Monte Carlo experiment to investigate the statistical properties of the MLE of the vector of static parameter $\theta$, when the DGP is our model. The procedure is the following: 1. Simulate the time series of $T = 10,000$ observations from Eqs. \eqref{model1eq1}, \eqref{model1eq2} and \eqref{model1eq3} for values of the static parameter vector (see the first column of Table \ref{dgpSD-MHmodelmc}, "Par") close to the estimates we obtain using real data (see Table \ref{avpar}), 2. Estimate the model, maximizing the log-likelihood function in Eq.~\eqref{likeGAS} according to the procedure \ref{section:estimation}, 3. Repeat 1) and 2) $S = 100$ times, 4. Collect the $S$ estimates to compute summary statistics. As Table \ref{dgpSD-MHmodelmc} suggests, the MLE can identify point estimates very close to the true values when the DGP is our model. The results are homogeneous in almost all coefficient estimates. The standard deviation in the simulation is low with respect to both the mean and the median. Moreover, the low quantile simulation range indicates that the coefficients are correctly identified.

\begin{table}[htt]
		\centering
      \textbf{Monte Carlo simulations of MLE finite sample performance}\par\medskip
	\begin{tabular}{ccccccccccccccccccc}
Par & True & Mean & Median & Std & $\Delta q$ \\ \hline
$\mu_1$ &  -1.465$\cdot 10^{-7}$ & -1.473$\cdot 10^{-7}$ & -1.471$\cdot 10^{-7}$ & 5.113$\cdot 10^{-8}$ & 1.747$\cdot 10^{-7}$ \\
$\mu_2$  &2.325$\cdot 10^{-2}$ & 2.335$\cdot 10^{-2}$ & 2.324$\cdot 10^{-2}$ & 3.045$\cdot 10^{-3}$ & 5.088$\cdot 10^{-3}$ \\    
$a_{1}$  &3.384$\cdot 10^{-2}$ & 3.379$\cdot 10^{-2}$ & 3.388$\cdot 10^{-2}$ & 1.722$\cdot 10^{-3}$ & 3.133$\cdot 10^{-3}$ \\   
$b_{1}$  & 8.271$\cdot 10^{-6}$ & 8.275$\cdot 10^{-6}$ & 8.277$\cdot 10^{-6}$ & 8.387$\cdot 10^{-8}$ & 2.686$\cdot 10^{-7}$ \\   
$c_{1}$  &-1,3470 & -1,3490 & -1,3480 & 109 & 513 \\               
$d_{1}$  &1.445 & 1.444 & 1.445 & 1.024$\cdot 10^{-2}$ & 3.196$\cdot 10^{-2}$ \\   
$\bar{b}_{10}$ &8.271$\cdot 10^{-6}$ & 8.275$\cdot 10^{-6}$ & 8.277$\cdot 10^{-6}$ & 8.387$\cdot 10^{-8}$ & 2.686$\cdot 10^{-7}$ \\ 
$\bar{d}_{10}$&0.374 & 0.373 & 0.374 & 1.643$\cdot 10^{-2}$ & 4.766$\cdot 10^{-2}$ \\   
$\bar{b}_{100}$&-3.472$\cdot 10^{-6}$ & -3.475$\cdot 10^{-6}$ & -3.485$\cdot 10^{-6}$ & 1.504$\cdot 10^{-7}$ & 4.733$\cdot 10^{-7}$ \\
$\bar{d}_{100}$&0.342 & 0.342 & 0.342 & 2.754$\cdot 10^{-2}$ & 4.161$\cdot 10^{-2}$ \\ \hline
$\alpha$ & 4.802$\cdot 10^{-3}$ & 2.807$\cdot 10^{-3}$ & 3.804$\cdot 10^{-3}$ & 2.399$\cdot 10^{-4}$ & 5.176$\cdot 10^{-4}$ \\   
$b_{0,0}$ &4.003$\cdot 10^{-6}$ & 4.003$\cdot 10^{-6}$ & 4.005$\cdot 10^{-6}$ & 2.043$\cdot 10^{-7}$ & 4.679$\cdot 10^{-7}$ \\ \hline
$\gamma$ & 9.542$\cdot 10^{-3}$ & 9.547$\cdot 10^{-3}$ & 9.544$\cdot 10^{-3}$ & 3.406$\cdot 10^{-4}$ & 9.965$\cdot 10^{-4}$ \\  
$\sigma_0$ & 5.083$\cdot 10^{-5}$ & 5.087$\cdot 10^{-5}$ & 5.082$\cdot 10^{-5}$ & 1.737$\cdot 10^{-6}$ & 5.870$\cdot 10^{-6}$ \\    	\hline 
	\end{tabular}
		\caption{\footnotesize True value of the parameters (True), mean, median, standard deviation (Std) and the difference between the quantile at 95\% and the quantile at 5\% ($\Delta q$) estimated using Monte Carlo simulations.} 
		\label{dgpSD-MHmodelmc}
\end{table}
 
\subsection{Filtering a misspecified dynamics}
\label{section:filteringmissdyn}

This section shows the ability of the SDAMH model as a predictive filter \cite{nelson1996asymptotically} of misspecified dynamics. Specifically, we simulate a model described by Eqs. (\ref{model1eq1}) and (\ref{model1eq2}), but the dynamics of the parameter $b_{0,t}$ is not described by Eq. (\ref{model1eq3}), but follows a different specification. More precisely, we test three deterministic and one stochastic dynamics of the parameter $b_{0,t}$, namely: 1. \text{\textbf{Fast Sine}} $\quad b^*_{0,t+1} = 0.5 + 0.5 \sin(2 \pi t / 200)$, 2. \text{\textbf{Step}} $\quad b^*_{0,t+1} = \mathbb{I} (t>500)$, 3. \text{\textbf{Ramp}} $\quad b^*_{0,t+1} = \mod(t/200)$, 4. \text{\textbf{AR(1)}} $\quad b^*_{0,t+1} = c + \phi b^*_{0,t} + u_t$ with $c = 0.05$ and $\phi = 0.9$. We then use our model to filter the generated paths of the time-varying parameter from the observations of $x_t$ and $r_t$ and we evaluate its performances by comparing with the true parameter dynamics. The Monte Carlo study is carried out using a sample size $T$ of 10,000 observations and $S = 100$ simulated paths~\footnote{For the AR(1) process, we generate only one sample of $b_{0,t}$, but $100$ simulations of the $(r_t, x_t)$ process.}. 
\begin{figure}[htt]
	\centering
       \textbf{Filtering misspecified dynamics on MCs}\par\medskip
	\subfigure[Fast Sine]{%
		\includegraphics[width=0.35\textwidth]{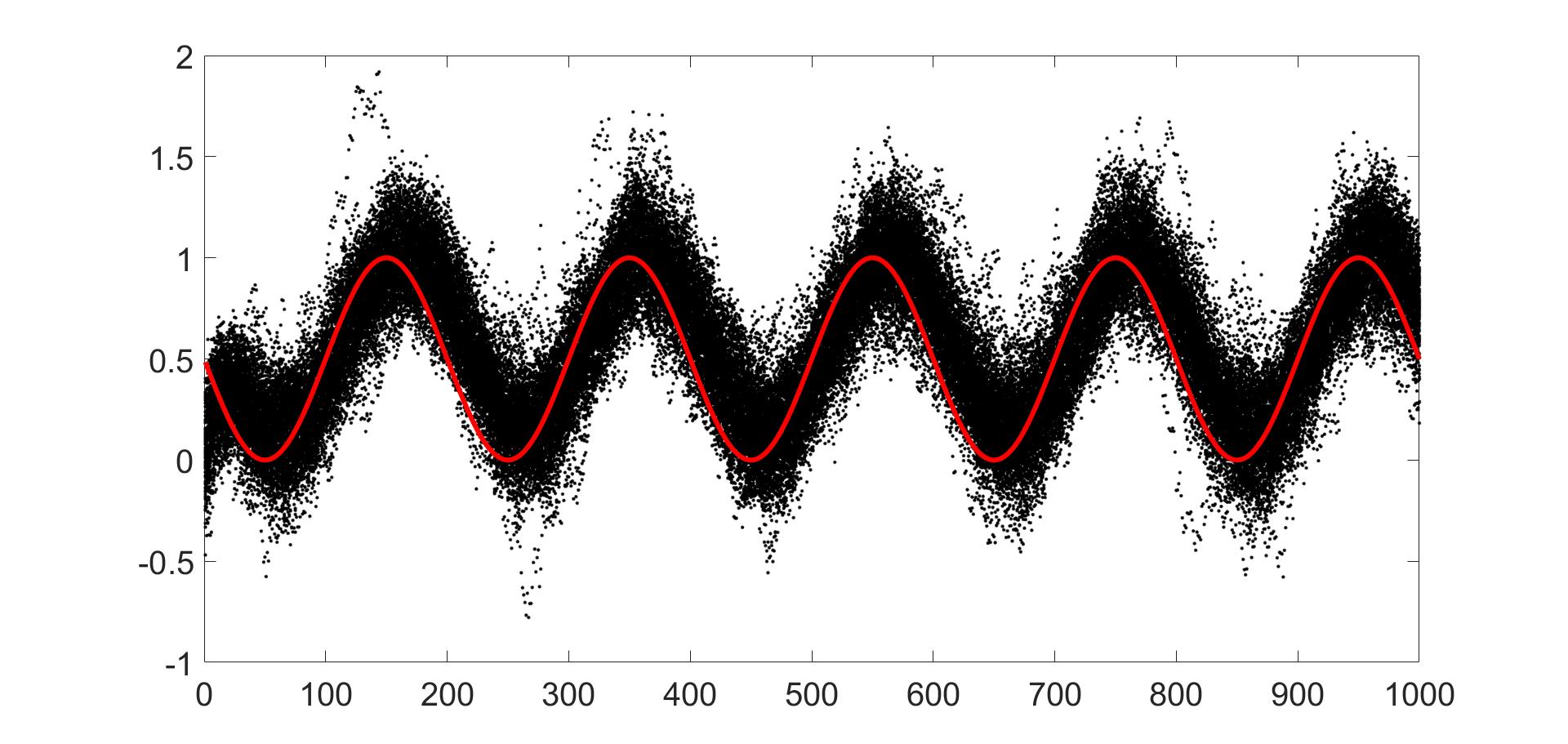}%
		\label{fig:q_time_all}%
	}\hfil
	\subfigure[Ramp]{%
		\includegraphics[width=0.35\textwidth]{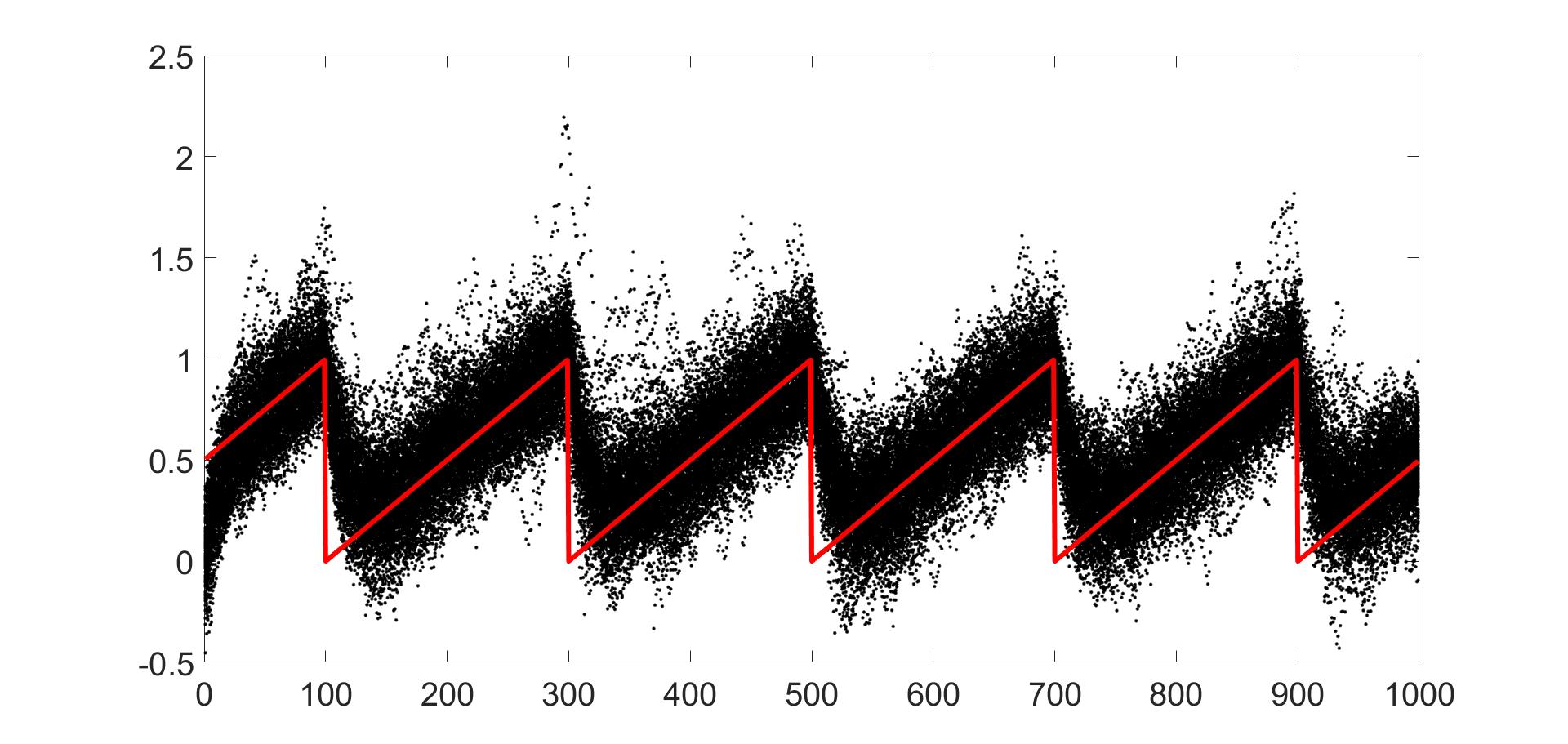}%
		\label{fig:q_time_sat}%
	}
	\subfigure[Step]{%
		\includegraphics[width=0.35\textwidth]{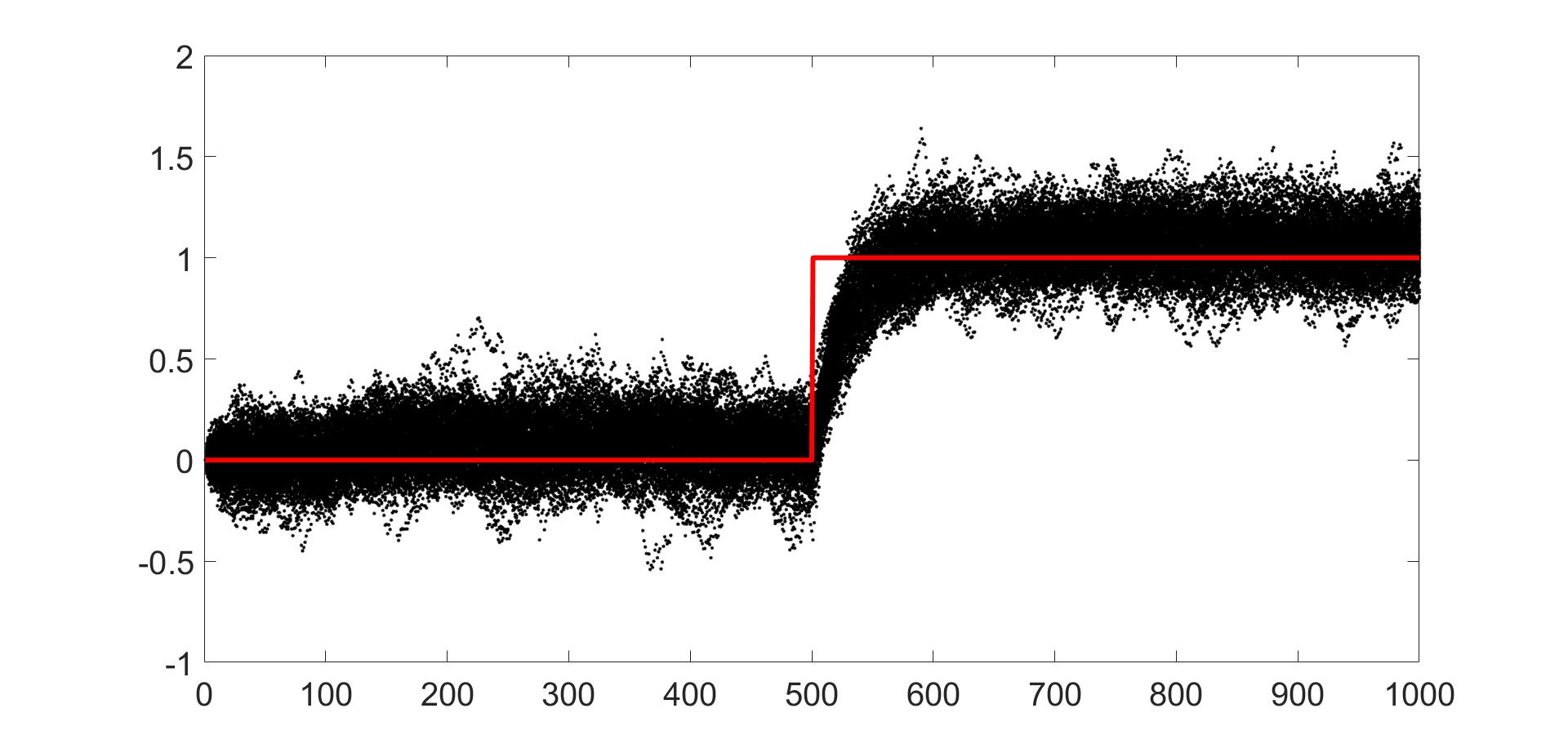}%
		\label{fig:q_time_unsat}%
	}\hfil
	\subfigure[AR(1)]{%
		\includegraphics[width=0.35\textwidth]{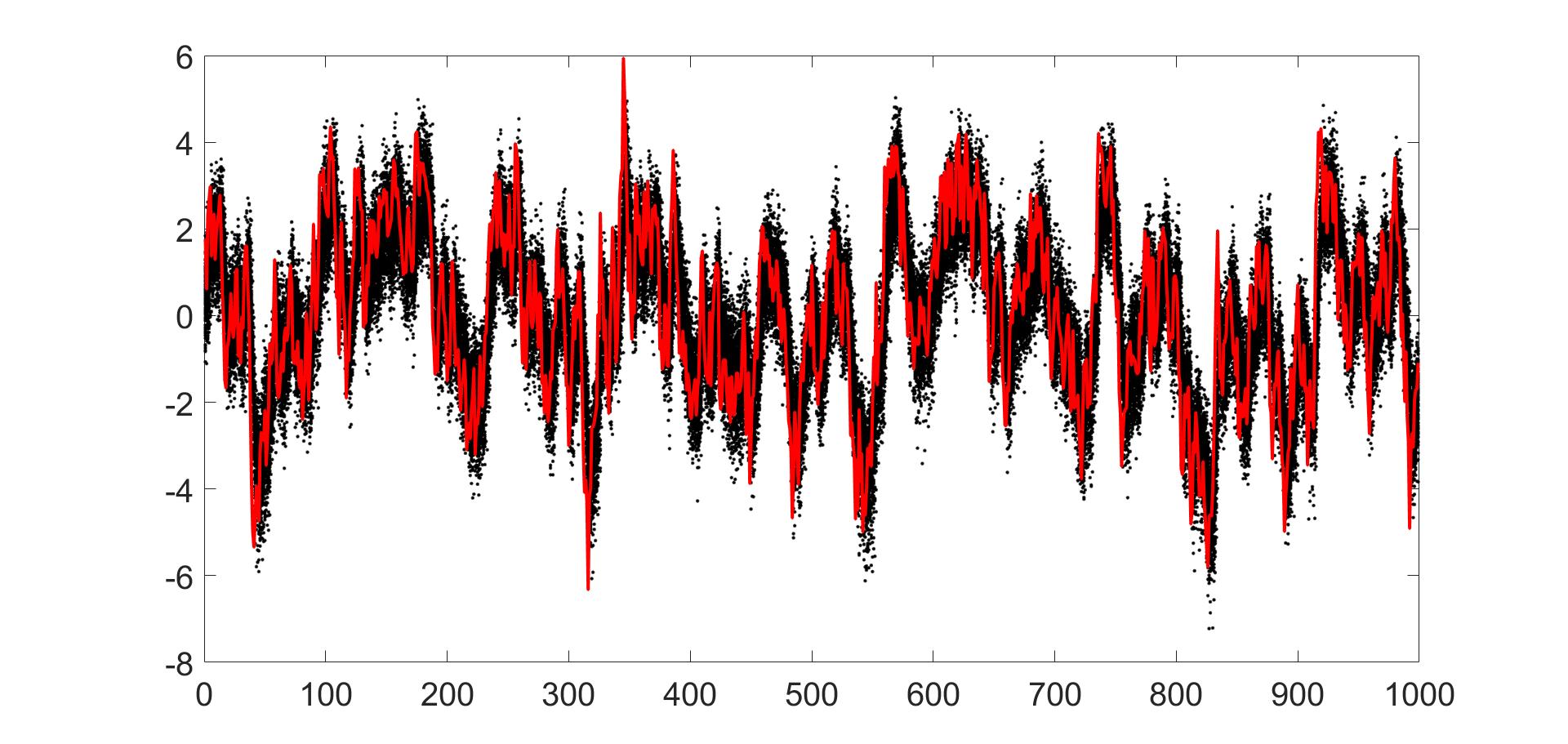}%
		\label{fig:q_time_sample}%
	}
	\caption{\footnotesize Temporal evolution of $b^*_{0,t}$ for the considered DGPs (red solid lines) and filtered dynamics from the SDAMH (black dots).}
 \label{fig1000}
\label{missdinfilter}
\end{figure}
Figure~\ref{missdinfilter} shows the simulated trajectories of $b_{0,t}$ (solid red line) together with the time series filtered using the SDAMH model (black dots) for the $S = 1,000$ Monte Carlo replicas. The filtered paths follow quite closely the true parameter dynamics, even if the DGP is not an SDAMH model. Moreover, the filtered paths are capable of tracing and capturing abrupt changes of $b^*_{0,t}$. Despite its integrated nature, the filter can also reproduce a misspecified stationary trajectory as shown in panel (d).

To quantify the accuracy of the model in capturing the behaviors of $b^*_{0,t}$, we compute the Mean Absolute Error (MAE), defined as $\text{MAE}(b_{0,t}) = \frac{1}{T}\sum_{t=1}^{T}\big|b_{0,t}-b^*_{0,t}\big|$. To consider a relative measure of deviation between the simulated and filtered path, we divide the MAE by the absolute value of the average of the simulated path ($|\bar{b}^*_{0,t}|$). This quantity is named $\text{MAE}^*$, and is computed as: $\text{MAE}^* = \frac{1}{|\bar{b}^*_{0,t}|}\text{MAE}.$ We then compute the average over the $S$ simulations using two different sample sizes: $T = 1,000$ and $T = 10,000$. The results are reported in Table \ref{accmeas}. These results show that the SDAMH works well in recovering the true pattern of the time-varying parameter and it can be considered an effective filter also for stochastic autoregressive processes, although in the case of stationary AR(1) process, the error is larger. This is consistent with the nonstationary characteristic of the SDAMH filter. Interestingly, in all cases, the MAE clearly shrinks as the sample size increases.

\begin{table}[htt]
	\centering
        \textbf{Relative Mean Absolute Error on MCs}\par\medskip
	\begin{tabular}{ccccccccccc}
		& T	       &Fast Sine & Step  & Ramp  & AR(1) \\ \hline 
		$\text{MAE}^*$  &1,000 & 0.178    & 0.220 & 0.216 & 0.251 \\
		$\text{MAE}^*$  &10,000& 0.085    & 0.093 & 0.89 & 0.103 \\
		\hline 
	\end{tabular}
	\caption{\footnotesize Mean Absolute Error divided by $|\bar{b}_{0,t}|$ ($\text{MAE}^*(b_{0,t})$) of the filtered paths averaged over the number of Monte Carlo simulations for two different sample size: $T = 1,000$ and $T = 10,000$.} 
	\label{accmeas}
\end{table}

\section{Model comparison and selection}
	\label{app:modelsel}
This appendix compares different models and selects the best one. Our comparison criterion is based on the skewness, kurtosis, and homoscedasticity of residuals, while the model selection is based on the Bayesian Information Criterion (BIC). We also compare the integrated and stationary specifications of the instantaneous impact DCS dynamics. The considered models are (i) the original Hasbrouck (H) model of Eq.~\eqref{hasbrouckmodel} with $p=5$ lags, (ii) the modified Hasbrouck (MH) model of Eq.~\eqref{modhasbrouckmodel} with $p=5$ lags, (iii) the AMH model of Eq.~\eqref{aggmodhsbrouckmodel}, and (iv) the SDAMH model of Eqs.~\eqref{model1eq1}, \eqref{model1eq2}, and \eqref{model1eq3}. For illustrative purposes, we report results for June 04, 2021 and MSFT, AMZN, CLSK, and FLGT. We selected these stocks because they represent well the typical behavior of small/large tick and high/low liquidity (in terms of average market orders executed per day) stocks. For the results for the entire sample of stocks and days, please refer to Online Appendix \ref{section:otherticks}. 
	
\subsection{Residuals diagnostic: skewness and kurtosis}
\label{section:rescheck}

In this part, we compute the skewness and kurtosis of the residuals to test whether the distributions of the data are far from the Gaussian assumption. For the residuals of the return equation we consider standardized residuals, whereas for the trade sign equation, we test quantile residuals. The reason is that in this case, we have residuals from an inverse logistic and, thus, of a generalized linear model (see \cite{agresti2015foundations} for more details). In the original H model, we test directly the residuals, since this is a linear model. 

Tables \ref{reschecksk} and \ref{rescheckk} show the results. For the return equation, since the skewness and the kurtosis of the original Hasbrouck's model are far from the Gaussian ones, we conclude that the model is highly misspecified. The inverse logistic transformation improves the residuals of the trade sign in the case of the MH model. The use of the aggregation technique (jointly with the non-linear transformation) strongly improves the quality of the results, but some issues are still present for the $r_t$ equation. Finally, with the introduction of the time-varying parameters in the SDAMH model, the analysis can be considered satisfactory.

\begin{table}[htt]
\centering
\textbf{Skewness}\par\medskip
\begin{tabular}{lcccccccc}
       & \multicolumn{2}{c}{\textbf{MSFT}} & \multicolumn{2}{c}{\textbf{AMZN}} & \multicolumn{2}{c}{\textbf{CLSK}} & \multicolumn{2}{c}{\textbf{FLGT}} \\ \hline
       & $r_t$           & $x_t$           & $r_t$           & $x_t$           & $r_t$           & $x_t$           & $r_t$           & $x_t$           \\ \hline
H      & -0.170          & -0.055          & 0.102           & -0.102          & -0.110          & -0.145           & 0.105           & -0.080          \\
MH     & -0.157          & -0.006          & 0.101           & -0.027          & -0.115          & -0.132           & 0.158           & -0.020          \\
AMH    & -0.089          & -0.007          & 0.103           & -0.023          & -0.120          & -0.121           & 0.131           & -0.018          \\
SDAMH  & -0.012          & -0.004          & 0.025           & -0.020          & -0.005          & -0.021           & 0.022           & -0.021          \\ \hline
\end{tabular}
\caption{\footnotesize Skewness of the standardized residuals of the return equation for the four investigated models, skewness of the standardized residuals of the trade sign equation for the H model, and the skewness of the quantile residuals for the remaining three models. Results for MSFT, AMZN , CLSK, and FLGT on June 04, 2021. } 
\label{reschecksk}
\end{table}

\begin{table}[htt]
\centering
\textbf{Kurtosis}\par\medskip
\begin{tabular}{lcccccccc}
       & \multicolumn{2}{c}{\textbf{MSFT}} & \multicolumn{2}{c}{\textbf{AMZN}} & \multicolumn{2}{c}{\textbf{CLSK}} & \multicolumn{2}{c}{\textbf{FLGT}} \\ \hline
       & $r_t$           & $x_t$           & $r_t$           & $x_t$           & $r_t$           & $x_t$           & $r_t$           & $x_t$           \\ \hline
H      & 11.505          & 3.637           & 10.152          & 7.004           & 9.850           & 4.122           & 8.512           & 8.870           \\
MH     & 10.536          & 3.564           & 9.128           & 3.782           & 9.325           & 3.175           & 7.150           & 3.655           \\
AMH    & 6.263           & 3.162           & 5.772           & 3.357           & 6.005           & 3.001           & 5.020           & 3.200           \\
SDAMH  & 3.338           & 3.092           & 3.104           & 2.925           & 3.150           & 3.120           & 2.955           & 3.045           \\ \hline
\end{tabular}
\caption{\footnotesize Kurtosis of the standardized residuals of the return equation for the four investigated models, kurtosis of the standardized residuals of the trade sign equation for the H model, and the kurtosis of the quantile residuals for the remaining three models. Results for MSFT, AMZN , CLSK, and FLGT on June 04, 2021. }
\label{rescheckk}
\end{table}

\subsection{Residuals diagnostic: homoscedasticity test}
\label{section:hetero}

To test whether our model is well specified, we perform a Lagrange Multiplier (LM) test with $p = \{1,2,3,4,5,7,10,15,20,50\}$ lags and, for each model, we report in Table \ref{tabhettest1} the number of times the null hypothesis of homoscedastic residuals is not rejected. For the non-linear models (AMH and SDAMH), we test the quantile residuals of the trade sign equation. The test shows that, while residuals of the Hasbrouck's and the AH model are heteroscedastic, especially for the return variable, the picture improves after lag aggregation and the non-linear transformation in the AMH model. The residual homoscedasticity hypothesis cannot be rejected for the DCS version where $b_{0,t}$ and $\sigma_t^2$ are time-varying. This strengthens the conclusion that our model is better specified than Hasbrouck's one. 

\begin{table}[htt]
\centering
        \textbf{Results of a Lagrange Multiplier test on residuals}\par\medskip
\begin{tabular}{lccccccccccc}
       & \multicolumn{2}{c}{\textbf{MSFT}} &  & \multicolumn{2}{c}{\textbf{AMZN}} &  & \multicolumn{2}{c}{\textbf{CLSK}} & \multicolumn{2}{c}{\textbf{FLGT}} \\ \hline
       & $r_t$           & $x_t$           &  & $r_t$             & $x_t$           &  & $r_t$           & $x_t$          &  & $r_t$             & $x_t$         \\ \hline
H      &     1           &     5           &  &       0           &        5        &  &     2           &    6           &  &     1             &     7         \\
AH     &     0           &     6           &  &       0           &        5        &  &     1           &    7           &  &     1             &     6         \\
AMH    &     2           &    10           &  &       2           &       10        &  &     4           &   9           &  &     3             &    9         \\ 
SDAMH  &     8           &    10           &  &       9           &       10        &  &     9           &   10           &  &    10             &    10         \\ \hline
\end{tabular}
	\caption{\footnotesize Number of times the null hypothesis is not rejected at 5\% for the LM test with lags $p = \{1,2,3,4,5,7,10,15,20,50\}$ on the residual equations for MSFT, AMZN, CLSK, and FLGT on June 04, 2021.}
	\label{tabhettest1}
\end{table}

Table \ref{tabhettest1} shows the results of the homoscedasticity test for two days from the sample and for MSFT, AMZN, CLSK, and FLGT. An interesting related analysis concerns the homoscedasticity test on aggregated residuals. We consider both aggregations on $100$ consecutive transactions and on a minute of physical time. Table \ref{tabhettest34} summarized the results and confirms that the SDAMH model provides homoscedastic aggregated residuals, pointing to the fact that a joint mid-quote return and trade sign model should account for liquidity and volatility time-variation (while, of course, the Hasbrouck's model cannot). It is worth mentioning that the idea that volatility clustering is related to the persistence of market impact and not to the clustering of volume or several trades has been proposed in \cite{gillemot2006there} by using surrogate data analysis. 
\newpage
\begin{table}[htt]
\centering
\textbf{LM test results on aggregated residuals}\par\medskip
\begin{tabular}{lcccclcccc}
 & \multicolumn{4}{c}{\textbf{100 trades}} &  & \multicolumn{4}{c}{\textbf{1 minute }} \\ 
 & \multicolumn{2}{c}{\textbf{MSFT}} & \multicolumn{2}{c}{\textbf{AMZN}} &  & \multicolumn{2}{c}{\textbf{MSFT}} & \multicolumn{2}{c}{\textbf{AMZN}} \\ \hline
    & $r_t$ & $x_t$ & $r_t$ & $x_t$ &  & $r_t$ & $x_t$ & $r_t$ & $x_t$ \\ \hline
H      & 2     & 6     & 2     & 0     &  & 4     & 6     & 1     & 0 \\
AH     & 0     & 3     & 2     & 1    &  & 0     & 4     & 3     & 2     \\
AMH    & 3    &  10   & 4    & 10    &  & 6     & 10    & 5    & 10     \\
SDAMH & 9   & 10    & 8    & 10     &  & 9     & 10     & 9    & 10    \\ \hline
\end{tabular}
	\caption{\footnotesize Number of times the null hypothesis cannot rejected at 5\% for the LM test with lags $p = \{1,2,3,4,5,7,10,15,20,50\}$ residuals aggregated over $100$ trades (left) and $1$ minute (right) for MSFT and AMZN on 04 June, 2021.}
	\label{tabhettest34}
\end{table}

\begin{table}[htt]
\centering
\textbf{LM test results on aggregated residuals}\par\medskip
\begin{tabular}{lcccclcccc}
 & \multicolumn{4}{c}{\textbf{100 trades}} &  & \multicolumn{4}{c}{\textbf{1 minute }} \\ 
 & \multicolumn{2}{c}{\textbf{CLSK}} & \multicolumn{2}{c}{\textbf{FLGT}} &  & \multicolumn{2}{c}{\textbf{CLSK}} & \multicolumn{2}{c}{\textbf{FLGT}} \\ \hline
    & $r_t$ & $x_t$ & $r_t$ & $x_t$ &  & $r_t$ & $x_t$ & $r_t$ & $x_t$ \\ \hline
H      & 2     & 6     & 2     & 0     &  & 2     & 4     & 1     & 1 \\
AH     & 0     & 3     & 2     & 1    &  & 1     & 4     & 3     & 1     \\
AMH    & 3    &  10   & 4    & 10    &  & 4     & 10    & 5    & 10     \\
SDAMH & 9   & 10    & 8    & 10     &  & 9     & 10     & 9    & 10    \\ \hline
\end{tabular}
	\caption{\footnotesize Number of times the null hypothesis cannot rejected at 5\% for the LM test with lags $p = \{1,2,3,4,5,7,10,15,20,50\}$ residuals aggregated over $100$ trades (left) and $1$ minute (right) for CLSK and FLGT on 04 June, 2021.}
	\label{tabhettest341}
\end{table}

\subsection{Model selection among constant parameter models}
\label{section:modelselstat}
	
In this section, we perform a model selection on the three constant parameter models (H, AH, and AMH) by using the Bayesian Information Criterion, defined as 
	
\begin{equation}\label{bic}
	BIC = K \log(T) - \log(\hat{\mathcal{L}}),
\end{equation}
	
where $K$ is an integer indicating the number of model parameters, $T$ is the length of the time series, and $\hat{\mathcal{L}}$ is the overall maximized likelihood function. Instead of reporting the value of the estimated BIC on each of the 22 sample days for the stocks, we report in Table \ref{biccomp} the number of times (days) the BIC of one of the proposed models is better than the others. We observe that the AMH model is better than both the H and AH models. In the range of 77-82 \% days, the model with the aggregation of lags overcomes the other two constant-parameter models. This shows that aggregation is a key element for better modeling of joint price and trade dynamics. 

\begin{table}[htt]
\centering
\textbf{Results of BIC on different static parameter models}\par\medskip
\begin{tabular}{lccccccccccc}
 & \multicolumn{2}{c}{\textbf{MSFT}} &  & \multicolumn{2}{c}{\textbf{AMZN}} &  & \multicolumn{2}{c}{\textbf{CLSK}} & \multicolumn{2}{c}{\textbf{FLGT}} \\ \hline
            & \# Days & \%     &  & \# Days & \%     &  & \# Days & \%     & \# Days & \% \\ \hline
H           & 2       & 9.1\%  &  & 2       & 9.1\%  &  & 3       & 13.6\% & 4       & 18.2\% \\
AH          & 3       & 13.6\% &  & 2       & 9.1\%  &  & 4       & 18.2\% & 2       & 9.1\%  \\
AMH         & 17      & 77.4\% &  & 18      & 81.8\% &  & 15      & 68.2\% & 16      & 72.7\% \\
Tot. Days   & 22      &        &  & 22      &        &  & 22      &        & 22      &        \\ \hline
\end{tabular}
	\caption{\footnotesize Number of times (days) a model is better than the other in terms of BIC. The results are in percentage.} 
	\label{biccomp}
\end{table}

\subsection{Integrated vs stationary score-driven specifications}
\label{section:modelselGAS}
	
 In this appendix, we take into consideration different DCS specifications of the AMH model with the aim to assess whether a stationary (autoregressive) or integrated score-driven model describes the data better. We compare the two different score-driven models using the BIC and the Out-of-Sample Likelihood (OSL). The latter quantity is defined as the logarithmic likelihood evaluated with the data for day $\tau$ and with the parameters estimated for day $\tau-1$ ($\theta^{(\tau-1)}$) and calculated as
	\begin{equation}\label{osl}
	\text{OSL} := \mathcal{L}(y_{t+1}^{(\tau)} \ | \mathcal{F}_t^{(\tau)}, \theta^{(\tau-1)}), 
	\end{equation}
	where $y_t^{(\tau)}$ is the vector of variables,  $\mathcal{F}_{t}^{(\tau)}$ is the available information set and $\theta^{(\tau)}$ is the vector of static parameter of day $\tau$, with $\tau = 2, \dots, 22$ and $t$ is the intraday trade time. The DCS versions proposed for the AMH model are the following	
	\begin{align}
	& \textbf{SD-AR}	&\quad b_{0,t+1} &= \omega + \beta b_{0,t} + \alpha x_t (r_t - \mu^{(1)}_{t}), \label{GAS1} \\
	& \textbf{SD-INT} &\quad b_{0,t+1} &=  b_{0,t} + \alpha x_t (r_t - \mu^{(1)}_{t}), \label{GAS2}
	\end{align}
Section~\ref{section:model} of the main paper introduces and discusses Eqs.~\eqref{GAS1} and~\eqref{GAS2}. Tables \ref{biccompgas_combined} and \ref{oslcompgas_combined} provide the results for the six sample securities not discussed in the main text.
\begin{table}[htt]
\centering
\begin{tabular}{cccccccccccc}
           & \textbf{AAPL} &  & \textbf{GOOG} &  & \textbf{INTC} &  & \textbf{BNFT} &  & \textbf{EXTR} &  & \textbf{MO} \\ \hline
SD-AR      & 0.00\%        &  & 0.00\%        &  & 4.55\%        &  & 9.09\%        &  & 4.55\%        &  & 0.00\%         \\
SD-INT     & 100\%         &  & 100\%         &  & 95.45\%       &  & 90.91\%       &  & 95.45\%       &  & 100\%          \\
\hline
\end{tabular}
\caption{\footnotesize Percentage of days out of 22 where a DCS specification performs better than the other one according to the BIC.}
\label{biccompgas_combined}
\end{table}

\begin{table}[htt]
\centering
\begin{tabular}{cccccccccccc}
           & \textbf{AAPL} &  & \textbf{GOOG} &  & \textbf{INTC} &  & \textbf{BNFT} &  & \textbf{EXTR} &  & \textbf{MO$^*$} \\ \hline
SD-AR      & 9.6\%         &  & 14.3\%        &  & 9.6\%         &  & 4.8\%         &  & 9.6\%         &  & 4.8\%         \\
SD-INT     & 90.4\%        &  & 85.7\%        &  & 90.4\%        &  & 95.2\%        &  & 90.4\%        &  & 95.2\%        \\
\hline
\end{tabular}
\caption{\footnotesize Percentage of days out of 21 where a DCS specification performs better than the other one according to the OSL.}
\label{oslcompgas_combined}
\end{table}

\section{Estimation and Initialization details}\label{app:estimation}

\subsection{Estimation}
\label{section:estimation}

Given the vector of unknown parameters $\theta \in \Theta \subset \mathbb{R}^{D} $, we maximize the log-likelihood function of Eq.~\eqref{likeGAS} to find maximum likelihood estimates (MLE) $\hat{\theta}$, as in Eq.~\eqref{maxprob}. The steps are the following:

\begin{enumerate}
	\item Estimate the parameters of the AH model using OLS and get $\hat{\theta}^{AH}$. We split $\hat{\theta}^{AH}$ in two components: a) the parameters of the $r_t$ equation $\hat{\theta}^{AH}_r$ and b) the parameters of the $x_t$ equation $\hat{\theta}^{AH}_x$. Thus, we have $\hat{\theta}^{AH}$ = $(\hat{\theta}^{AH}_r, \hat{\theta}^{AH}_x)$.
        
    \item Estimate the parameters of the AMH model by initializing the likelihood function with a parameter vector $\hat{\theta}^{AMH}_0 = (\hat{\theta}^{AH}_r, 2 \cdot \hat{\theta}^{AH}_x)$. The return equations of the AH and AMH model are the same. For this reason, we initialize $\hat{\theta}^{AMH}_r$ with $\hat{\theta}^{AH}_r$. Concerning the trade sign equation, we approximate $x_t$ with $2\pi_t - 1=\tanh{\left(\mu_t^{(2)}/2\right)}$. When $\mu_t^{(2)} \simeq 0.0$, by linearization we obtain that $2\pi_t - 1 \simeq \mu_t^{(2)}/2$ and then $x_t \simeq \mu_t^{(2)}/2$. Under this regime, we guess the following linear relation between the AH and AMH trade sign parameters: $\hat{\theta}_x^{AH} \simeq \hat{\theta}^{AMH}_x/2$, hence the initialization rule.

    \item  Maximize the log-likelihood of the SDAMH model initializing the constant parameters with the ones estimated for the AMH model. At this stage, the initial value of the time-varying parameters $b_{0,t}$ and $\sigma_t$ enter the procedure. We initialize them with the constant instantaneous impact $b_0$ and volatility $\sigma_1$ estimated for the AMH model. 

\end{enumerate}
The maximization of the log-likelihood is carried out using a standard quasi-Newton optimization method.

\section{The role of the market state on the information content of a trade}
\label{section:valresmhm}

 In this appendix, we explore in more detail the role of the market state in determining the LRCIRF and thus the information content of a trade. We do this in three ways: first, we restrict the regression analysis of Section \ref{section:cirf} to an intraday period when the integrated dynamics of the instantaneous impact parameter $b_{0_t}$ is less evident. In the second approach, we perform the regression of LRCIRF on the market state by considering a non-linear constant parameter model, namely the AMH model. The first analysis can also be seen as a robustness check of the results of Section \ref{section:cirf}. We restricted the regression analysis to the period from 10:30 onward and, in doing so, we removed from the analysis the period when the highest values of $b_{0,t}$ are observed (see Figure~\ref{b0tv}) restricting the effects of the integrated dynamics of the time-varying parameter on the regression. The results, reported in Table \ref{regtablow}, indicate that the sign of the relevant coefficients is statistically significant and consistent with the unconditional results of Table \ref{regtab}. In fact, it is $\gamma_1 > 0$, $\gamma_2 < 0$ and $\gamma_3 > 0$. The $R^2$ remains high in all cases, ranging from $0.293$ to $0.445$. Thus, the role of the state is significant also when the instantaneous impact is relatively small. However, we notice some differences. For the regression of Eq. \eqref{eqreg1} the value of $\gamma_1$ is on average small when compared with the previous estimates but still statistically significant. Despite the reduction in the variability range of $b_{0,t}$, the effect of the state on the cumulative long-term impulse response function ($\gamma_2$) is roughly the same and still significant. On the other hand, the coefficients $\gamma_{3}$ of the regression of Eq. \eqref{eqreg4} are lower (in absolute value) than those estimated in Section \ref{section:cirf}.

  \newpage   

	\begin{table}[ht]
		\centering
              \textbf{Regression results - I}\par\medskip
                 		\resizebox{150mm}{40mm}{
			\begin{tabular}{llllllllllllllll}
				& \multicolumn{4}{c}{\textbf{MSFT}}      & \multicolumn{4}{c}{\textbf{AMZN}}               \\ 
				\hline
	              & $\gamma_0$ & $\gamma_1$ & $\gamma_2$  & $\gamma_{3}$ & $\gamma_0$ & $\gamma_1$ & $\gamma_2$ & $\gamma_{3}$    \\ \hline
        Est   &  1.937$\cdot 10^{-5}$    & 0.411    &  -2.061$\cdot 10^{-3}$   & -  &    4.993$\cdot 10^{-5}$   &  0.731 &  -5.030$\cdot 10^{-3}$   & -   \\
				$ \# \{\text{p} > 5 \% \}$   & 2        &  3 & 2 & -   &      3    &    2  &  3  &-          \\ 
				$R^2$& 0.445& & & &  0.403 & &  \\\hline
	Est   &1.937$\cdot 10^{-6}$     & -   &  -  & -2.023$\cdot 10^{-3}$   & 3.474$\cdot 10^{-5}$  &    - &  -  &  -3.795$\cdot 10^{-3}$   \\
				$ \# \{\text{p} > 5 \% \}$ & 4 &  - &-  & 3  &    4 & -  &  -  &  2       \\ 
				$R^2$& 0.365 & & & &  0.397 & &  \\
								& \multicolumn{4}{c}{\textbf{CLSK}}      & \multicolumn{4}{c}{\textbf{FLGT}}               \\ 
				\hline
	              & $\gamma_0$ & $\gamma_1$ & $\gamma_2$  & $\gamma_{3}$ & $\gamma_0$ & $\gamma_1$ & $\gamma_2$ & $\gamma_{3}$    \\ \hline
        Est   &  6.072$\cdot 10^{-5}$    & 1.093    &  -1.943$\cdot 10^{-3}$   & -  &    3.659$\cdot 10^{-5}$   &  0.625 &  -3.470$\cdot 10^{-3}$   & -   \\
				$ \# \{\text{p} > 5 \% \}$   & 3        &  3 & 3 & -   &      4    &    2  &  3  &-          \\ 
				$R^2$& 0.325& & & &  0.301 & &  \\\hline
	Est   &3.318$\cdot 10^{-4}$     & -   &  -  & -7.565$\cdot 10^{-3}$   & 4.982$\cdot 10^{-4}$  &    - &  -  &  -2.984$\cdot 10^{-3}$   \\
				$ \# \{\text{p} > 5 \% \}$ & 5 &  - &-  & 3  &    4 & -  &  -  &  3       \\ 
				$R^2$& 0.298 & & & &  0.293 & &  \\
				\hline 
		\end{tabular}}
		\caption{\footnotesize Results of regressions of models of Eqs.~\eqref{eqreg1} (top) and \eqref{eqreg4} (bottom).Temporal average of the point estimates (Est), number of times we do not reject the null hypothesis at 5\% level $ \# \{\text{p} > 5 \% \}$, and average $R^2$ in case of "low" range of $b_{0,t}$.}
		\label{regtablow}
	\end{table}
 
To deepen our understanding of the role of the state on the information content of trade (and thus LRCIRF), we performed a regression analysis for the AMH model, for which the instantaneous impact parameter $b_0$ is constant. Being a non-linear model, the IRF depends on the history of the process and the regression we perform, namely
 \begin{equation}
 \quad \text{LRCIRF}_{t} = \gamma_{0} + \gamma_{2} |\mu^{(*)}_{t-1}| +  \eta_{t}. \label{eqreg3}
 \end{equation}
aims at measuring the strength and significance of their relationship. Table \ref{regtabmhm} shows that $\gamma_{2}$ is negative and significant for both stocks, indicating a role for the state in the long-term cumulative response function also when the model neglects the impact of time variations. Finally, the small number of times the null hypothesis of no dependence is rejected and high $R^2$ strengthens the evidence that the state has a relevant role in determining the information content of stock trades. 	

\begin{table}[ht]
		\centering
              \textbf{Regression results - II}\par\medskip
   \begin{tabular}{llllllllllllllll}
				& \multicolumn{4}{c}{\textbf{MSFT}}      & \multicolumn{4}{c}{\textbf{AMZN}}       \\ 
				\hline
				& $\gamma_0$ & $\gamma_1$ & $\gamma_2$  & $\gamma_{3}$ & $\gamma_0$ & $\gamma_1$ & $\gamma_2$ & $\gamma_{3}$    \\ \hline
				Est   &   1.719$\cdot 10^{-5}$   & -   &  -1.687$\cdot 10^{-3}$  & - &   2.016$\cdot 10^{-5}$  &  -  &  -1.892$\cdot 10^{-3}$   & -   \\
				$ \# \{\text{p} > 5 \% \}$ & 3 &  -  & 2 & -   &      2   &    -   &  2  &-          \\ 
				$R^2$& 0.379 & & & &  0.359 & &  \\
								& \multicolumn{4}{c}{\textbf{CLSK}}      & \multicolumn{4}{c}{\textbf{FLGT}}       \\ 
				\hline
				& $\gamma_0$ & $\gamma_1$ & $\gamma_2$  & $\gamma_{3}$ & $\gamma_0$ & $\gamma_1$ & $\gamma_2$ & $\gamma_{3}$    \\ \hline
				Est   &   5.243$\cdot 10^{-5}$   & -   &  -2.645$\cdot 10^{-3}$  & - &   4.362$\cdot 10^{-5}$  &  -  &  -3.384$\cdot 10^{-3}$   & -   \\
				$ \# \{\text{p} > 5 \% \}$ & 5 &  -  & 4 & -   &      3   &    -   &  3  &-          \\ 
				$R^2$& 0.278 & & & &  0.274 & &  \\
				\hline 
		\end{tabular}
		\caption{\footnotesize Results of regressions of Eq.~\eqref{eqreg4} for the AMH model. Temporal average of the point estimates (Est), number of times we do not reject the null hypothesis at 5\% level $ \# \{\text{p} > 5 \% \}$, and average $R^2$.}
		\label{regtabmhm}
	\end{table}

As a final robustness check, in order to examine the role of trade size without suffering from attenuation bias caused by pooling buy and sell orders, we run separate regressions conditioned on trade direction:
\begin{align*}
\text{LRCIRF}_t^{\text{Buy}} &= \gamma_0^+ + \gamma_1^+ b_{0,t} + \gamma_2^+ |\mu_{t-1}^{(*)}| + \gamma_4^+ |v_t^{\text{Buy}}| + \eta_t^+, \\
\text{LRCIRF}_t^{\text{Sell}} &= \gamma_0^- + \gamma_1^- b_{0,t} + \gamma_2^- |\mu_{t-1}^{(*)}| + \gamma_4^- |v_t^{\text{Sell}}| + \eta_t^-,
\end{align*}
where $v_t^{\text{Buy}}$ and $v_t^{\text{Sell}}$ denote trade volumes for buy and sell orders, respectively. Table~\ref{tab:volume_impact} for buy trades (the results for sell trades are similar and are omitted here for brevity) confirms that the trade size has no systematic incremental explanatory power over the time-varying instantaneous impact $b_{0,t}$ and the market state $|\mu_{t-1}^{(*)}|$.

\begin{table}[!ht]
\centering
\small
\textbf{Trade size regressions}\par\medskip
\resizebox{\textwidth}{!}{
\begin{tabular}{lcccccccc}
\hline
& \multicolumn{4}{c}{\textbf{AAPL}} & \multicolumn{4}{c}{\textbf{AMZN}} \\
\hline
& $\gamma_0^+$ & $\gamma_1^+$ & $\gamma_2^+$ & $\gamma_4^+$ & $\gamma_0^+$ & $\gamma_1^+$ & $\gamma_2^+$ & $\gamma_4^+$ \\
Est & 2.635$\cdot 10^{-5}$ & 0.4273 & -1.117$\cdot 10^{-3}$ & 3.028$\cdot 10^{-5}$ & 3.163$\cdot 10^{-5}$ & 0.453 & -1.716$\cdot 10^{-3}$ & 4.827$\cdot 10^{-5}$ \\
$\#\{p>5\%\}$ & 2 & 2 & 2 & 16 & 1 & 2 & 2 & 17 \\
$R^2$ & 0.426 & & & & 0.355 & & & \\
\hline
& \multicolumn{4}{c}{\textbf{BNFT}} & \multicolumn{4}{c}{\textbf{CLSK}} \\
\hline
& $\gamma_0^+$ & $\gamma_1^+$ & $\gamma_2^+$ & $\gamma_4^+$ & $\gamma_0^+$ & $\gamma_1^+$ & $\gamma_2^+$ & $\gamma_4^+$ \\
Est & 6.263$\cdot 10^{-6}$ & 0.793 & -2.801$\cdot 10^{-3}$ & 2.037$\cdot 10^{-5}$ & 4.532$\cdot 10^{-5}$ & 0.525 & -1.435$\cdot 10^{-3}$ & 4.718$\cdot 10^{-5}$ \\
$\#\{p>5\%\}$ & 1 & 0 & 1 & 18 & 1 & 2 & 2 & 18 \\
$R^2$ & 0.395 & & & & 0.307 & & & \\
\hline
& \multicolumn{4}{c}{\textbf{EXTR}} & \multicolumn{4}{c}{\textbf{FLGT}} \\
\hline
& $\gamma_0^+$ & $\gamma_1^+$ & $\gamma_2^+$ & $\gamma_4^+$ & $\gamma_0^+$ & $\gamma_1^+$ & $\gamma_2^+$ & $\gamma_4^+$ \\
Est & 2.743$\cdot 10^{-5}$ & 1.265 & -0.171$\cdot 10^{-5}$ & 1.346$\cdot 10^{-5}$ & 4.223$\cdot 10^{-5}$ & 0.453 & -1.890$\cdot 10^{-3}$ & 3.637$\cdot 10^{-5}$ \\
$\#\{p>5\%\}$ & 3 & 0 & 3 & 19 & 2 & 1 & 2 & 17 \\
$R^2$ & 0.318 & & & & 0.477 & & & \\
\hline
& \multicolumn{4}{c}{\textbf{GOOG}} & \multicolumn{4}{c}{\textbf{INTC}} \\
\hline
& $\gamma_0^+$ & $\gamma_1^+$ & $\gamma_2^+$ & $\gamma_4^+$ & $\gamma_0^+$ & $\gamma_1^+$ & $\gamma_2^+$ & $\gamma_4^+$ \\
Est & 9.677$\cdot 10^{-6}$ & 0.865 & -6.386$\cdot 10^{-5}$ & -3.498$\cdot 10^{-6}$ & 5.146$\cdot 10^{-5}$ & 1.339 & -4.567$\cdot 10^{-5}$ & -2.263$\cdot 10^{-5}$ \\
$\#\{p>5\%\}$ & 0 & 0 & 2 & 17 & 1 & 1 & 1 & 20 \\
$R^2$ & 0.526 & & & & 0.304 & & & \\
\hline
& \multicolumn{4}{c}{\textbf{MO*}} & \multicolumn{4}{c}{\textbf{MSFT}} \\
\hline
& $\gamma_0^+$ & $\gamma_1^+$ & $\gamma_2^+$ & $\gamma_4^+$ & $\gamma_0^+$ & $\gamma_1^+$ & $\gamma_2^+$ & $\gamma_4^+$ \\
Est & 5.883$\cdot 10^{-5}$ & 1.203 & -4.800$\cdot 10^{-5}$ & 2.793$\cdot 10^{-6}$ & 4.001$\cdot 10^{-5}$ & 0.507 & -1.536$\cdot 10^{-3}$ & 3.534$\cdot 10^{-5}$ \\
$\#\{p>5\%\}$ & 4 & 0 & 5 & 18 & 1 & 1 & 2 & 16 \\
$R^2$ & 0.312 & & & & 0.453 & & & \\
\hline
\end{tabular}}
\caption{\footnotesize Results of the trade size augmented regressions estimated for buy trades. Temporal average of the point estimates (Est), number of times we do not reject the null hypothesis at level 5\% $\#\{p > 5\% \}$, and average $R^2$. }
\label{tab:volume_impact}
\end{table}

\section{Derivation and numerical validation of the permanent impact coefficient}
	\label{section:derivpif}

\subsection{Derivation of the coefficient}
 
We start considering the SDAMH model given by Eqs. \eqref{model1eq1}, \eqref{model1eq2}, and \eqref{model1eq3}. To shorten the notation, we define $\bar{r}_t := \sum_{j=t-M}^{t-1} r_j / M$ and $\bar{x}_t := \sum_{j=t-M}^{t-1} x_j / M$. Replacing Eq. \eqref{model1eq1} in Eq. in \eqref{sumrsumx}, we obtain 
\begin{equation}
\begin{split}
\hat{\beta}_t^{(s)} := \frac{\bar{r}_t}{\bar{x}_t} =  a_1 \frac{\bar{r}_{t-1}}{\bar{x}_t} + \frac{\frac{1}{M}  \sum_{j=t-M}^{t-1} b_{0,j} x_{j}}{\bar{x}_t} + b_1 \frac{\bar{x}_{t-1}}{\bar{x}_t} + \bar{b}_{10} \frac{\frac{1}{L_1-1}  \sum_{i = 2}^{L_1} \bar{x}_{t-i}}{\bar{x}_t} +  \bar{b}_{100} \frac{\frac{1}{L_2-L_1-1} \sum_{i = L_1+1}^{L_2} \bar{x}_{t-i}}{\bar{x}_t}\,.
\end{split}
\label{eq:derivation}
\end{equation}
Relying on the fact that the return process $r_t$ typically shows weak evidence of serial correlation, from now on we assume that the local average of $r_t$ over intervals of $M$ observations does not depend on time, i.e., $\bar{r}_{t-i}=\bar{r}$ for any $i\geq 0$. It readily follows that the r.h.s. in the first line in Eq. (\ref{eq:derivation}) boils down to 
\[
 a_1 \frac{\bar{r}}{\bar{x}_t} = a_1  \hat{\beta}_t^{(s)}\,.
\]

Consistent with the long-memory property of the order flow, the local average of the process $x_t$ assumes different values when computed at different intervals intraday. We re-name the first term in the second line of Eq. (\ref{eq:derivation}) and approximate it as follows  
\begin{equation}
\bar{b}_{0,t}^{(s)} := \frac{\frac{1}{M}\sum_{j=t-M}^{t-1} b_{0,j} x_{j}}{\bar{x}_t} \simeq  \frac{\frac{1}{M} \bar{b}_{0,t} \sum_{j=t-M}^{t-1}  x_{j}}{\bar{x}_t} = \bar{b}_{0,t},
\end{equation}
where $\bar{b}_{0,t}$ is the time-average of $b_{0,j}$; $\bar{b}_{0,t}^{(s)}$ is the order-flow-weighted average of $b_{0,j}$. Both averages are computed over the interval from $t-M$ to $t-1$~\footnote{Applying the law of iterated expectations and assuming that $b_{0,t}$ is independent from $x_t$, it is easy to show that $\bar{b}^{(s)}_{0,t} = \bar{b}_{0,t}$. In the case of real data where $b_{0,t}$ depends on $x_t$, the two quantities are similar. In our sample, the mean absolute deviation is $3.262 \cdot 10^{-5}$ for MSFT, $1.370 \cdot 10^{-5}$ for AMZN, $2.218 \cdot 10^{-5}$ for BLSK, and $1.344 \cdot 10^{-5}$ for FLGT and the maximum absolute deviation is $1.973 \cdot 10^{-4}$ for MSFT, $1.444 \cdot 10^{-4}$ for AMZN, $1.353 \cdot 10^{-4}$ for BLSK, and $1.160 \cdot 10^{-4}$ for FLGT.}. Eventually, after defining
\[
\omega^{(1)}_t := \frac{\bar{x}_{t-1}}{\bar{x}_t}\,,\quad 
\omega ^{(L_1)}_t := \frac{1}{L_1-1}\frac{\sum_{i=2}^{L_1}\bar{x}_{t-i}}{\bar{x}_t}\,,\quad
\omega ^{(L_2)}_t := \frac{1}{L_2-L_1}\frac{\sum_{i=L_1+1}^{L_2}\bar{x}_{t-i}}{\bar{x}_t}\,,
\]

Eq. \eqref{eq:derivation} reduces to
\begin{equation}
\hat \beta^{(s)}_t \simeq   a_1\hat\beta^{(s)}_t  + \bar{b}_{0,t} + b_1 \omega_t^{(1)} + \bar{b}_{10} \omega_t^{(L_1)} + \bar{b}_{100} \omega_t^{(L_2)}, 
\end{equation}
from which we get the permanent price impact function for the SDAMH model as
\begin{equation}\label{ppagg00}
\hat \beta^{(s)}_t \simeq  \frac{\bar{b}_{0,t} + b_1 \omega_t^{(1)} + \bar{b}_{10} \omega_t^{(L_1)} + \bar{b}_{100} \omega_t^{(L_2)} }{1-a_1}. 
\end{equation}

Notice that, coherently with the nonlinearity of the SDAMH model and the presence of a time-varying parameter, the permanent impact coefficient of Eq. \eqref{ppagg00} depends on time both through the market state ($\omega_t^{(1)}$, $\omega_t^{(L_1)}$ and $\omega_t^{(L_2)}$) and through the dynamic instantaneous impact coefficient ($\bar{b}_{0,t}$). It is important to recall that the daily estimates of $b_0$ for the AMH model are similar to the unconditional daily average of $b_{0,t}$. For this, the permanent impact function of the AMH model can be deduced from its score-driven version, replacing the average of the estimated $\bar{b}_{0,t}$ in the intraday intervals of $M$ observations with $b_0$ in Eq. \eqref{ppagg00}. We finally get the permanent impact function of the AMH model as

\begin{equation}
\hat \beta^{(s)}_t \simeq  \frac{b_0 + b_1\omega_t^{(1)} + \bar{b}_{10} \omega_t^{(L_1)} + \bar{b}_{100} \omega_t^{(L_2)}}{1-a_1}. 
\end{equation}

It is easy to show (following the same procedure) that for a linear SVAR($p$) model, the estimated coefficient is given by

\begin{equation}
\hat \beta^{(s)} =  \frac{\sum_{i=0}^{p}b_i}{1-\sum_{i=1}^{p}a_i}. 
\end{equation}

\subsection{Monte Carlo simulated permanent impact function}
\label{section:mcpif}

In this appendix, we use Monte Carlo simulations to assess whether the SDAMH approach provides estimates of the permanent impact which are coherent with the ones obtained using the standard regression approach. The permanent impact is often estimated (see, for example, \cite{cartea2016incorporating}) by performing the linear regression of price changes over a given time interval versus the contemporaneous aggregated order flow, that is, the difference between the volume of buy and sell market orders. Since our model considers price returns and trade signs, we compare the SDAMH estimation with the one obtained from the linear regression of aggregated return versus the contemporaneous net order flow sign, i.e. the difference between the number of buy and sell trades. Moreover, we will consider aggregation in trade time rather than in physical time, since Hasbrouck's model and all the modifications we propose are defined in trade time. For the Monte Carlo experiment, we simulate the $r_t$ and $x_t$ SDAMH models of Eqs. \eqref{model1eq1}, \eqref{model1eq2} and \eqref{model1eq3} using parameters similar to the estimates obtained on real data \footnote{The parameters are the monthly estimates of MSFT (see Table \eqref{avpar}).} and then we compute the permanent impact coefficient on the simulations $S$ using the two approaches. For the SDAMH model, we estimate the static parameters on a simulated series of observations $T = 10,000$, maximizing the log-likelihood function of the model. Then, we obtain the filtered estimates of the time-varying parameter $b_{0,t}$. Considering aggregation windows of $M = 101$ observations~\footnote{Since the $x_t \in \{-1,+1\}$, we use $M$ odd to avoid issues in the estimation of $\hat \beta^{(s)}_t$ when aggregating data.}, we compute the permanent impact coefficients $\hat\beta^{(s)}_t$ of Eq. \eqref{piamh} to get a series of $T/M$ permanent impacts. Then, we compute the average of $\hat\beta^{(s)}_t$'s over the $T/M$ intervals. We repeat the experiment $S = 1,000$ times. In the end, we get the average permanent impact coefficient of the SDAMH computed on the $S$ replications, named $\beta^{(SD)}$. Although SDAMH provides time-varying estimates of permanent impact by nature, averaging over the $T/M$ intervals in each simulation allows us to derive an average value of the permanent impact coefficient of the SDAMH model, which can be compared to the static estimate of the regression model. For the regression approach, we consider the simulated series with $T = 10,000$, we aggregate data on intervals $M = 101$, and we use regression of the price return in each interval versus the contemporaneous difference between buy and sell trades to estimate the permanent impact coefficient. Finally, we name $\beta^{(reg)}$ the permanent impact coefficient of the regression computed as the average on the $S = 1,000$ simulations. Table \ref{pifSDAMHc} shows the estimates of the permanent impact and some summary statistics on Monte Carlo simulations of the two approaches. From the table, we note that the two estimates show similar mean and median (Med) values. Moreover, the two coefficients are properly identified: they exhibit low standard deviation and quantile range values. However, the standard deviation and the quantile range of the $\beta^{(SD)}$ are lower than the ones of the $\beta^{(reg)}$ indicating that the SDAMH provides more accurate estimates than the ones the regression approach on aggregated data.
\newpage

\begin{table}[ht]
	\centering
               \textbf{Monte Carlo results for the permanent impact function}\par\medskip
	\begin{tabular}{ccccccccccc}
		 & Mean & Med & Std & $\Delta q$ \\  \hline
		$\beta^{(SD)}$ & 7.511 $\cdot 10^{-6}$ & 7.508 $\cdot 10^{-6}$ &  6.500 $\cdot 10^{-7}$   &  2.250 $\cdot 10^{-7}$      \\
		$\beta^{(reg)}$& 7.406$\cdot 10^{-6}$  &  7.423 $\cdot 10^{-6}$ & 1.352 $\cdot 10^{-6}$   & 3.010 $\cdot 10^{-7}$  \\ \hline
	\end{tabular}
	\caption{\footnotesize Summary statistics of the permanent impact for SDAMH model ($\beta^{(SD)}$) and for regression model ($\beta^{(reg)}$): Mean, Median (Med), standard deviation (Std), the difference between the quantile in 95\% and the quantile at 5\% ($\Delta q$).} 
	\label{pifSDAMHc}
\end{table}

\section{Results for the other tickers}
	\label{section:otherticks}

In this section, we report for completeness the results for all other analyzed tickers. Specifically, in Section \ref{section:otherticksest} we report the estimated coefficients of the score-driven model, and in Section \ref{section:otherticksimpvol} the filtered dynamics of the time-varying impact and volatility for June, 2021.

\subsection{SDAHM parameter estimates}\label{section:otherticksest}

Comparing Tables \ref{avpar} and \ref{tab:avpar3}, for high liquidity stocks, and Tables \ref{avpar1} and \ref{tab:avpar2}, for less liquid stocks confirms consistent findings between the four stocks reported in the main text and the remaining six in the sample. The estimates for (BNFT, EXTR, MO$^*$) and (AAPL, GOOG, INTC) reveal distinct characteristics in the price and sign dynamics.
\begin{table}[h!]
\centering
\begin{tabular}{lrrrrrr}
\textbf{} & \textbf{AAPL} & & \textbf{GOOG} & & \textbf{INTC} & \\
\textbf{} & \textbf{Est} & \textbf{Std} & \textbf{Est} & \textbf{Std} & \textbf{Est} & \textbf{Std} \\ \hline
$\mu_1$   & 5.052$\cdot 10^{-7}$   & 6.255$\cdot 10^{-7}$ & 1.959$\cdot 10^{-7}$ & 1.528$\cdot 10^{-6}$ & 3.284$\cdot 10^{-7}$ & 7.564$\cdot 10^{-7}$ \\
          & (8.114)   &  & (11.336)  &  & (2.508)  &  \\
$a_1$    & -0.098  & 0.012 & -0.065  & 0.026 & -0.084 & 0.008 \\
          & (34.603)   &  & (5.363)  &  & (21.048)  &  \\
$b_1$    & 1.416$\cdot 10^{-5}$  & 2.158$\cdot 10^{-6}$  & 3.791$\cdot 10^{-5}$  & 7.367$\cdot 10^{-6}$  & 3.722$\cdot 10^{-5}$ & 4.983$\cdot 10^{-6}$ \\
          & (61.319)   &  & (23.670)  &  & (79.856)  &  \\
$b_{10}$   & 9.020$\cdot 10^{-7}$   & 8.101$\cdot 10^{-7}$   & -5.001$\cdot 10^{-6}$   & 2.446$\cdot 10^{-6}$   & 5.996$\cdot 10^{-6}$ & 9.598$\cdot 10^{-7}$ \\
          & (7.865)   &  & (16.761)  &  & (14.658)  &  \\
$b_{100}$  & -3.239$\cdot 10^{-6}$  & 2.529$\cdot 10^{-6}$  & -1.088$\cdot 10^{-6}$  & 3.427$\cdot 10^{-6}$  & -1.507$\cdot 10^{-6}$ & 2.105$\cdot 10^{-6}$ \\
          & (40.274)   &  & (10.017)  &  & (2.880)  &  \\
$\mu_2$   & -0.020  & 0.048  & 0.012  & 0.097  & -0.030  & 0.055 \\
          & (4.329)   &  & (35.400)  &  & (5.106)  &  \\
$c_1$    & -1647.415  & 1105.073  & -644.049  & 500.687  & -1650.583  & 679.074 \\
          & (53.805)   &  & (53171.237)  &  & (7.899)  &  \\
$d_1$    & 1.772  & 0.142  & 1.821  & 0.148  & 2.018  & 0.144 \\
          & (39.207)   &  & (49.287)  &  & (16.085)  &  \\
$d_{10}$   & 0.317  & 0.067  & 0.446  & 0.118  & 0.224  & 0.088 \\
          & (79.637)   &  & (11.939)  &  & (14.837)  &  \\
$d_{100}$  & 0.443  & 0.142  & 0.098  & 0.179  & 0.103  & 0.091 \\
          & (83.029)   &  & (39.508)  &  & (4.991)  &  \\
$b_{0,0}$    & 3.056$\cdot 10^{-5}$ & 6.252$\cdot 10^{-6}$  & 7.914$\cdot 10^{-5}$  & 9.360$\cdot 10^{-6}$  & 6.290$\cdot 10^{-5}$ & 8.883$\cdot 10^{-6}$ \\
          & (64.610)   &  & (10.908)  &  & (12.237)  &  \\
$\alpha$ & 0.003  & 0.003  & 0.005  & 0.002  & 0.002  & 0.001 \\
          & (26.386)   &  & (7.336)  &  & (11.752)  &  \\
$\gamma$ & 0.005  & 0.004  & 0.010  & 0.009  & 0.004  & 0.002 \\
          & (20.919)   &  & (15.098)  &  & (28.390)  &  \\
$\sigma_0$    & 5.884$\cdot 10^{-5}$ & 4.053$\cdot 10^{-5}$  & 0.000  & 7.423$\cdot 10^{-5}$  & 0.000 & 2.136$\cdot 10^{-5}$ \\
          & (37.086)   &  & (20.080)  &  & (27.264)  &  \\
\hline
\end{tabular}
\caption{Temporal daily average of the coefficients ("Est"), standard deviation computed on the 22 trading samples, and average absolute $t$-stats (in brackets) of the estimates. Results reported for high liquidity assets: AAPL, GOOG, and INTC.}
\label{tab:avpar3}
\end{table}
\begin{table}[h!]
	\centering
	\begin{tabular}{lrrrrrr}
		\textbf{} & \textbf{BNFT} & & \textbf{EXTR} & & \textbf{MO$^*$} & \\ 
        \textbf{} & \textbf{Est} & \textbf{Std} & \textbf{Est} & \textbf{Std} & \textbf{Est} & \textbf{Std} \\ \hline
		$\mu_1$   & -1.134$\cdot 10^{-5}$   & 3.171$\cdot 10^{-5}$ & 6.482$\cdot 10^{-6}$ & 1.784$\cdot 10^{-5}$ & 3.558$\cdot 10^{-7}$ & 2.056$\cdot 10^{-6}$ \\
		& (27.034)   &  & (10.003)  &  & (9.635)  &  \\
		$a_1$    & -0.127  & 0.044 & -0.098  & 0.039 & -0.111 & 0.019 \\
		& (12.338)   &  & (28.937)  &  & (7.670)  &  \\
		$b_1$    & -1.606$\cdot 10^{-4}$  & 4.478$\cdot 10^{-5}$  & -1.515$\cdot 10^{-4}$  & 5.061$\cdot 10^{-5}$  & -2.600$\cdot 10^{-5}$ & 8.507$\cdot 10^{-6}$ \\
		& (18.194)   &  & (33.750)  &  & (13.710)  &  \\
		$b_{10}$   & 1.603$\cdot 10^{-5}$   & 2.124$\cdot 10^{-5}$   & 4.639$\cdot 10^{-5}$   & 1.572$\cdot 10^{-5}$   & 9.625$\cdot 10^{-6}$ & 1.992$\cdot 10^{-6}$ \\
		& (31.435)   &  & (22.275)  &  & (12.242)  &  \\
		$b_{100}$  & -1.334$\cdot 10^{-5}$  & 9.501$\cdot 10^{-5}$  & -3.178$\cdot 10^{-5}$  & 3.725$\cdot 10^{-5}$  & -6.235$\cdot 10^{-6}$ & 6.107$\cdot 10^{-6}$ \\
		& (28.954)   &  & (7.212)  &  & (8.380)  &  \\
		$\mu_2$   & -0.039  & 0.209  & -0.096  & 0.217  & -0.015  & 0.064 \\
		& (48.045)   &  & (14.334)  &  & (11.107)  &  \\
		$c_1$    & -147  & 266  & -312  & 206  & -10.997  & 655 \\
		& (6.732)   &  & (6.371)  &  & (11.448)  &  \\
		$d_1$    & 1.608  & 0.182  & 1.834  & 0.169  & 1.729  & 0.150 \\
		& (23.847)   &  & (26.913)  &  & (41.027)  &  \\
		$d_{10}$   & 0.185  & 0.165  & 0.149  & 0.157  & 0.0275  & 0.083 \\
		& (25.961)   &  & (14.808)  &  & (15.373)  &  \\
		$d_{100}$  & 0.012  & 0.364  & -0.034  & 0.333  & 0.169  & 0.218 \\
		& (26.976)   &  & (13.349)  &  & (3.782)  &  \\
		$b_{0,0}$    & 4.814$\cdot 10^{-4}$ & 2.965$\cdot 10^{-4}$  & 2.576$\cdot 10^{-4}$  & 4.980$\cdot 10^{-5}$  & 6.545$\cdot 10^{-5}$ & 1.428$\cdot 10^{-5}$ \\
		& (18.433)   &  & (28.264)  &  & (56.452)  &  \\
		$\alpha$ & 0.017  & 1.339$\cdot 10^{-2}$  & 0.003  & 5.177$\cdot 10^{-3}$  & 0.003  & 1.807$\cdot 10^{-3}$ \\
		& (9.225)   &  & (15.758)  &  & (19.570)  &  \\
		$\gamma$ & 0.017  & 1.353$\cdot 10^{-2}$  & 0.011  & 4.305$\cdot 10^{-3}$  & 0.007  & 2.457$\cdot 10^{-3}$ \\
		& (19.563)   &  & (25.689)  &  & (6.976)  &  \\
		$\sigma_0$    & 1.133$\cdot 10^{-4}$ & 8.425$\cdot 10^{-4}$  & 2.084$\cdot 10^{-4}$  & 4.873$\cdot 10^{-4}$  & 1.547$\cdot 10^{-4}$ & 3.067$\cdot 10^{-5}$ \\
		& (27.858)   &  & (30.167)  &  & (16.743)  &  \\
		\hline
	\end{tabular}
	\caption{Temporal daily average of the coefficients ("Est"), standard deviation computed on the 22 trading samples, and average absolute $t$-stats (in brackets) of the estimates. Results reported for low liquidity assets: BNFT, EXTR, and MO*.}
	\label{tab:avpar2}
\end{table}

One may argue that the trading environment is much richer than what is captured by past returns and trade signs. The depth of the market at various price levels, the volume of past trades, and the imbalance of order flow, among others, may contribute significantly to determining the evolution of the market and unraveling the information content of trade. To address this concern, we have augmented the model specification by adding two external regressors: the bid-ask spread ($SP_t$) and the depth imbalance ($DI_t$). This extension allows the model to capture more nuanced aspects of the trading environment, specifically variables that are well-known to influence market behavior. The augmented model reads as follows:
\begin{align*}
r_t & = \mu^{(1)}_{t} + \phi ~DI_t + v_{1,t}, \\
\pi_t  &= Logit^{-1} \big( \mu^{(2)}_{t} \big), \\
b_{0,t+1} & =  b_{0,t} + \alpha x_t (r_t - \mu^{(1)}_{t}) + \theta ~SP_t, \\
\log(\sigma^2_{t+1}) & = \log(\sigma^2_{t}) + \gamma \left[\frac{(r_t - \mu^{(1)}_{t})^2}{\sigma^2_{t}} - 1\right].
\end{align*}
Tables~\ref{tab:sdamhm_msft_amzn}, \ref{tab:sdamhm_clsk_flgt}, \ref{tab:sdamhm_aapl_intc}, \ref{tab:sdamhm_goog_bnft}, and \ref{tab:sdamhm_extr_mo}  report point estimates. Across the entire sample, the incremental explanatory power of $DI_t$ and $SP_t$ is negligible, with coefficients $\phi$ and $\theta$ remaining statistically insignificant at the 5\% level across most trading days. Signed order flow and its multi-scale memory act as a sufficient statistic for liquidity dynamics, confirming the parsimony of our baseline SDAMH model.
\begin{table}[htt]
    \centering
    \par\medskip
    \resizebox{165mm}{35mm}{
    \begin{tabular}{cccccccccccccccccc}
    \hline
    &&&\textbf{MSFT}&&&&&&\textbf{AMZN}&& \\ \hline
               & Est & Std         &   & Est & Std &              &Est & Std &          & Est & Std \\ \hline
    $\mu_{1}$&-1.221 $\cdot 10^{-7}$ & 5.865 $\cdot 10^{-7}$ & $\mu_{2}$& 0.029 & 0.053 & $\mu_{1}$& 1.383 $\cdot 10^{-6}$ & 1.701 $\cdot 10^{-6}$ & $\mu_{2}$&-0.035 & 0.092 & \\
               &(0.025) &  &                                      &(0.006)  &  &                   &(0.006) &  &                                          &(0.001)  &  \\ 
    $a_1$         &0.025 & 0.024   &$c_1$          &  -7.408  & 1.722     & $a_1$   &-0.107 & 0.043 &$c_1$& -3.313 & 1.162 \\
                   &(0.005) &  &                     & (0.008) &  &                                  &(0.009) &         & & (0.002) &  \\
    $b_1$    &9.121 $\cdot 10^{-6}$ & 9.269 $\cdot 10^{-7}$  &$d_1$& 1.113 & 0.075   & $b_1$& 1.195 $\cdot 10^{-5}$ & 2.432 $\cdot 10^{-6}$ &$d_1$& 1.691 & 0.136 \\
            &(0.000) &                                       && (0.005) &  &            &(0.000) &                                      && (0.002) &  \\
    $\bar{b}_{10}$&-2.418 $\cdot 10^{-6}$ & 5.426 $\cdot 10^{-7}$ & $\bar{d}_{10}$& 0.307 & 0.070 &    $\bar{b}_{10}$& -3.206$\cdot 10^{-6}$ & 1.965 $\cdot 10^{-6}$ &    $\bar{d}_{10}$& 0.337 & 0.154 \\
                          &(0.000) &                                                   && (0.006) & &                   &(0.002) &    &&                                (0.002) &  \\
    $\bar{b}_{100}$& -2.382 $\cdot 10^{-6}$ & 1.300 $\cdot 10^{-6}$ & $\bar{d}_{100}$& 0.227 & 0.196 & $\bar{b}_{100}$&  -2.515 $\cdot 10^{-6}$ & 4.835 $\cdot 10^{-6}$  & $\bar{d}_{100}$& 0.230 & 0.310 \\
            &(0.003) &                                                   && (0.002) & &                        &(0.002) &             && (0.003)  &  \\ 
    $\phi$& 6.377$\cdot 10^{-6}$ & 7.933 $\cdot 10^{-7}$  & $\theta$  &  2.243 $\cdot 10^{-6}$ & 1.261 $\cdot 10^{-6}$&     $\phi$&8.739$\cdot 10^{-6}$ &1.481$\cdot 10^{-6}$ & $\theta$ & 7.457$\cdot 10^{-9}$ & 1.946$\cdot 10^{-9}$ \\
        &(0.072) &  & &(0.075)  & &  & (0.108)  &  & &(0.096)      \\ 
    $\gamma$& 1.213$\cdot 10^{-2}$ & 9.506$\cdot 10^{-3}$  & $\sigma_{0}$& 5.312 $\cdot 10^{-5}$ & 2.252 $\cdot 10^{-5}$ &    $\gamma$&1.195$\cdot 10^{-2}$ &1.549$\cdot 10^{-2}$& $\sigma_0$&  9.983$\cdot 10^{-5}$& 5.274$\cdot 10^{-5}$ \\
        &(0.003) & & &(0.002)&   &        &(0.000)  &&&(0.005)&   \\ 
    $\alpha$& 2.815$\cdot 10^{-2}$ & 5.654 $\cdot 10^{-2}$  & $b_{0,0}$  & 4.521$\cdot 10^{-6}$ & 1.490 $\cdot 10^{-6}$ &     $\alpha$&4.070$\cdot 10^{-2}$ & 2.535$\cdot 10^{-2}$  & $b_{0,0}$ & 1.465$\cdot 10^{-3}$& 1.213$\cdot 10^{-2}$  \\
        &(0.002) &  & &(0.004)  & &  & (0.003)  &  & &(0.001)      \\ 
    \hline
    \end{tabular}
 }
    \caption{\footnotesize Point estimates of the augmented SDAMHM model. Temporal daily averages of the coefficients (Est), standard deviations computed over trading days (Std), and average $p$-values in parentheses. Results reported for MSFT and AMZN.} 
    \label{tab:sdamhm_msft_amzn}
\end{table}
\begin{table}[htt]
\centering
\par\medskip
\renewcommand{\arraystretch}{1.15}
\resizebox{0.98\textwidth}{!}{
\begin{tabular}{lrrlrr@{\qquad}lrrlrr}
\hline
\multicolumn{6}{c}{\textbf{CLSK}} & \multicolumn{6}{c}{\textbf{FLGT}} \\ \hline
 & Est & Std & & Est & Std & & Est & Std & & Est & Std \\ \hline

$\mu_{1}$ & 1.170 $\cdot 10^{-5}$ & 2.786 $\cdot 10^{-5}$ & $\mu_{2}$ & -0.071 & 0.098 &
$\mu_{1}$ & -7.524 $\cdot 10^{-6}$ & 2.461 $\cdot 10^{-5}$ & $\mu_{2}$ & 0.008 & 0.122 \\
& (0.433) & & & (0.003) & & & (0.319) & & & (0.013) & \\

$a_1$ & -0.053 & 0.042 & $c_1$ & -542.855 & 165.152 &
$a_1$ & -0.030 & 0.046 & $c_1$ & -822.948 & 201.817 \\
& (0.028) & & & (0.000) & & & (0.004) & & & (0.000) & \\

$b_1$ & 1.879 $\cdot 10^{-4}$ & 2.798 $\cdot 10^{-5}$ & $d_1$ & 1.581 & 0.192 &
$b_1$ & 1.263 $\cdot 10^{-4}$ & 3.567 $\cdot 10^{-5}$ & $d_1$ & 1.635 & 0.158 \\
& (0.000) & & & (0.000) & & & (0.000) & & & (0.000) & \\

$\bar{b}_{10}$ & -3.054 $\cdot 10^{-5}$ & 2.657 $\cdot 10^{-5}$ & $\bar{d}_{10}$ & 0.190 & 0.161 &
$\bar{b}_{10}$ & 5.710 $\cdot 10^{-7}$ & 1.931 $\cdot 10^{-5}$ & $\bar{d}_{10}$ & 0.278 & 0.129 \\
& (0.006) & & & (0.002) & & & (0.005) & & & (0.008) & \\

$\bar{b}_{100}$ & -7.230 $\cdot 10^{-5}$ & 8.369 $\cdot 10^{-5}$ & $\bar{d}_{100}$ & 0.270 & 0.231 &
$\bar{b}_{100}$ & -2.166 $\cdot 10^{-5}$ & 7.954 $\cdot 10^{-5}$ & $\bar{d}_{100}$ & 0.195 & 0.264 \\
& (0.001) & & & (0.020) & & & (0.005) & & & (0.000) & \\

$\phi$ & -7.951 $\cdot 10^{-5}$ & 2.959 $\cdot 10^{-5}$ & $\theta$ & 6.557 $\cdot 10^{-6}$ & 5.327 $\cdot 10^{-6}$ &
$\phi$ & -2.073 $\cdot 10^{-5}$ & 1.838 $\cdot 10^{-5}$ & $\theta$ & 4.560 $\cdot 10^{-7}$ & 4.670 $\cdot 10^{-7}$ \\
& (0.550) & & & (0.626) & & & (0.290) & & & (0.730) & \\

$\gamma$ & 0.048 & 0.022 & $\sigma_0$ & 0.002 & 5.173 $\cdot 10^{-4}$ &
$\gamma$ & 0.042 & 0.025 & $\sigma_0$ & 9.397 $\cdot 10^{-4}$ & 4.945 $\cdot 10^{-4}$ \\
& (0.049) & & & (0.004) & & & (0.000) & & & (0.016) & \\

$\alpha$ & 0.004 & 0.011 & $b_{0,0}$ & 3.351 $\cdot 10^{-4}$ & 1.582 $\cdot 10^{-4}$ &
$\alpha$ & 0.001 & 0.004 & $b_{0,0}$ & 1.642 $\cdot 10^{-4}$ & 8.604 $\cdot 10^{-5}$ \\
& (0.005) & & & (0.047) & & & (0.000) & & & (0.024) & \\

\hline
\end{tabular}
}
\caption{\footnotesize Point estimates of the augmented SDAMHM model. Temporal daily averages of the coefficients (Est), standard deviations computed over trading days (Std), and average $p$-values in parentheses. Results reported for CLSK and FLGT.}
\label{tab:sdamhm_clsk_flgt}
\end{table}
\begin{table}[htt]
\centering
\par\medskip
\renewcommand{\arraystretch}{1.15}
\resizebox{0.98\textwidth}{!}{
\begin{tabular}{lrrlrr@{\qquad}lrrlrr}
\hline
\multicolumn{6}{c}{\textbf{AAPL}} & \multicolumn{6}{c}{\textbf{INTC}} \\ \hline
 & Est & Std & & Est & Std & & Est & Std & & Est & Std \\ \hline

$\mu_{1}$ & 3.375 $\cdot 10^{-7}$ & 5.507 $\cdot 10^{-7}$ & $\mu_{2}$ & -0.017 & 0.032 &
$\mu_{1}$ & 3.238 $\cdot 10^{-7}$ & 2.204 $\cdot 10^{-6}$ & $\mu_{2}$ & -0.015 & 0.055 \\
& (0.681) & & & (0.000) & & & (0.060) & & & (0.037) & \\

$a_1$ & -0.038 & 0.019 & $c_1$ & -1.579 $\cdot 10^{4}$ & 6.059 $\cdot 10^{4}$ &
$a_1$ & -0.010 & 0.008 & $c_1$ & -1.289 $\cdot 10^{4}$ & 3.670 $\cdot 10^{4}$ \\
& (0.000) & & & (0.000) & & & (0.050) & & & (0.000) & \\

$b_1$ & 1.396 $\cdot 10^{-5}$ & 1.144 $\cdot 10^{-6}$ & $d_1$ & 1.691 & 0.260 &
$b_1$ & 2.932 $\cdot 10^{-5}$ & 2.966 $\cdot 10^{-6}$ & $d_1$ & 2.209 & 0.226 \\
& (0.000) & & & (0.000) & & & (0.000) & & & (0.000) & \\

$\bar{b}_{10}$ & 6.185 $\cdot 10^{-7}$ & 1.145 $\cdot 10^{-7}$ & $\bar{d}_{10}$ & 0.329 & 0.086 &
$\bar{b}_{10}$ & 1.033 $\cdot 10^{-5}$ & 2.129 $\cdot 10^{-6}$ & $\bar{d}_{10}$ & 0.386 & 0.136 \\
& (0.000) & & & (0.000) & & & (0.002) & & & (0.000) & \\

$\bar{b}_{100}$ & -4.020 $\cdot 10^{-6}$ & 2.023 $\cdot 10^{-6}$ & $\bar{d}_{100}$ & 0.233 & 0.100 &
$\bar{b}_{100}$ & 6.993 $\cdot 10^{-7}$ & 3.391 $\cdot 10^{-6}$ & $\bar{d}_{100}$ & 0.058 & 0.079 \\
& (0.233) & & & (0.000) & & & (0.647) & & & (0.000) & \\

$\phi$ & -1.283 $\cdot 10^{-5}$ & 2.765 $\cdot 10^{-6}$ & $\theta$ & -5.477 $\cdot 10^{-6}$ & 3.138 $\cdot 10^{-5}$ &
$\phi$ & 4.004 $\cdot 10^{-5}$ & 5.376 $\cdot 10^{-6}$ & $\theta$ & 8.109 $\cdot 10^{-7}$ & 9.536 $\cdot 10^{-6}$ \\
& (0.428) & & & (0.765) & & & (0.345) & & & (0.764) & \\

$\gamma$ & 0.018 & 0.016 & $\sigma_0$ & 9.903 $\cdot 10^{-5}$ & 2.325 $\cdot 10^{-5}$ &
$\gamma$ & 0.027 & 0.015 & $\sigma_0$ & 1.011 $\cdot 10^{-4}$ & 1.451 $\cdot 10^{-4}$ \\
& (0.237) & & & (0.006) & & & (0.098) & & & (0.008) & \\

$\alpha$ & 0.023 & 0.018 & $b_{0,0}$ & 2.121 $\cdot 10^{-6}$ & 6.334 $\cdot 10^{-5}$ &
$\alpha$ & 0.045 & 0.025 & $b_{0,0}$ & -6.355 $\cdot 10^{-6}$ & 5.137 $\cdot 10^{-5}$ \\
& (0.011) & & & (0.000) & & & (0.001) & & & (0.000) & \\

\hline
\end{tabular}
}
\caption{\footnotesize Point estimates of the augmented SDAMHM model. Temporal daily averages of the coefficients (Est), standard deviations computed over trading days (Std), and average $p$-values in parentheses. Results reported for AAPL and INTC.}
\label{tab:sdamhm_aapl_intc}
\end{table}

\begin{table}[htt]
\centering
\par\medskip
\renewcommand{\arraystretch}{1.15}
\resizebox{0.98\textwidth}{!}{
\begin{tabular}{lrrlrr@{\qquad}lrrlrr}
\hline
\multicolumn{6}{c}{\textbf{GOOG}} & \multicolumn{6}{c}{\textbf{BNFT}} \\ \hline
 & Est & Std & & Est & Std & & Est & Std & & Est & Std \\ \hline

$\mu_{1}$ & 2.025 $\cdot 10^{-7}$ & 2.144 $\cdot 10^{-6}$ & $\mu_{2}$ & 0.006 & 0.067 &
$\mu_{1}$ & -9.975 $\cdot 10^{-6}$ & 3.392 $\cdot 10^{-5}$ & $\mu_{2}$ & -0.037 & 0.199 \\
& (0.803) & & & (0.017) & & & (0.202) & & & (0.038) & \\

$a_1$ & -0.049 & 0.024 & $c_1$ & -4.976 $\cdot 10^{3}$ & 1.652 $\cdot 10^{3}$ &
$a_1$ & -0.020 & 0.014 & $c_1$ & -1.128 $\cdot 10^{3}$ & 628.435 \\
& (0.016) & & & (0.000) & & & (0.008) & & & (0.000) & \\

$b_1$ & 1.143 $\cdot 10^{-5}$ & 8.734 $\cdot 10^{-6}$ & $d_1$ & 1.808 & 0.287 &
$b_1$ & 1.364 $\cdot 10^{-4}$ & 2.353 $\cdot 10^{-5}$ & $d_1$ & 1.709 & 0.202 \\
& (0.009) & & & (0.000) & & & (0.000) & & & (0.000) & \\

$\bar{b}_{10}$ & -9.526 $\cdot 10^{-7}$ & 3.813 $\cdot 10^{-6}$ & $\bar{d}_{10}$ & 0.354 & 0.135 &
$\bar{b}_{10}$ & 2.592 $\cdot 10^{-5}$ & 2.627 $\cdot 10^{-5}$ & $\bar{d}_{10}$ & 0.284 & 0.169 \\
& (0.006) & & & (0.000) & & & (0.006) & & & (0.002) & \\

$\bar{b}_{100}$ & -1.296 $\cdot 10^{-6}$ & 4.474 $\cdot 10^{-6}$ & $\bar{d}_{100}$ & 0.065 & 0.132 &
$\bar{b}_{100}$ & -3.218 $\cdot 10^{-5}$ & 1.082 $\cdot 10^{-4}$ & $\bar{d}_{100}$ & 0.020 & 0.042 \\
& (0.001) & & & (0.011) & & & (0.008) & & & (0.003) & \\

$\phi$ & 2.146 $\cdot 10^{-6}$ & 4.846 $\cdot 10^{-6}$ & $\theta$ & 4.204 $\cdot 10^{-9}$ & 4.265 $\cdot 10^{-9}$ &
$\phi$ & -1.288 $\cdot 10^{-4}$ & 3.398 $\cdot 10^{-5}$ & $\theta$ & 7.600 $\cdot 10^{-6}$ & 1.078 $\cdot 10^{-5}$ \\
& (0.625) & & & (0.863) & & & (0.776) & & & (0.397) & \\

$\gamma$ & 0.015 & 0.015 & $\sigma_0$ & 1.623 $\cdot 10^{-4}$ & 2.290 $\cdot 10^{-5}$ &
$\gamma$ & 0.023 & 0.050 & $\sigma_0$ & 5.033 $\cdot 10^{-4}$ & 1.303 $\cdot 10^{-4}$ \\
& (0.564) & & & (0.022) & & & (0.511) & & & (0.003) & \\

$\alpha$ & 0.002 & 0.003 & $b_{0,0}$ & 1.319 $\cdot 10^{-5}$ & 1.777 $\cdot 10^{-5}$ &
$\alpha$ & 0.018 & 0.008 & $b_{0,0}$ & 1.202 $\cdot 10^{-4}$ & 7.441 $\cdot 10^{-5}$ \\
& (0.023) & & & (0.000) & & & (0.030) & & & (0.010) & \\

\hline
\end{tabular}
}
\caption{\footnotesize Point estimates of the augmented SDAMHM model. Temporal daily averages of the coefficients (Est), standard deviations computed over trading days (Std), and average $p$-values in parentheses. Results reported for GOOG and BNFT.}
\label{tab:sdamhm_goog_bnft}
\end{table}

\begin{table}[htt]
\centering
\par\medskip
\renewcommand{\arraystretch}{1.15}
\resizebox{0.98\textwidth}{!}{
\begin{tabular}{lrrlrr@{\qquad}lrrlrr}
\hline
\multicolumn{6}{c}{\textbf{EXTR}} & \multicolumn{6}{c}{\textbf{MO$^{*}$}}\\ 
\hline
 & Est & Std & & Est & Std & & Est & Std & & Est & Std \\ \hline

$\mu_{1}$ & 1.813 $\cdot 10^{-5}$ & 2.982 $\cdot 10^{-5}$ & $\mu_{2}$ & -0.095 & 0.210 &
$\mu_{1}$ & -3.241 $\cdot 10^{-7}$ & 2.400 $\cdot 10^{-6}$ & $\mu_{2}$ & -0.012 & 0.054 \\
& (0.289) & & & (0.017) & & & (0.154) & & & (0.045) & \\

$a_1$ & -0.110 & 0.042 & $c_1$ & -1.874 $\cdot 10^{3}$ & 742.218 &
$a_1$ & -0.069 & 0.023 & $c_1$ & -9.022 $\cdot 10^{3}$ & 2.862 $\cdot 10^{3}$ \\
& (0.003) & & & (0.000) & & & (0.000) & & & (0.000) & \\

$b_1$ & 1.842 $\cdot 10^{-4}$ & 2.960 $\cdot 10^{-5}$ & $d_1$ & 1.968 & 0.197 &
$b_1$ & 4.534 $\cdot 10^{-5}$ & 3.767 $\cdot 10^{-6}$ & $d_1$ & 1.983 & 0.212 \\
& (0.000) & & & (0.000) & & & (0.000) & & & (0.000) & \\

$\bar{b}_{10}$ & 4.064 $\cdot 10^{-5}$ & 1.790 $\cdot 10^{-5}$ & $\bar{d}_{10}$ & 0.347 & 0.194 &
$\bar{b}_{10}$ & 9.485 $\cdot 10^{-6}$ & 2.647 $\cdot 10^{-6}$ & $\bar{d}_{10}$ & 0.230 & 0.107 \\
& (0.005) & & & (0.010) & & & (0.001) & & & (0.004) & \\

$\bar{b}_{100}$ & -1.385 $\cdot 10^{-5}$ & 5.472 $\cdot 10^{-5}$ & $\bar{d}_{100}$ & 0.113 & 0.322 &
$\bar{b}_{100}$ & -6.740 $\cdot 10^{-6}$ & 7.523 $\cdot 10^{-6}$ & $\bar{d}_{100}$ & 0.092 & 0.213 \\
& (0.009) & & & (0.009) & & & (0.004) & & & (0.002) & \\

$\phi$ & -2.100 $\cdot 10^{-4}$ & 2.945 $\cdot 10^{-5}$ & $\theta$ & 5.923 $\cdot 10^{-6}$ & 5.292 $\cdot 10^{-6}$ &
$\phi$ & -5.349 $\cdot 10^{-5}$ & 7.190 $\cdot 10^{-6}$ & $\theta$ & 2.046 $\cdot 10^{-7}$ & 1.602 $\cdot 10^{-7}$ \\
& (0.578) & & & (0.662) & & & (0.650) & & & (0.979) & \\

$\gamma$ & 0.033 & 0.018 & $\sigma_0$ & 4.367 $\cdot 10^{-4}$ & 2.416 $\cdot 10^{-4}$ &
$\gamma$ & 0.021 & 0.009 & $\sigma_0$ & 1.125 $\cdot 10^{-4}$ & 1.254 $\cdot 10^{-4}$ \\
& (0.432) & & & (0.003) & & & (0.270) & & & (0.003) & \\

$\alpha$ & 0.003 & 0.017 & $b_{0,0}$ & 7.555 $\cdot 10^{-5}$ & 5.983 $\cdot 10^{-5}$ &
$\alpha$ & 0.014 & 0.015 & $b_{0,0}$ & 2.886 $\cdot 10^{-5}$ & 1.646 $\cdot 10^{-5}$ \\
& (0.013) & & & (0.008) & & & (0.011) & & & (0.008) & \\

\hline
\end{tabular}
}
\caption{\footnotesize 
Point estimates of the augmented SDAMHM model. 
Temporal daily averages of the coefficients (Est), standard deviations computed over trading days (Std), and average $p$-values in parentheses. Results reported for EXTR and MO$^{*}$.}
\label{tab:sdamhm_extr_mo}
\end{table}

\subsection{Time-varying impact and volatility}
	\label{section:otherticksimpvol}

\begin{figure}[htt]
	\centering
         \textbf{All sample time-varying estimates of the instantaneous impact}\par\medskip
	\subfigure[June, 2021 - AAPL]
	{\includegraphics[width=7cm]{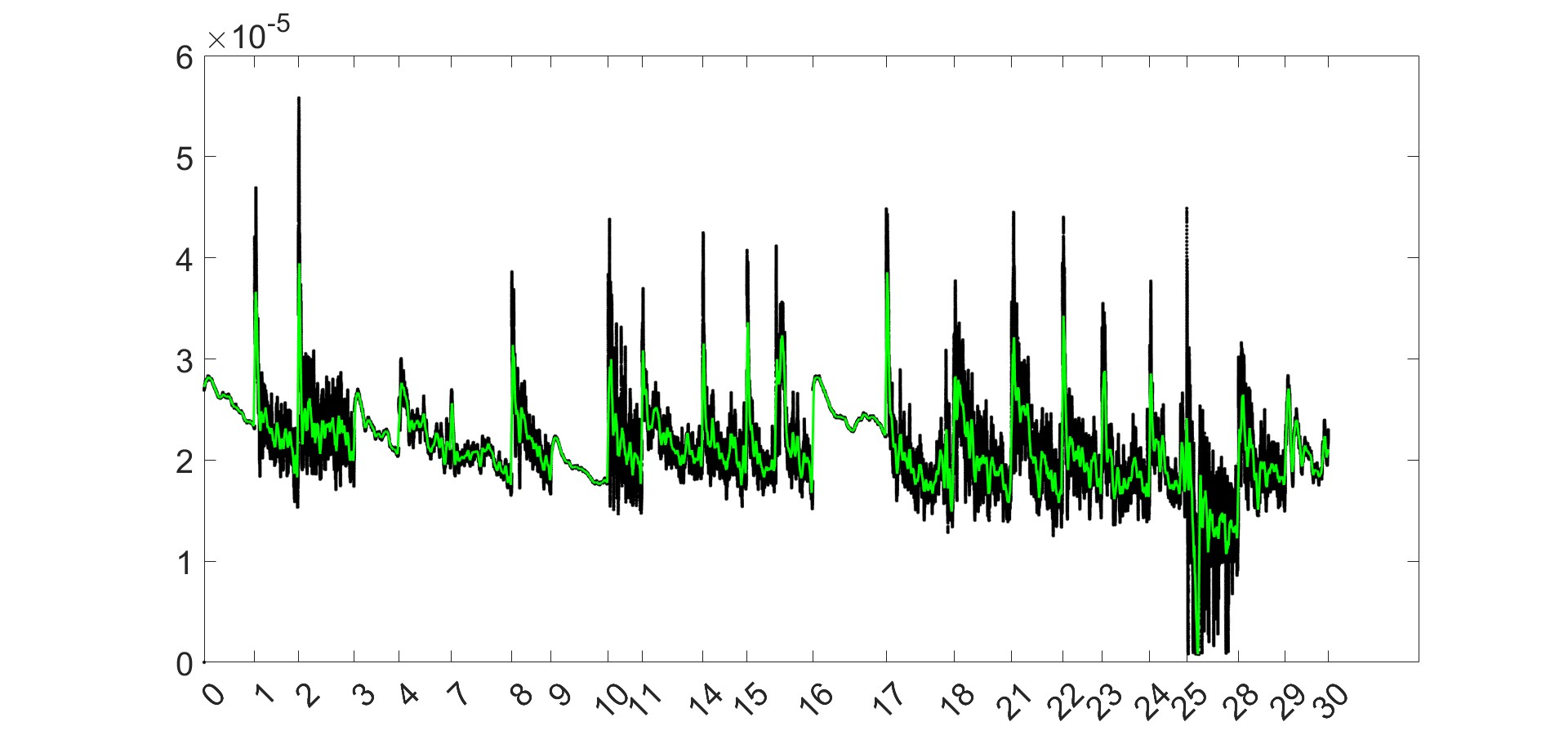}} 
        \hspace{1mm}
	\subfigure[June, 2021 - BNFT]
	{\includegraphics[width=7cm]{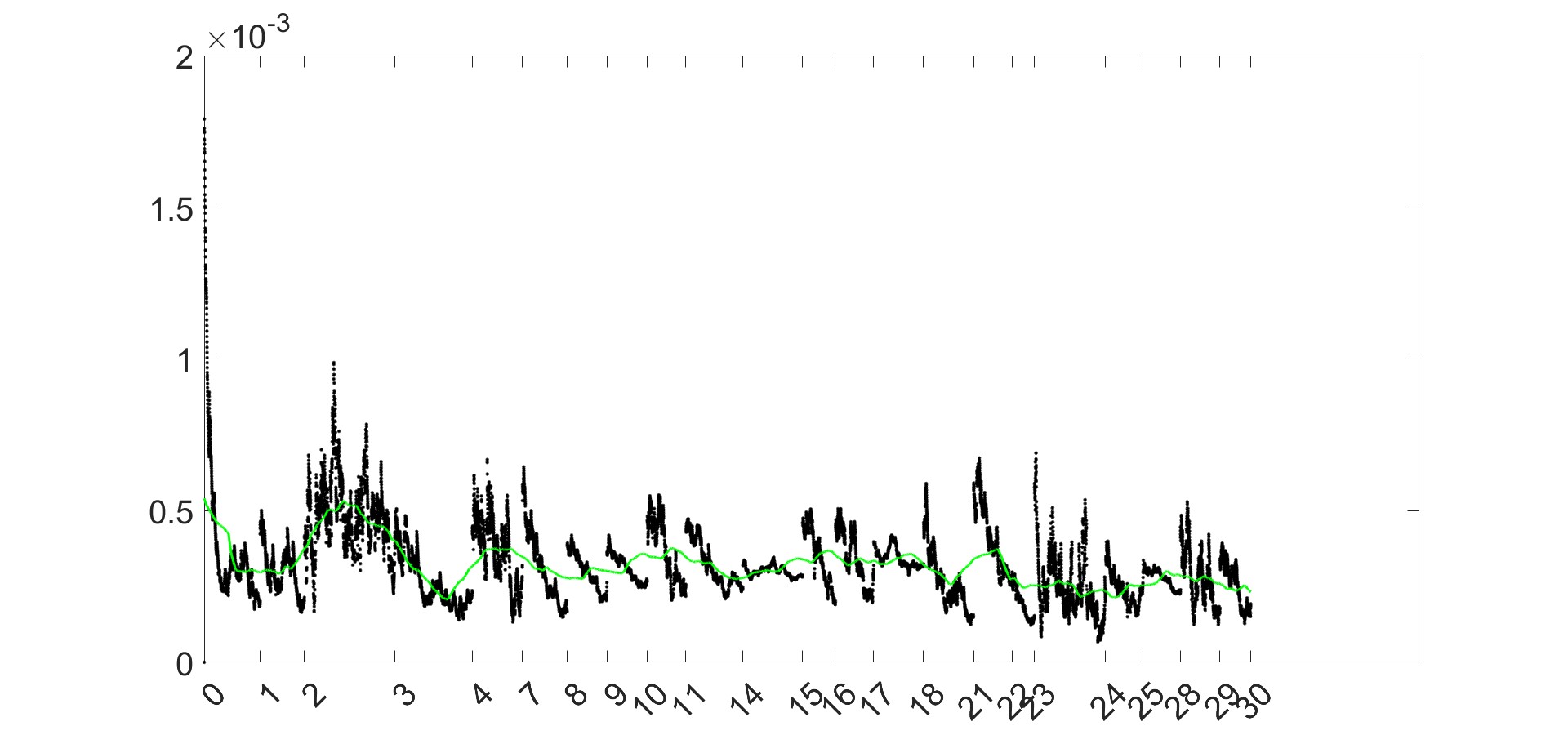}}
         \hspace{1mm}
	\subfigure[June, 2021 - GOOG]
	{\includegraphics[width=7cm]{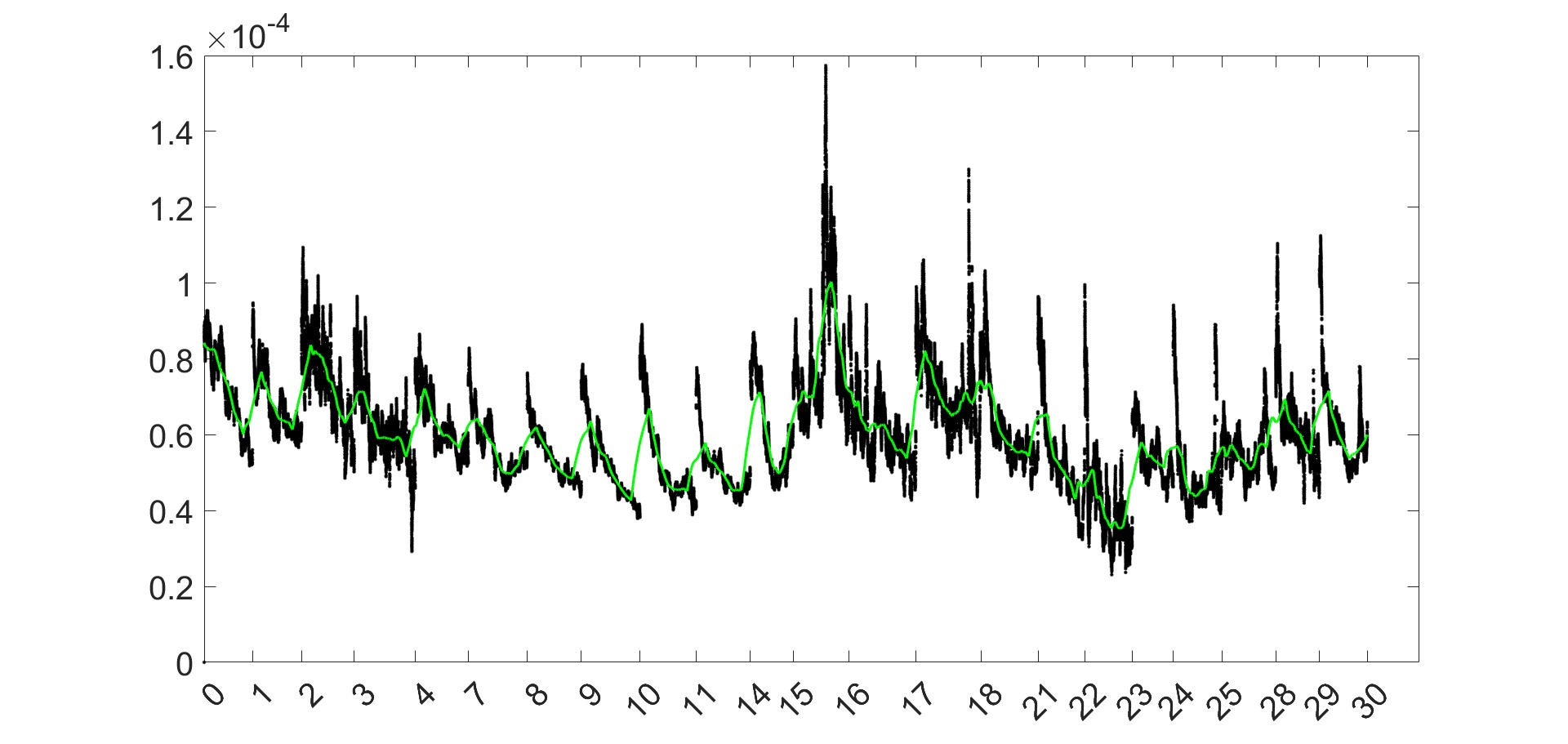}}
         \hspace{1mm}
	\subfigure[June, 2021 - EXTR]
	{\includegraphics[width=7cm]{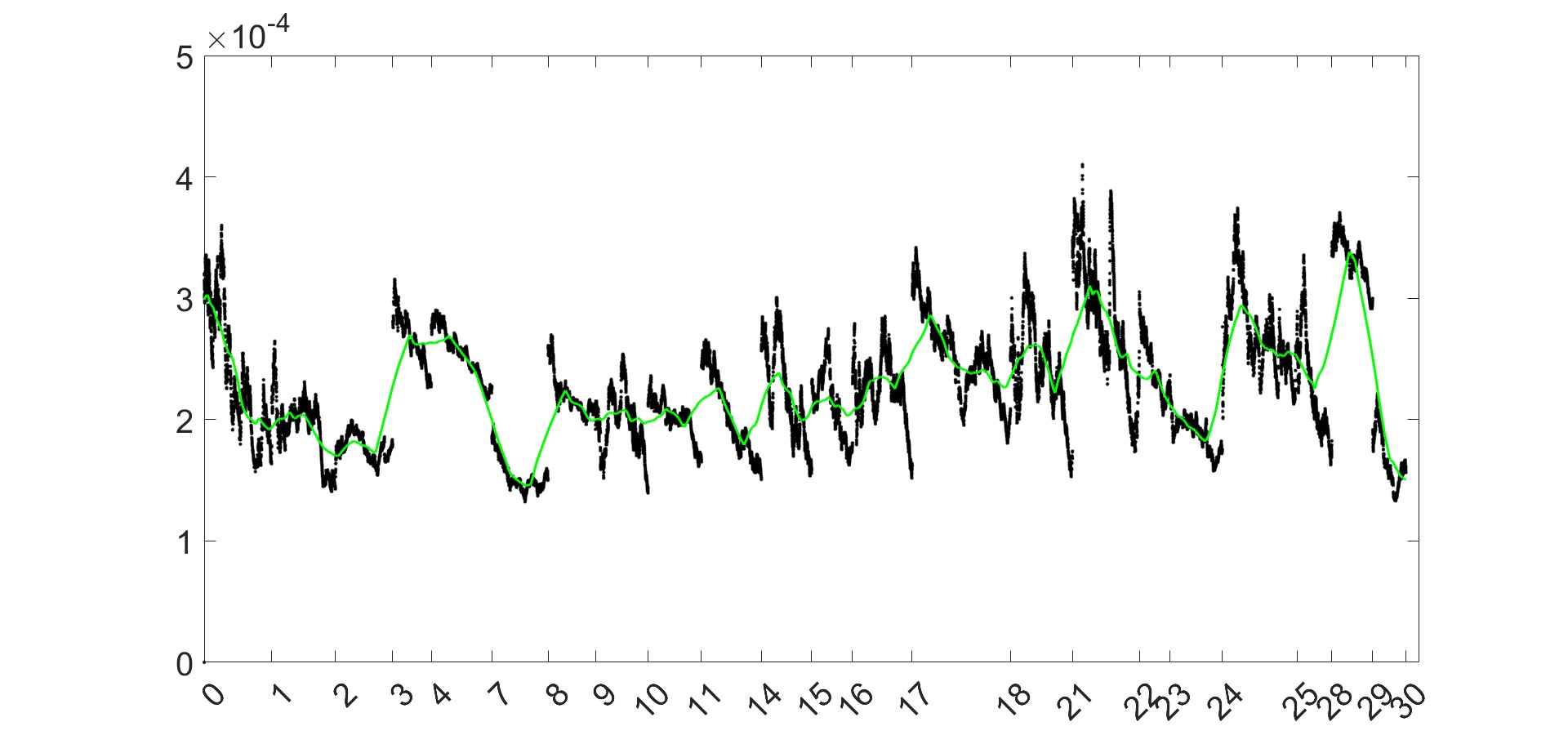}}
          \hspace{1mm}
 	\subfigure[June, 2021 - INTC]
	{\includegraphics[width=7cm]{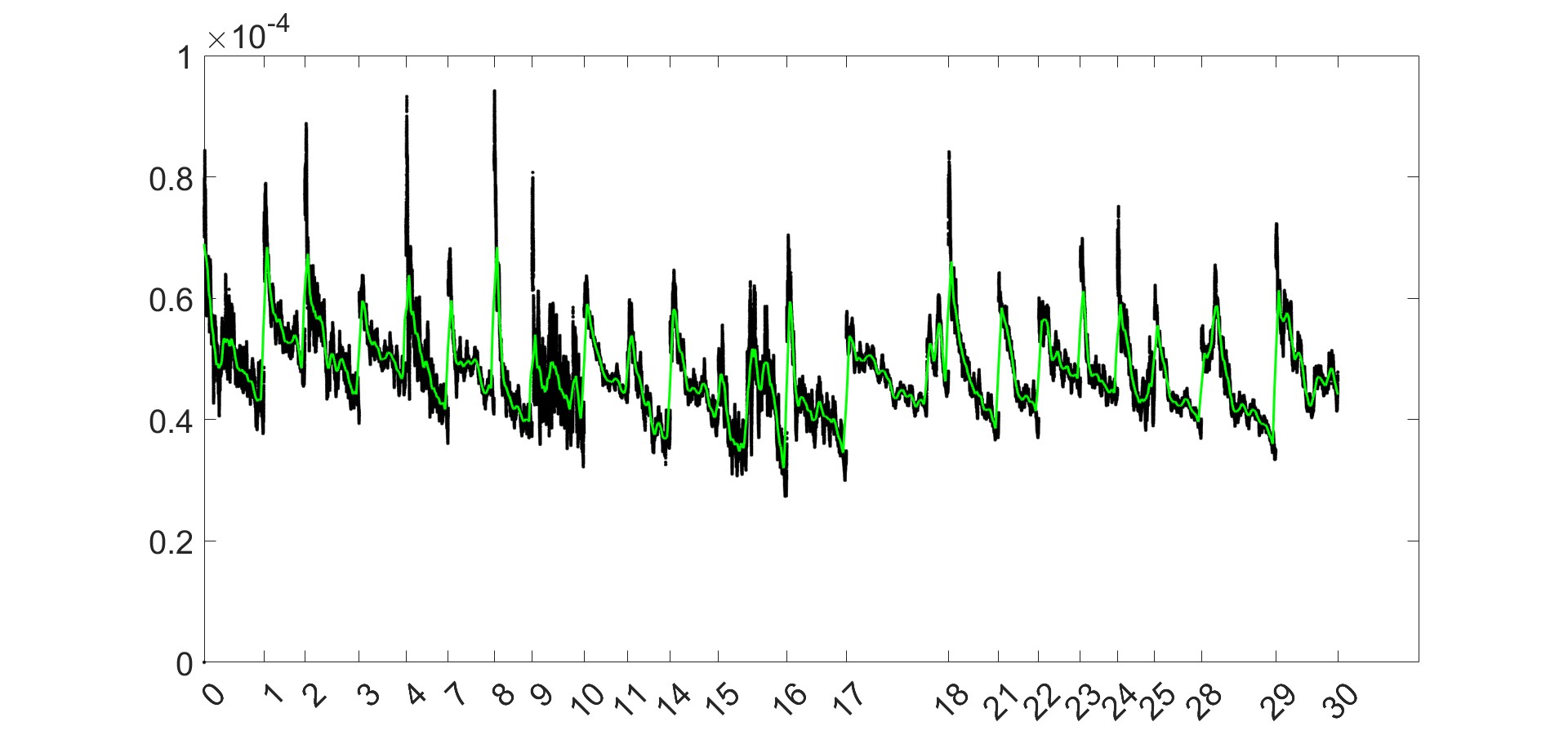}}
         \hspace{1mm}
	\subfigure[June, 2021 - MO$^*$]
	{\includegraphics[width=7cm]{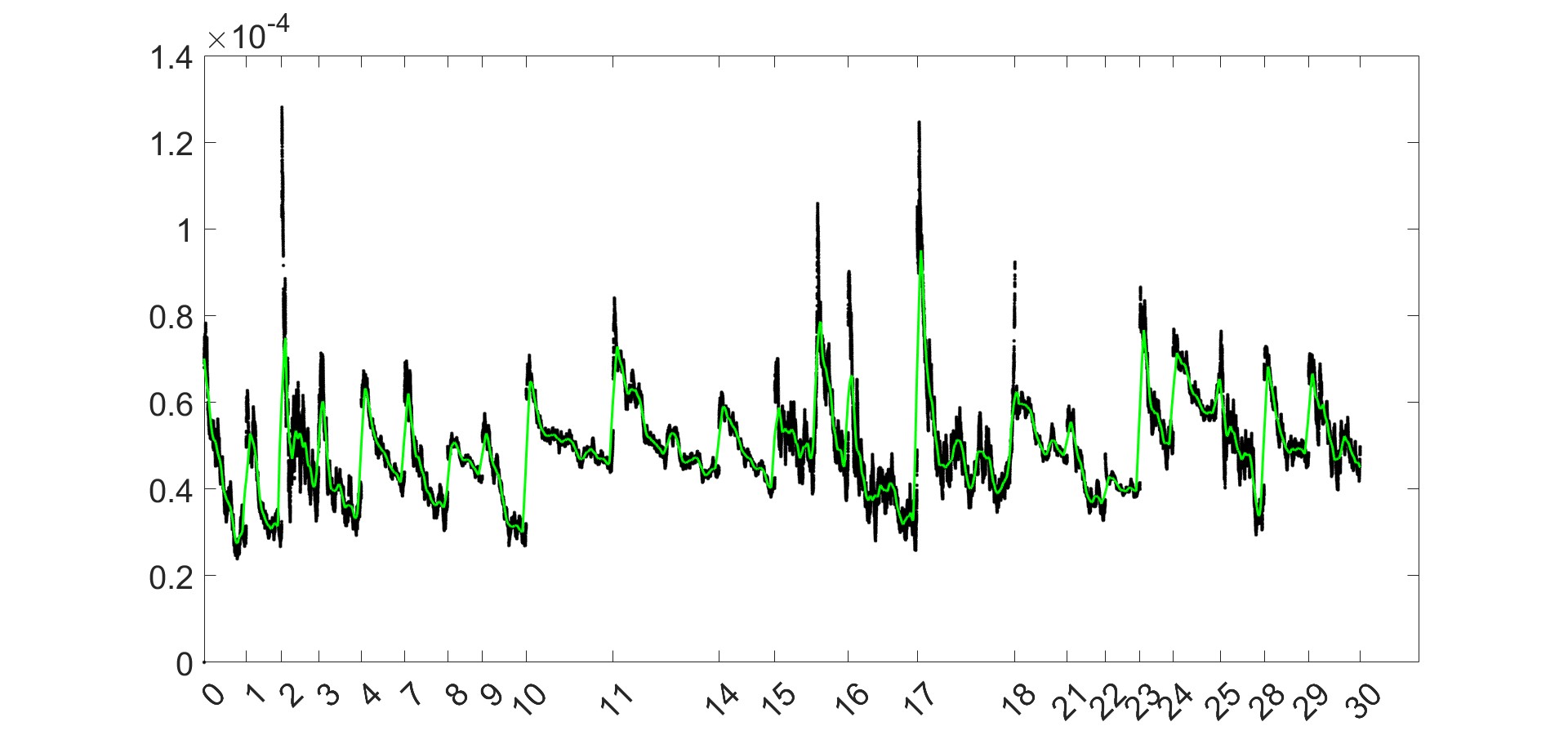}}
	\caption{\footnotesize Estimates of the time-varying instantaneous impact parameter of the SDAMH model (black lines) and its moving average (green solid lines). Overview of June, 2021 trading month.}
	\label{b0tv2}
\end{figure}

\begin{figure}[htt]
	\centering
         \textbf{All sample time-varying estimates of the instantaneous impact}\par\medskip
	\subfigure[June, 2021 - AAPL]
	{\includegraphics[width=7cm]{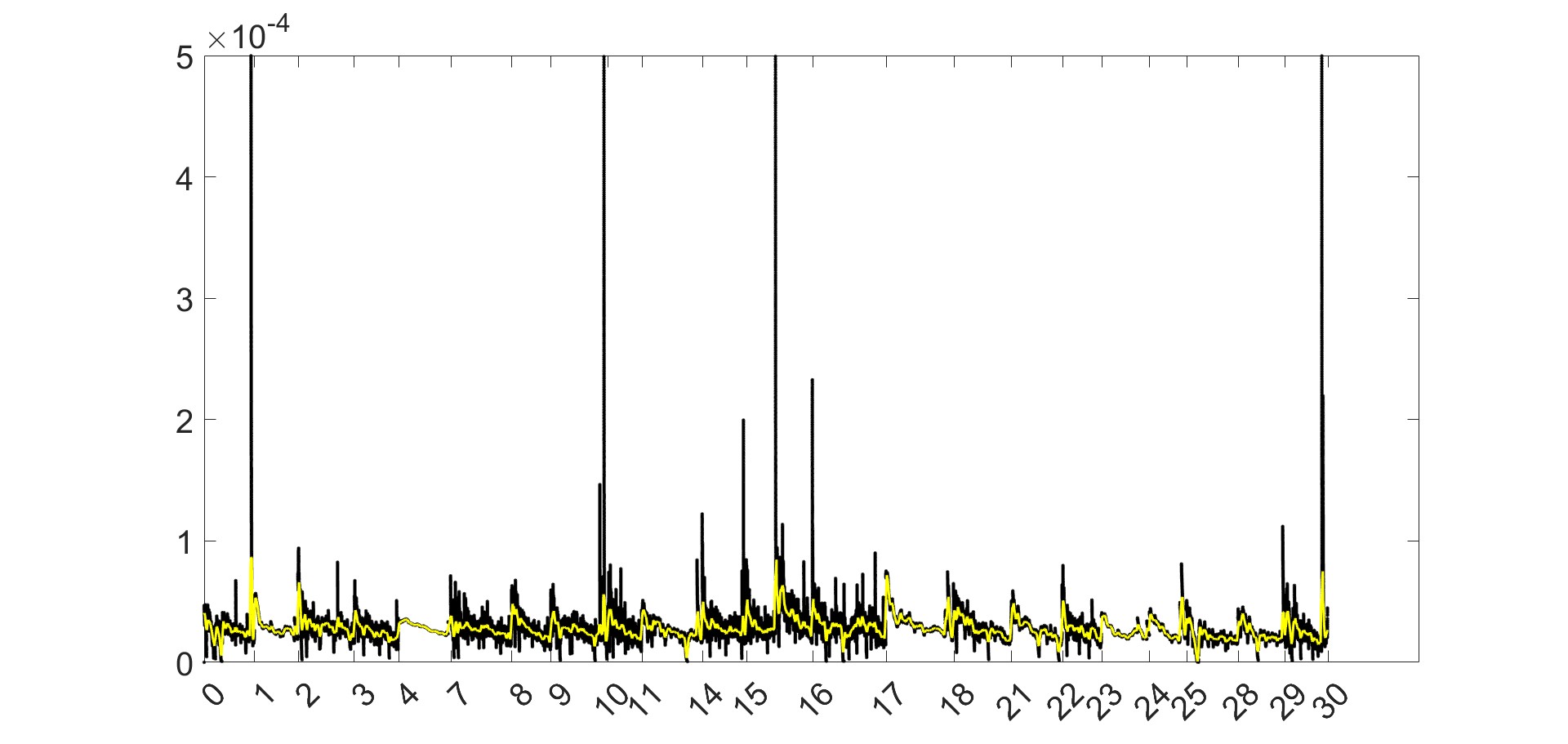}} 
        \hspace{1mm}
	\subfigure[June, 2021 - BNFT]
	{\includegraphics[width=7cm]{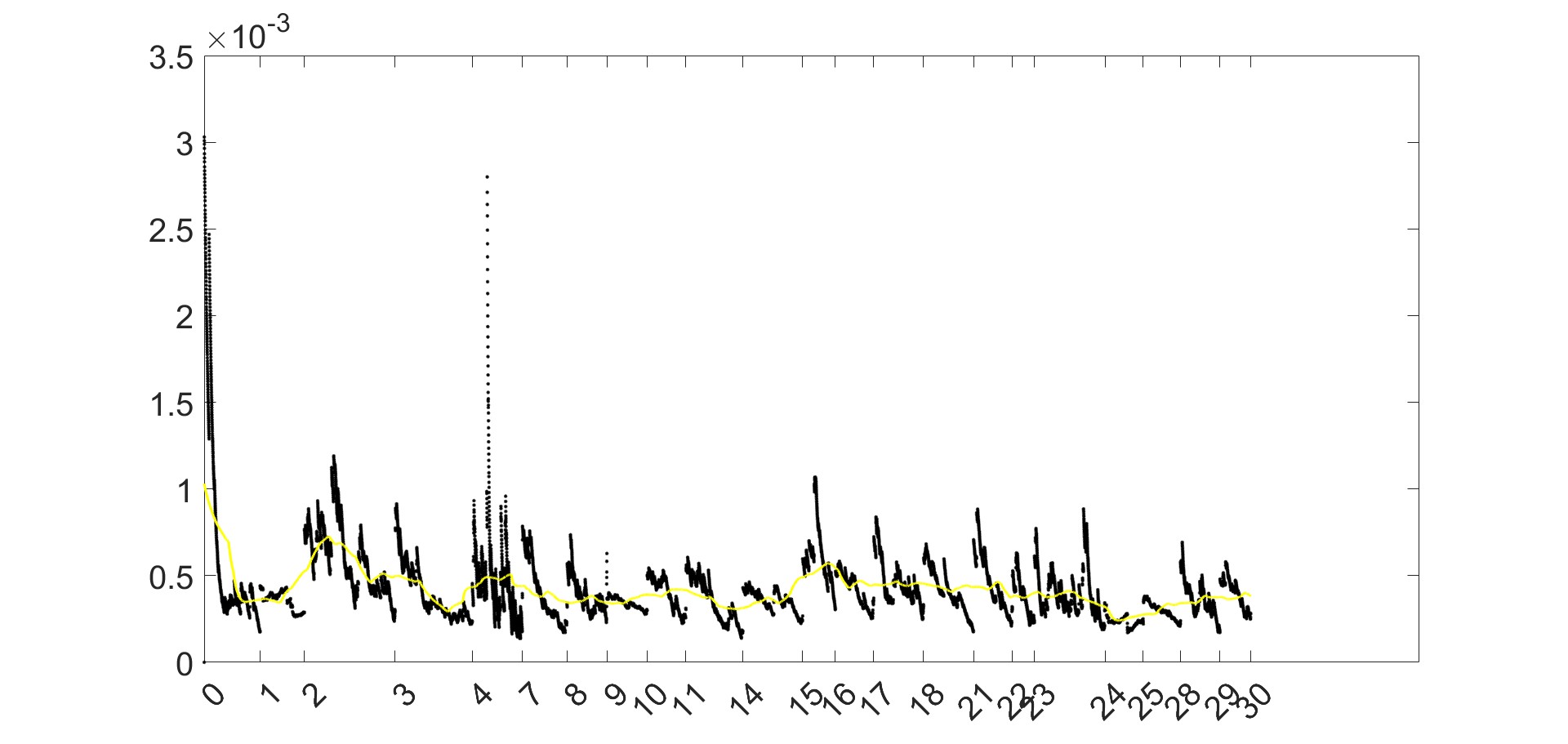}}
         \hspace{1mm}
	\subfigure[June, 2021 - GOOG]
	{\includegraphics[width=7cm]{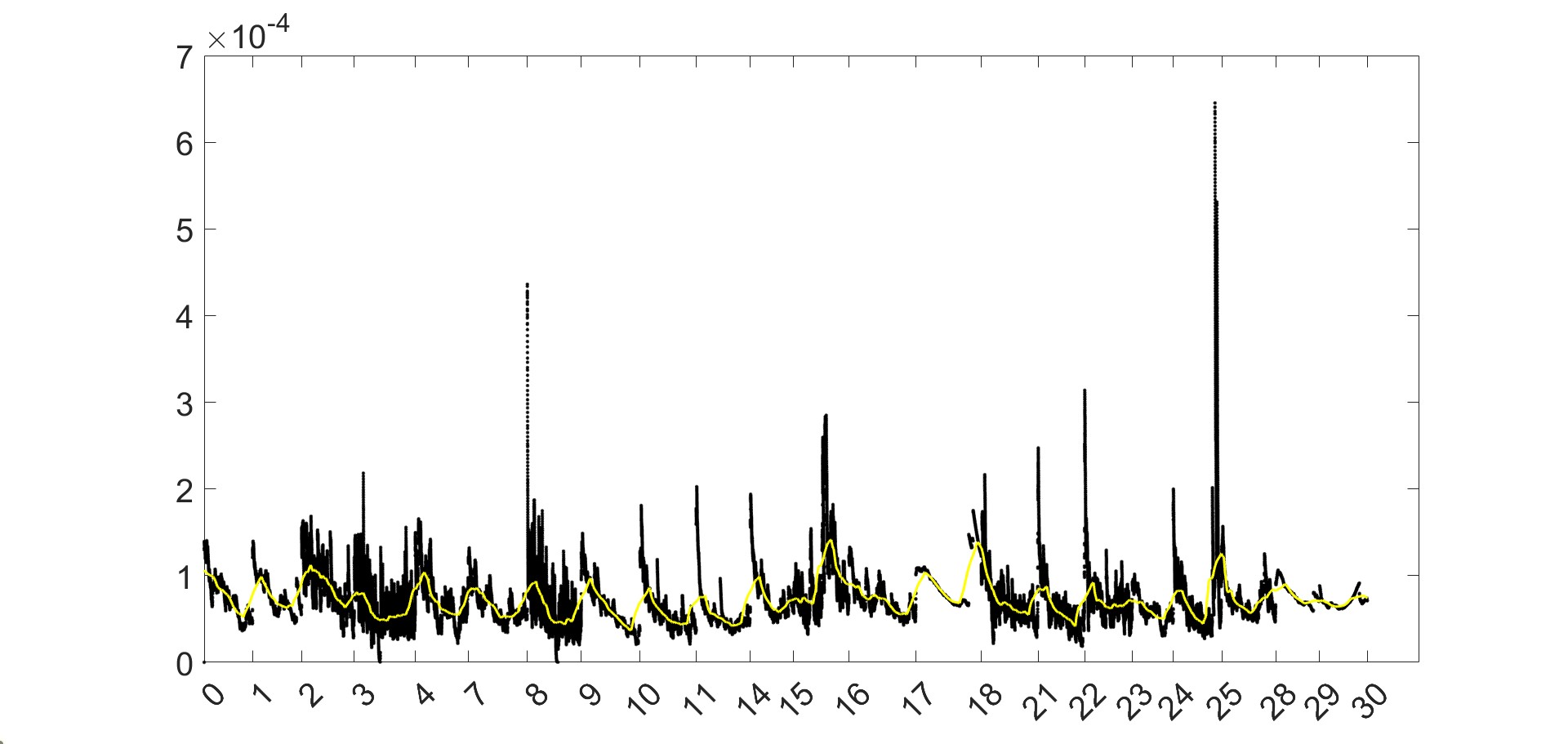}}
         \hspace{1mm}
	\subfigure[June, 2021 - EXTR]
	{\includegraphics[width=7cm]{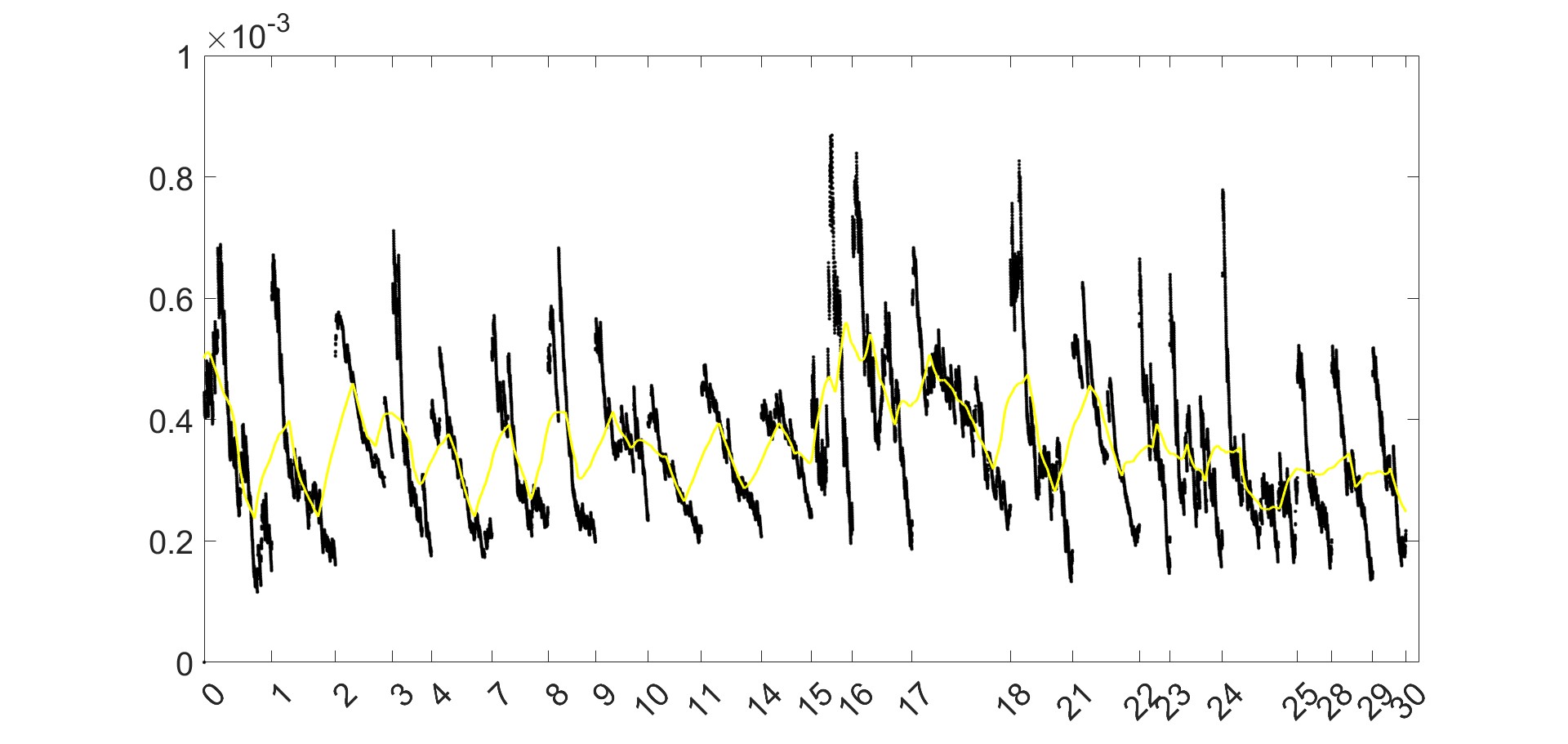}}
          \hspace{1mm}
 	\subfigure[June, 2021 - INTC]
	{\includegraphics[width=7cm]{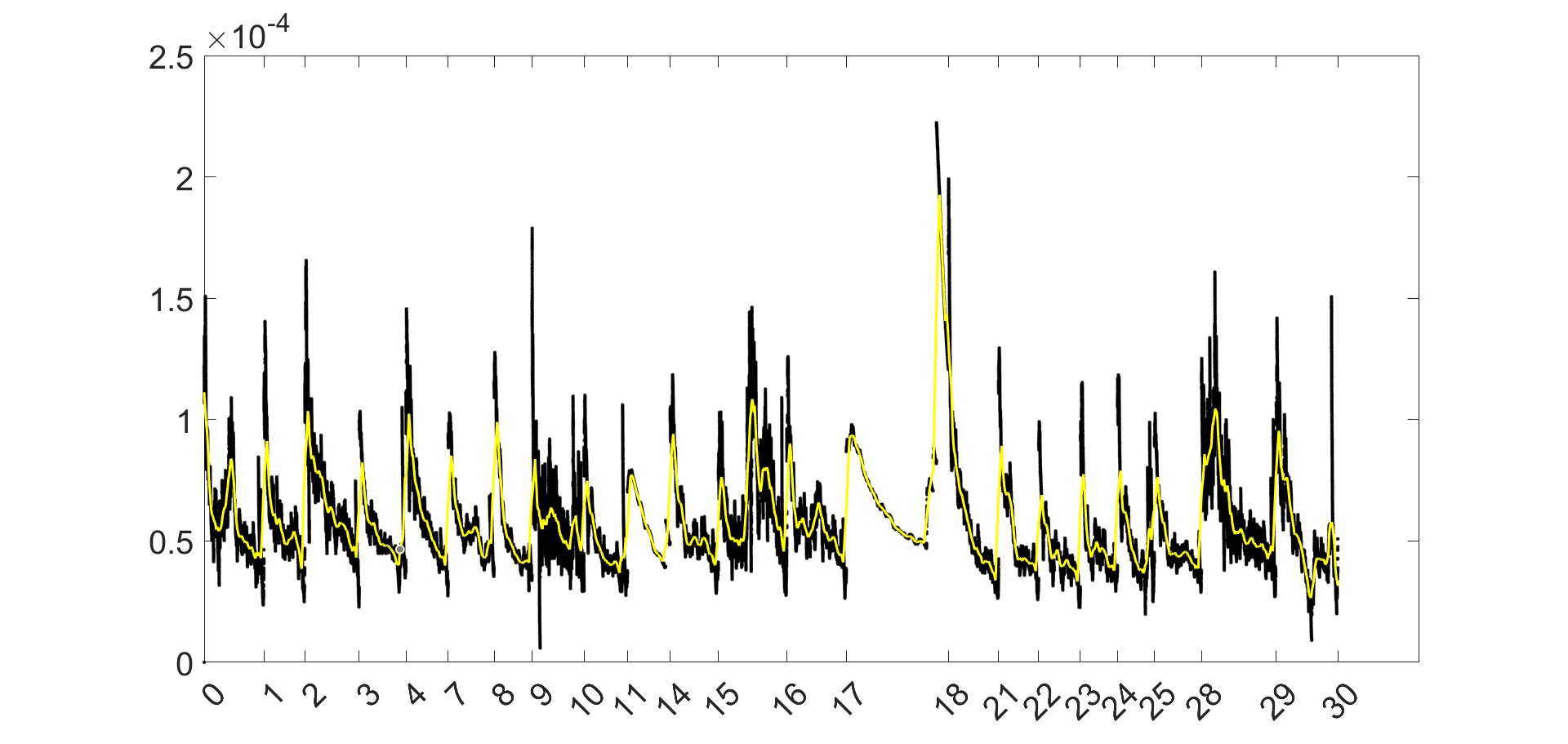}}
         \hspace{1mm}
	\subfigure[June, 2021 - MO$^*$]
	{\includegraphics[width=7cm]{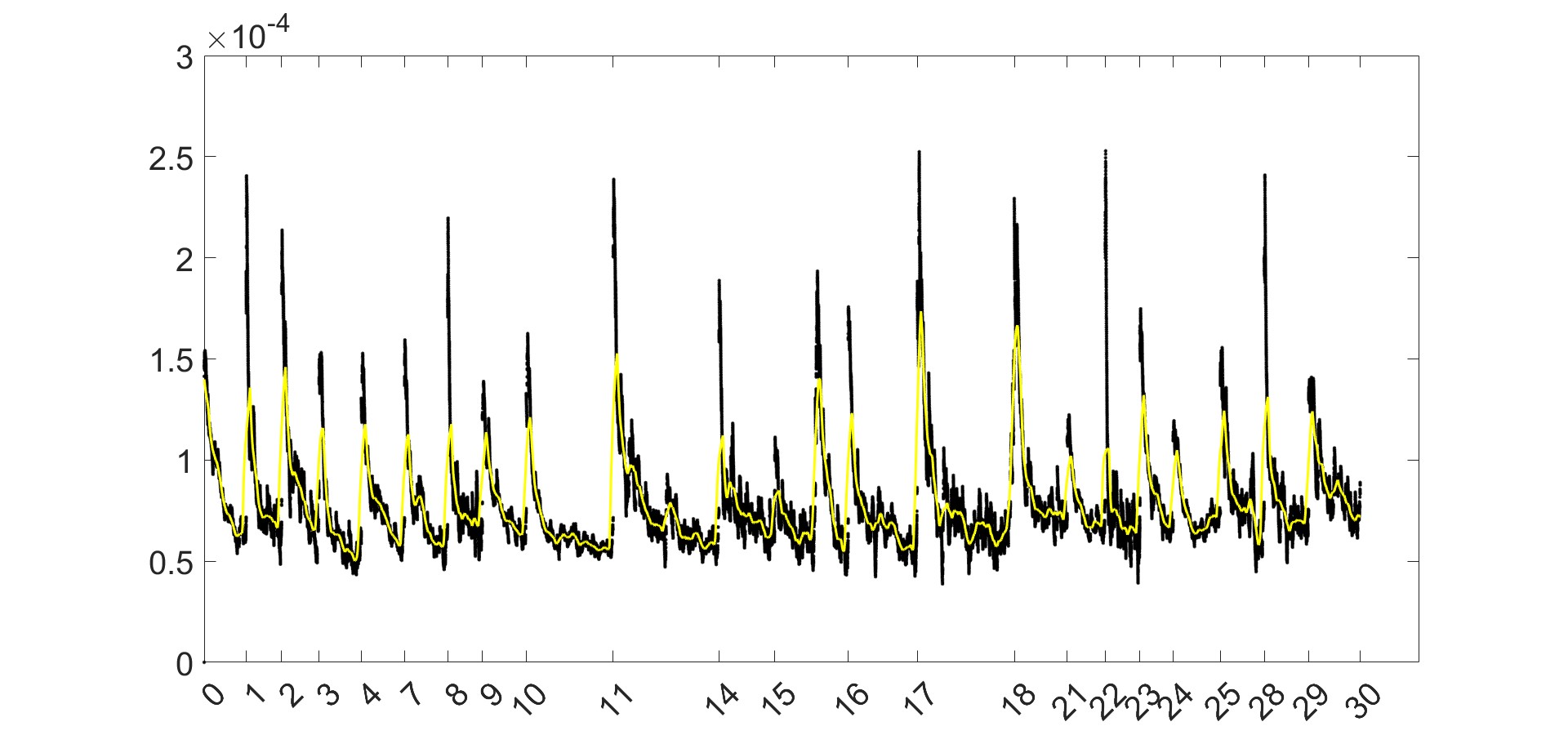}}
	\caption{\footnotesize Estimates of the time-varying volatility parameter of the SDAMH model (black lines) and its moving average (yellow solid lines). Overview of June, 2021 trading month.}
	\label{b0tv3}
\end{figure}

In Figure~\ref{b0tv2}, we observe the filtered instantaneous impact parameter for June, 2021 across a range of low -- BNFT, EXTR, and MO* -- and high -- AAPL, GOOG, and INTC -- liquidity stocks. The black lines trace the instantaneous impact parameter over time, while the green lines represent a moving average that helps smooth out short-term fluctuations, allowing us to better observe the underlying trend. What is immediately clear across all panels is the persistent intra-day pattern, which mirrors the dynamics detailed in the main section of the paper. For the more liquid stocks -- AAPL, GOOG, and INTC -- the instantaneous impact exhibits a pattern that is consistent with their high liquidity. These stocks display a relatively stable mean level, punctuated by brief spikes that correspond to periods of more market activity. These spikes are most often observed around key market times, such as not only the opening sessions but also during the day, as it may happen during the FOMC announcement. Even for these highly liquid assets, occasional structural breaks can be seen—most notably on days of significant market events, where the instantaneous impact deviates sharply from its long-term average (see FOMC announcement). In contrast, the less liquid stocks -- BNFT, EXTR, and MO$^*$ -- demonstrate similar volatile behavior. The instantaneous impact for these stocks, while still showing an intra-day pattern, fluctuates during the days, with frequent and sometimes large departures from the mean. This aligns with the characteristics discussed in the main section, where less liquid markets tend to react more sensitively to individual trades.  Moving on to Figure~\ref{b0tv3}, which depicts the time-varying volatility parameter over the same period and for the same set of stocks, we observe many similarities in the underlying dynamics between volatility and the instantaneous impact. The black lines here represent the filtered volatility parameter, while the yellow lines trace its moving average. Similar to the impact parameter, all stocks exhibit an intra-day volatility pattern, with a general decline throughout the trading day, interrupted by spikes during periods of increased market activity. For the more liquid stocks -- AAPL, GOOG, and INTC -- the volatility show several pronounced spikes while the pattern confirms the findings in the main part of the paper. The less liquid stocks display a larger volatility pattern. All these stocks show frequent and sharp spikes. The patterns observed here are in line with the analysis in the main section of the paper, where lower liquidity exacerbates the market's response to trades, leading to more volatile price movements and a higher probability of significant deviations from the mean level. Both figures suggest a strong relation between the intra-day patterns of the instantaneous impact and volatility. Moreover, the spikes in volatility often mirror those in the instantaneous impact, suggesting that increased volatility may be both a cause and a consequence of sharp movements in the impact parameter. This raises again interesting questions about the interplay between these two time-varying quantities -- whether they are jointly driven by a common latent factor, or whether one drives the other. While this remains an open question, the consistency of these patterns across stocks with different liquidity levels underscores the robustness of the SDAMH model in capturing these dynamics. Ultimately, both figures confirm the broader narrative presented in the paper: that the SDAMH model successfully captures the key time-varying nature of both impact and volatility.
 
\newpage

\end{appendices}

\end{document}